\documentclass[journal,12pt,onecolumn,draftclsnofoot]{IEEEtran}
\usepackage{amsmath,amsthm,amssymb}

\usepackage{mathrsfs,mathtools,mathabx}
\usepackage{bm}
\usepackage{graphicx}
\usepackage{dsfont,enumitem,stackrel}
\usepackage{color}
\usepackage{array}
\usepackage{cases}
\usepackage{algorithm,algpseudocode}
\usepackage{algpseudocode}
\usepackage{pgfplots}
\pgfplotsset{compat=newest}
\usepackage{accents}
\usepackage{subfigure}
\usepackage{float}
\usepackage{cite}

\usepackage[most]{tcolorbox}
\usepackage{multicol} 

\usepackage{pgfplotstable}
\usepackage{xparse}
\usepackage{xcolor}

\usepackage{tikz}
\usetikzlibrary{spy}
\usetikzlibrary{arrows,decorations.pathreplacing,patterns, shapes, positioning,calc}
\usepgfplotslibrary{fillbetween}
\usetikzlibrary{intersections}

\usepackage[bookmarks=false, hidelinks]{hyperref}
\usepackage{booktabs} 
\usepackage[nolist,withpage]{acronym}
\usepackage{etoolbox}

\makeatletter
\newif\if@in@acrolist
\AtBeginEnvironment{acronym}{\@in@acrolisttrue}
\newrobustcmd{\LU}[2]{\if@in@acrolist#1\else#2\fi}

\newcommand{\ACF}[1]{{\@in@acrolisttrue\acf{#1}}}

\definecolor{darkcerulean}{rgb}{0.03, 0.27, 0.49}
\definecolor{chestnut}{rgb}{0.8, 0.36, 0.36}
\definecolor{airforceblue}{rgb}{0.36, 0.54, 0.66}
\definecolor{cadmiumorange}{rgb}{0.93, 0.53, 0.18}
\definecolor{bleudefrance}{rgb}{0.19, 0.55, 0.91}
\definecolor{carolinablue}{rgb}{0.6, 0.73, 0.89}
\definecolor{blue(ncs)}{rgb}{0.0, 0.53, 0.74}
\definecolor{brown(web)}{rgb}{0.65, 0.16, 0.16}
\definecolor{pearl}{rgb}{0.94, 0.92, 0.84}
\definecolor{burntumber}{rgb}{0.54, 0.2, 0.14}
\definecolor{asparagus}{rgb}{0.53, 0.66, 0.42}
\definecolor{cssgreen}{rgb}{0.0, 0.5, 0.0}
\definecolor{cadmiumgreen}{rgb}{0.0, 0.42, 0.24}
\definecolor{cadmiumorange}{rgb}{0.93, 0.53, 0.18}
\definecolor{amaranth}{rgb}{0.9, 0.17, 0.31}
\definecolor{bluegray}{rgb}{0.4, 0.6, 0.8}
\definecolor{cadmiumgreen}{rgb}{0.0, 0.42, 0.24}
\definecolor{amaranth}{rgb}{0.9, 0.17, 0.31}
\definecolor{amethyst}{rgb}{0.6, 0.4, 0.8}
\definecolor{amber}{rgb}{1.0, 0.75, 0.0}
\definecolor{azure}{rgb}{0.0, 0.5, 1.0}
\definecolor{babyblue}{rgb}{0.54, 0.81, 0.94}
\definecolor{bazaar}{rgb}{0.6, 0.47, 0.48}
\definecolor{celestialblue}{rgb}{0.29, 0.59, 0.82}
\definecolor{darklavender}{rgb}{0.45, 0.31, 0.59}
\definecolor{bluebell}{rgb}{0.64, 0.64, 0.82}
\definecolor{chamoisee}{rgb}{0.63, 0.47, 0.35}
\definecolor{darkcerulean}{rgb}{0.03, 0.27, 0.49}
\definecolor{iris}{rgb}{0.35, 0.31, 0.81}
\definecolor{dodgerblue}{rgb}{0.12, 0.56, 1.0}
\definecolor{celestialblue}{rgb}{0.29, 0.59, 0.82}
\definecolor{jazzberryjam}{rgb}{0.50, 0.62, 0.37}
\definecolor{cadetgrey}{rgb}{0.57, 0.64, 0.69}

\definecolor{burntsienna}{rgb}{0.81, 0.40, 0.26}
\definecolor{burntumber}{rgb}{0.54, 0.2, 0.14}
\definecolor{bulgarianrose}{rgb}{0.28, 0.02, 0.03}
\definecolor{burgundy}{rgb}{0.5, 0.0, 0.13}
\definecolor{cordovan}{rgb}{0.54, 0.25, 0.27}
\definecolor{eggplant}{rgb}{0.38, 0.25, 0.32}

\definecolor{brickred}{rgb}{0.8, 0.25, 0.33}
\definecolor{chestnut}{rgb}{0.8, 0.36, 0.36}
\definecolor{airforceblue}{rgb}{0.36, 0.54, 0.66}
\definecolor{cadmiumorange}{rgb}{0.93, 0.53, 0.18}
\definecolor{bleudefrance}{rgb}{0.19, 0.55, 0.91}
\definecolor{carolinablue}{rgb}{0.6, 0.73, 0.89}
\definecolor{blue(ncs)}{rgb}{0.0, 0.53, 0.74}
\definecolor{dodgerblue}{rgb}{0.12, 0.56, 1.0}
\definecolor{cssgreen}{rgb}{0.0, 0.5, 0.0}
\definecolor{cadmiumgreen}{rgb}{0.0, 0.42, 0.24}
\definecolor{cadmiumorange}{rgb}{0.93, 0.53, 0.18}
\definecolor{amaranth}{rgb}{0.9, 0.17, 0.31}
\definecolor{bluegray}{rgb}{0.4, 0.6, 0.8}
\definecolor{cadmiumgreen}{rgb}{0.0, 0.42, 0.24}
\definecolor{amber}{rgb}{1.0, 0.75, 0.0}
\definecolor{azure}{rgb}{0.0, 0.5, 1.0}
\definecolor{babyblue}{rgb}{0.54, 0.81, 0.94}
\definecolor{bazaar}{rgb}{0.6, 0.47, 0.48}
\definecolor{celestialblue}{rgb}{0.29, 0.59, 0.82}
\definecolor{darklavender}{rgb}{0.45, 0.31, 0.59}
\definecolor{bluebell}{rgb}{0.64, 0.64, 0.82}
\definecolor{chamoisee}{rgb}{0.63, 0.47, 0.35}
\definecolor{darkcerulean}{rgb}{0.03, 0.27, 0.49}
\definecolor{iris}{rgb}{0.35, 0.31, 0.81}
\definecolor{jazzberryjam}{rgb}{0.65, 0.04, 0.37}
\definecolor{charcoal}{rgb}{0.21, 0.27, 0.31}
\newtheorem{prop}{Proposition}
\newtheorem{theorem}{Theorem}
\newtheorem{remark}{Remark}	
\newtheorem{lem}{Lemma}	
\newtheorem{cor}{Corollary}
\theoremstyle{definition}

\pgfmathdeclarefunction{erfapprox}{1}{%
  \pgfmathparse{%
    (#1)==0
      ? 0
      : (#1/sqrt((#1)^2)) * (1 - exp(-((#1)^2*(4/pi + 0.147*(#1)^2))/(1+0.147*(#1)^2)))
  }%
}

\pgfmathdeclarefunction{mu}{1}{%
  \pgfmathparse{%
    0.9*(1 - erfapprox((#1)/(2*sqrt(2)))) + 
    2.3*(1 - erfapprox((3*#1)/(2*sqrt(2)))) + 
    3.1*(1 - erfapprox((5*#1)/(2*sqrt(2)))) + 
    3.3*(1 - erfapprox((7*#1)/(2*sqrt(2)))) + 
    2.9*(1 - erfapprox((9*#1)/(2*sqrt(2)))) + 
    1.9*(1 - erfapprox((11*#1)/(2*sqrt(2)))) + 
    0.3*(1 - erfapprox((13*#1)/(2*sqrt(2)))) + 
    (-1.9)*(1 - erfapprox((15*#1)/(2*sqrt(2)))) + 
    (-4.7)*(1 - erfapprox((17*#1)/(2*sqrt(2))))
  }%
}

\begin{document}
%
%\onecolumn
% paper title

\title{Towards Optimal Constellation Design for \\
Digital  Over-the-Air Computation}

\author{ Saeed Razavikia,~\IEEEmembership{Member,~IEEE}, Deniz Gündüz,~\IEEEmembership{Fellow,~IEEE}, Carlo Fischione,~\IEEEmembership{Fellow,~IEEE}

 \thanks{S. Razavikia and C. Fischione are with the School of Electrical Engineering and Computer Science, KTH Royal Institute of Technology, Stockholm, Sweden (e-mail: sraz@kth.se, carlofi@kth.se). C. Fischione is also with Digital Futures of KTH. 
 
Deniz Gündüz is with the Department of Electrical and Electronic Engineering, Imperial College London, SW7 2AZ, London, U.K. (e-mail:
d.gunduz@imperial.ac.uk)

%S. Razavikia was supported jointly by the Wallenberg AI, Autonomous Systems and Software Program (WASP), and the Ericsson Research Foundation.

This research was initiated at KTH Royal Institute of Technology and completed at the Information Processing and Communications Laboratory, Imperial College London, during the first author’s visit.

 }
 }

\maketitle

\begin{abstract}
Over-the-air computation (OAC) has emerged as a key technique for efficient function computation over multiple-access channels (MACs) by exploiting the waveform superposition property of the wireless domain.  While conventional OAC methods rely on analog amplitude modulation, their performance is often limited by noise sensitivity and hardware constraints, motivating the use of digital modulation schemes.  This paper proposes a novel digital modulation framework optimized for computation over additive white Gaussian noise (AWGN) channels. 
The design is formulated as an additive mapping problem to determine the optimal constellation that minimizes the mean-squared error (MSE) under a transmit power constraint.  We express the optimal constellation design as a system of nonlinear equations and establish the conditions guaranteeing the uniqueness of its solution.  In the high signal-to-noise-ratio (SNR) regime, we derive closed-form expressions for the optimal modulation parameters using the generalized Lambert function, providing analytical insight into the system’s behavior. Furthermore, we discuss extensions of the framework to higher-dimensional grids corresponding to multiple channel uses, to non-Gaussian noise models, and to computation over real-valued domains via hybrid digital–analog modulation. 
\end{abstract}


%===================================================
\section{Introduction}
%===================================================

Traditionally, communication and computation have been regarded as separate tasks. The rapid growth of data generated by edge devices, often powered by artificial intelligence, poses significant challenges due to the limited processing power and energy of such devices. To overcome this, computation or inference tasks are increasingly offloaded to resource-rich edge servers via wireless links~\cite{Guangxu2020edge,hellstrom2022wireless,gunduz2021communicate,perez2025waveforms}. Rather than viewing interference in multiple-access channels (MACs) as a nuisance, the MAC can be regarded as an analog adder whose waveform superposition naturally forms linear aggregates (e.g., sums/means). This observation leads to over-the-air computation (OAC): edge nodes shape and align their transmissions so the receiver directly estimates the desired function from the superimposed signal~\cite{nazer2007Computation,goldenbaum2014nomographic}. OAC thereby narrows the gap between communication and computation and has enabled diverse applications, including federated learning~\cite{razavikia2024blind,amiri2020federated,amiri2020machine,yang2020federated}, distributed inference~\cite{yilmaz2025private}, on-chip computing~\cite{guirado2023whype}, and wireless control~\cite{park2021optimized}.

Despite its advantages, OAC is mostly constrained to analog amplitude modulation, which limits compatibility with existing wireless hardware and is highly sensitive to noise, fading, and interference~\cite{csahin2023over,csahin2023survey}. Digital modulation, with its superior error detection and correction properties and widespread adoption, offers an appealing alternative.

Recent studies have focused on utilizing digital modulations, such as binary phase-shift keying or frequency-shift keying, to mitigate the limitations of analog OAC~\cite{zhu2020one,csahin2023distributed,csahin2023over,csahin2024over,qiao2024massive}. Many of these methods rely on type-based multiple-access principles, where histograms of received symbols are used to approximate output functions~\cite{mergen2006type,perez2025waveforms,qiao2024massive}. However, most approaches are restricted to specific functions, such as sum or sign computation, and often require trade-offs such as increased bandwidth usage~\cite{csahin2023over}. To address these constraints, the authors in~\cite{saeed2023ChannelComp,razavi2025revisit,razavi2025revisitII} proposed \textit{ChannelComp}, a general framework for arbitrary finite function computation over MACs using digital modulation. This framework extends beyond the sum function, offering compatibility with various digital modulation schemes and supporting a wide range of applications. To further reduce complexity, \textit{SumComp} was introduced in~\cite{Razavikia2024Ring}, which computes sums via integer-based digital modulations. SumComp utilizes a two-dimensional grid representation over the ring of integers, enabling compatibility with legacy constellations such as multi-level quadrature amplitude modulation (QAM) and hexagonal QAM. In \cite{liu2025digital,xie2023joint,yan2025remac}, channel coding has been incorporated to enhance reliability.

However, these studies essentially try to adapt or optimize certain common constellations for the computation task. The fundamental question of the optimal constellation design for a given computation task remains an open problem. In conventional communication systems, designing optimal constellations is a well-studied problem~\cite{forney1998modulation,makowski2006optimality,zhang2019constellation}. However, to the best of our knowledge, a systematic approach to designing optimal constellations for OAC, where the objective is function computation rather than message transmission, remains largely unexplored.  In this paper, we take a step toward identifying the optimal constellation design for OAC. Utilizing a flexible two-dimensional structured grid framework, we aim to design constellations that minimize the mean-squared error (MSE) of function computation under an additive noisy MAC. To this end, we formulate an optimization problem that minimizes the MSE under an average power constraint on the constellation diagram. The resulting optimality conditions yield a system of nonlinear equations. To ensure uniqueness and obtain the optimal solution to the proposed optimization problem, we analyze the geometry of this system of equations and establish conditions that define the optimality regions. Subsequently, the optimal solution is guaranteed by numerically solving the system.

For scenarios with a high signal-to-noise ratio (SNR), we derive a closed-form solution for the modulation parameters using a generalized Lambert function. This analytical result offers insights into the structure and behavior of the optimal constellation diagram. Furthermore, we extend the discussion to demonstrate how this scheme can be generalized to an $N$-dimensional grid configuration and other noise distributions, such as the Cauchy distribution. In addition, we show that the proposed scheme enables computation over real-valued domains through a hybrid digital–analog modulation approach. Numerical experiments validate the superiority of the proposed constellation design over standard modulation schemes and confirm the accuracy of the analytical solutions.

In short, our contributions can be summarized as follows:

\begin{itemize}
    \item \textbf{Problem formulation:}  Given the modulation order and SNR, we determine the optimal constellation points for computing sums over AWGN channels. Unlike prior works that adapt standard modulations (e.g., QAM) for OAC, we formulate this task as an explicit optimization problem over a structured two-dimensional grid, thereby enabling systematic design of optimal constellations for computation rather than communication. 
    \item \textbf{Optimal parameters of the proposed solution:} We use both the maximum likelihood~(ML) and maximum a posteriori~(MAP) criteria to estimate the sum of the transmitted signals.  Then, we derive the MSE value for the given estimator and determine the optimality conditions for the encoder parameters. The geometry of the resulting nonlinear system of equations is analyzed to establish the uniqueness of the optimal solution. 
    \item \textbf{Closed-form solution:} For high SNR scenarios, a closed-form solution is derived using a generalized Lambert function, offering analytical insight into the structure of the optimal design. We further show that this closed-form solution asymptotically converges to the numerically obtained optimal solution in the high SNR region.
    \item \textbf{Extensions:} We also provide a brief discussion on how to generalize the proposed framework to an $N$-dimensional grid and other noise distributions such as the Cauchy distribution. This can help extend the proposed framework to compute over multiple communication resources, such as frequency or time.  Furthermore, we introduce a hybrid digital–analog modulation scheme that enables computation over real-valued domains—representing, to the best of our knowledge, the first hybrid modulation design developed specifically for OAC. 
    \item \textbf{Numerical validation:} We validate all the analytical results through numerical experiments.  We compare the performance of the proposed constellation diagrams with standard QAM modulations. Numerical experiments demonstrate the superiority of the proposed constellation design and validate the accuracy of the analytical derivations.
\end{itemize}

% -----------------------------------
\subsection{Paper Organization}
% -----------------------------------

The signal model, problem statement, and the encoding and decoding procedures are detailed in Section~\ref{sec:system}. In Section~\ref{sec:MainResults},  the main analytical results, including the optimal modulation constellations for ML and MAP decoders, are presented. Numerical results evaluating the performance of optimal constellation diagrams are presented in Section~\ref{sec:computaion}. Extensions to heavy-tailed noise and $N$-dimensional grids are presented in Section~\ref{sec:general}, and the paper concludes with final remarks in Section~\ref{sec:conclusion}.

%---------------------------------------------------
\subsection{Notation}
%---------------------------------------------------

 The sets of integers, real numbers, and complex numbers are denoted by \(\mathbb{Z}\), \(\mathbb{R}\), and \(\mathbb{C}\), respectively. Also, $\mathbb{Z}^{+}$, $\mathbb{R}^{+}$ denote the set of positive integers and positive real numbers, respectively.  For an integer $n$, $[n]$ represents the set $\{1,2,\dots, n\}$. $\mathbb{Z}_Q$ denotes a finite set of $Q$ integers, i.e., $\mathbb{Z}_Q = \{0,1,2,\ldots,Q-1\}$.  Scalars are represented by lowercase letters, such as $x$, while operators are indicated using calligraphic notation, e.g., $\mathscr{X}$.  For a complex scalar $x\in \mathbb{C}$, $\mathfrak{Re}({x})$ and  $\mathfrak{Im}({x})$ denote the real and imaginary parts of $x$, respectively. The notation \(\mathcal{CN}(0, \sigma^2)\) signifies a zero-mean circularly symmetric complex Gaussian distribution, where both the real and imaginary components follow a Gaussian distribution with variance \(\sigma^2/2\), denoted by \(\mathcal{N}(0, \sigma^2/2)\).

%======================
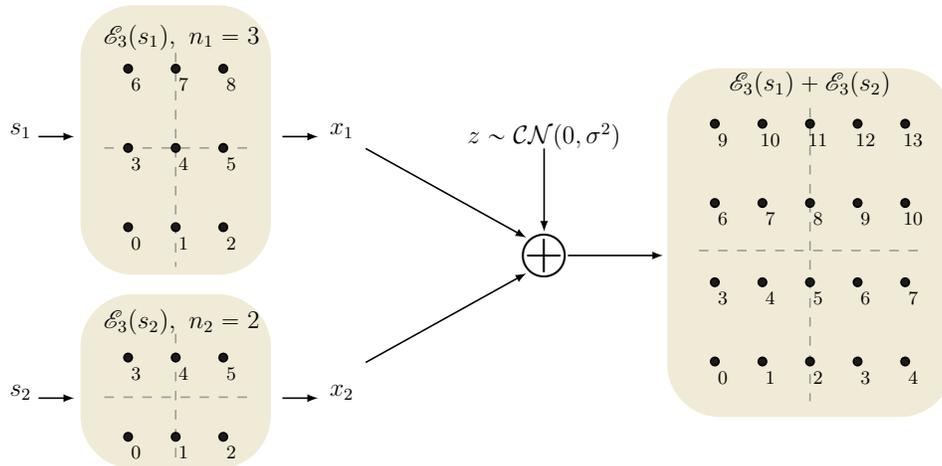
\begin{figure}[t]
    \centering 
\label{fig:system}
\definecolor{eggshell}{rgb}{0.94, 0.92, 0.84}
\definecolor{flavescent}{rgb}{0.97, 0.91, 0.56}
\definecolor{lightapricot}{rgb}{0.99, 0.84, 0.69}
\definecolor{peach-orange}{rgb}{1.0, 0.8, 0.6}
\definecolor{pearl}{rgb}{0.94, 0.92, 0.84}
\definecolor{peach-yellow}{rgb}{0.98, 0.87, 0.68}
\tikzset{every picture/.style={line width=0.75pt}} %set default line width to 0.75pt        
\scalebox{0.8}{
    \begin{tikzpicture}[x=0.75pt,y=0.75pt,yscale=-1,scale=0.75]

\begin{scope}[shift={(-5cm,-3cm)},]

\def\ncols{3}
\def\nrows{3}

% Origin offset (matches node placement above)
\newcommand{\xoffset}{240pt}
\newcommand{\yoffset}{25pt}
    
\newcommand{\dx}{30pt}  % Horizontal spacing
\newcommand{\dy}{50pt}  % Vertical spacing

% Calculate width and height
\pgfmathsetlengthmacro{\totalwidth}{(\ncols - 1)*\dx}
\pgfmathsetlengthmacro{\totalheight}{(\nrows - 1)*\dy}

% Draw background rectangle
\pgfmathsetlengthmacro{\xstart}{\xoffset - 30pt}
\pgfmathsetlengthmacro{\xend}{\xoffset + \totalwidth + 30pt}
\pgfmathsetlengthmacro{\ystart}{\yoffset }
\pgfmathsetlengthmacro{\yend}{\yoffset + \totalheight + 55pt}
\draw[draw opacity=0, fill=pearl, rounded corners=25pt] 
    (\xstart,\yend) rectangle (\xend,\ystart-15pt);

% Dashed horizontal axis (middle row)
\pgfmathsetlengthmacro{\ymid}{\yoffset + \dy*(\nrows - 1 - 1.5)} % center between rows 1 and 2
\draw [color={rgb, 255:red, 155; green, 155; blue, 155 }, dash pattern=on 4.5pt off 4.5pt]
    (\xoffset - 15pt,\ymid+0.5*\totalwidth +20pt) -- (\xoffset + \totalwidth + 15pt,\ymid+0.5*\totalwidth +20pt);

% Dashed vertical axis (middle column)
\pgfmathsetlengthmacro{\xmid}{\xoffset + \dx*2} % middle column (index 2)
\draw [color={rgb, 255:red, 155; green, 155; blue, 155 }, dash pattern=on 4.5pt off 4.5pt]
    (\xmid-0.5*\totalheight+20pt ,\yoffset+15pt) -- (\xmid-0.5*\totalheight+20pt,\yoffset + \totalheight + 50pt);

\foreach \y [count=\row from 0] in {0,1,2} {
    \foreach \x [count=\col from 0] in {0,1,2} {
        % Compute label as col + 3*row
        \pgfmathtruncatemacro{\label}{\col + 3*\row}
        
        % Compute positions
        \pgfmathsetlengthmacro{\Xcoord}{240pt + \x*\dx}
        \pgfmathsetlengthmacro{\Ycoord}{(3-\row)*\dy} % Reverse row indexing if top-down

        \draw[fill=black!90] (\Xcoord,\Ycoord) node{} circle (3.5);
        \pgfmathsetlengthmacro{\LabelY}{\Ycoord + 10pt}
        \pgfmathsetlengthmacro{\LabelX}{\Xcoord + 5pt}
        \draw (\LabelX,\LabelY) node{\footnotesize $\label$};
    }
}

\draw (365,40) node   {$\mathscr{E}_3(s_1), ~n_1=3$};
\end{scope}

\begin{scope}[shift={(-5cm,2cm)},]

\def\ncols{3}
\def\nrows{2}

% Origin offset (matches node placement above)
\newcommand{\xoffset}{240pt}
\newcommand{\yoffset}{55pt}
    
\newcommand{\dx}{30pt}  % Horizontal spacing
\newcommand{\dy}{50pt}  % Vertical spacing

% Calculate width and height
\pgfmathsetlengthmacro{\totalwidth}{(\ncols - 1)*\dx}
\pgfmathsetlengthmacro{\totalheight}{(\nrows - 1)*\dy}

% Draw background rectangle
\pgfmathsetlengthmacro{\xstart}{\xoffset - 30pt}
\pgfmathsetlengthmacro{\xend}{\xoffset + \totalwidth + 30pt}
\pgfmathsetlengthmacro{\ystart}{\yoffset }
\pgfmathsetlengthmacro{\yend}{\yoffset + \totalheight + 55pt}
\draw[draw opacity=0, fill=pearl, rounded corners=25pt] 
    (\xstart,\yend) rectangle (\xend,\ystart-5pt);

% Dashed horizontal axis (middle row)
\pgfmathsetlengthmacro{\ymid}{\yoffset + \dy*(\nrows - 1 - 1.5)} % center between rows 1 and 2
\draw [color={rgb, 255:red, 155; green, 155; blue, 155 }, dash pattern=on 4.5pt off 4.5pt]
    (\xoffset - 15pt,\ymid+0.5*\totalwidth +55pt) -- (\xoffset + \totalwidth + 15pt,\ymid+0.5*\totalwidth +55pt);

% Dashed vertical axis (middle column)
\pgfmathsetlengthmacro{\xmid}{\xoffset + \dx*2} % middle column (index 2)
\draw [color={rgb, 255:red, 155; green, 155; blue, 155 }, dash pattern=on 4.5pt off 4.5pt]
    (\xmid-0.5*\totalheight -5pt,\yoffset+20pt) -- (\xmid-0.5*\totalheight-5pt,\yoffset + \totalheight + 50pt);

\foreach \y [count=\row from 0] in {0,1} {
    \foreach \x [count=\col from 0] in {0,1,2} {
        % Compute label as col + 3*row
        \pgfmathtruncatemacro{\label}{\col + 3*\row}
        
        % Compute positions
        \pgfmathsetlengthmacro{\Xcoord}{240pt + \x*\dx}
        \pgfmathsetlengthmacro{\Ycoord}{(3-\row)*\dy-10pt} % Reverse row indexing if top-down

        \draw[fill=black!90] (\Xcoord,\Ycoord) node{} circle (3.5);
        \pgfmathsetlengthmacro{\LabelY}{\Ycoord + 10pt}
        \pgfmathsetlengthmacro{\LabelX}{\Xcoord + 5pt}
        \draw (\LabelX,\LabelY) node{\footnotesize $\label$};
    }
}

\draw (365,90) node   {$\mathscr{E}_3(s_2),~n_2=2$};
\end{scope}

% \draw (260pt, 115pt) node {\LARGE $\bigoplus$};

% \draw [-latex]  (470pt,115pt) -- (490pt,115pt) ;

\begin{scope}[shift={(8cm,-0.9cm)}]

% Number of columns and rows
\def\ncols{5}
\def\nrows{4}

% Origin offset (matches node placement above)
\newcommand{\xoffset}{240pt}
\newcommand{\yoffset}{25pt}
    
\newcommand{\dx}{30pt}  % Horizontal spacing
\newcommand{\dy}{50pt}  % Vertical spacing

% Calculate width and height
\pgfmathsetlengthmacro{\totalwidth}{(\ncols - 1)*\dx}
\pgfmathsetlengthmacro{\totalheight}{(\nrows - 1)*\dy}

% Draw background rectangle
\pgfmathsetlengthmacro{\xstart}{\xoffset - 30pt}
\pgfmathsetlengthmacro{\xend}{\xoffset + \totalwidth + 30pt}
\pgfmathsetlengthmacro{\ystart}{\yoffset - 20pt}
\pgfmathsetlengthmacro{\yend}{\yoffset + \totalheight + 35pt}
\draw[draw opacity=0, fill=pearl, rounded corners=25pt] 
    (\xstart,\yend) rectangle (\xend,\ystart-15pt);

% Dashed horizontal axis (middle row)
\pgfmathsetlengthmacro{\ymid}{\yoffset + \dy*(\nrows - 1 - 1.5)} % center between rows 1 and 2
\draw [color={rgb, 255:red, 155; green, 155; blue, 155 }, dash pattern=on 4.5pt off 4.5pt]
    (\xoffset - 10pt,\ymid+5pt) -- (\xoffset + \totalwidth + 10pt,\ymid+5pt);

% Dashed vertical axis (middle column)
\pgfmathsetlengthmacro{\xmid}{\xoffset + \dx*2} % middle column (index 2)
\draw [color={rgb, 255:red, 155; green, 155; blue, 155 }, dash pattern=on 4.5pt off 4.5pt]
    (\xmid,\yoffset - 10pt) -- (\xmid,\yoffset + \totalheight + 25pt);

\foreach \y [count=\row from 0] in {0,1,2,3} {
    \foreach \x [count=\col from 0] in {0,1,2,3,4} {
        % Compute label as col + 3*row
        \pgfmathtruncatemacro{\label}{\col + 3*\row}
        
        % Compute positions
        \pgfmathsetlengthmacro{\Xcoord}{240pt + \x*\dx}
        \pgfmathsetlengthmacro{\Ycoord}{25pt + (3-\row)*\dy} % Reverse row indexing if top-down

        \draw[fill=black!90] (\Xcoord,\Ycoord) node{} circle (3.5);
        \pgfmathsetlengthmacro{\LabelY}{\Ycoord + 10pt}
        \pgfmathsetlengthmacro{\LabelX}{\Xcoord + 5pt}
        \draw (\LabelX,\LabelY) node{\footnotesize $\label$};
    }
}

\draw (400,0) node   {$\mathscr{E}_3(s_1)+\mathscr{E}_3(s_2)$};
\end{scope}

\draw [-latex] (55, 10)  -- (85,10);
\draw [-latex] (260, 10)  -- (290,10);
%\draw (300, -40) node { $p_1$};%
%\draw [-latex]  (300,-30) -- (300,-5) ;%
%\draw (300, 10) node { $\bigotimes$};
%\draw [-latex] (310, 10)  -- (340,10);
\draw (310, 5) node {$x_1$};
\draw (40, 5) node {$s_1$};

\draw (480,110) node {\LARGE $\bigoplus$};
\draw[-latex]    (330,20) -- (465,95) ;
\draw[-latex]    (330,200) -- (465,125) ;

\draw [-latex] (55, 230)  -- (85,230);
%\draw (300, 180) node { $p_2$};
%\draw [-latex]  (300,190) -- (300,215) ;
%\draw (300, 230) node { $\bigotimes$};
\draw [-latex] (260, 230)  -- (290,230);
\draw (310, 225) node {$x_2$};
\draw (40, 225) node {$s_2$};

\draw[-latex]    (500,110) -- (580,110) ;
\draw[-latex]    (480,20) -- (480,90) ;
\draw (480,10) node {$z\sim \mathcal{CN}(0,\sigma^2)$};
    \end{tikzpicture}}

\caption{Illustration of the proposed encoding structure for digital OAC with $K=2$ nodes. The two grids on the left represent the individual encoder mappings \( \mathscr{E}_3(s_1) \) and \( \mathscr{E}_3(s_2) \), with modulation levels  \((q_1,n_1) = (3,3) \) and \( (q_2,n_1) = (3,2) \). Each node maps its input symbol \( s_k \in \{0, \dots, qn_k - 1\} \) onto a structured two-dimensional grid in the complex plane. The rightmost grid depicts the superimposed constellation diagram resulting from the addition \( \mathscr{E}_3(s_1) + \mathscr{E}_3(s_2) \) at the CP.}

    \label{fig:systemconst}
\end{figure}
%======================

\section{Problem Formulation}\label{sec:system}

\subsection{Channel Model}

We consider a MAC consisting of $K$ transmitter nodes and a single receiver node, referred to as the computation point (CP).  Each transmitter observes a random variable $s_k\in \mathbb{Z}_{Q_k}$, where $Q_k$ denotes the size of alphabet $\mathbb{Z}_{Q_k}$\footnote{For clarity of presentation, the proposed scheme is first described over the integer domain. However, the framework is not limited to discrete alphabets; as shown in Section~\ref{sec:hybrid}, it can be naturally extended to the real domain using hybrid digital–analog modulation.}.  The symbols $\{s_k\}_{k=1}^K$ are assumed mutually independent across both nodes and channel uses, with each uniformly distributed over $\mathbb{Z}_{Q_k}$. Transmitters transmit simultaneously over a MAC to enable the CP to compute a predetermined function $f(s_1,\ldots,s_K)$.  We consider $f$ to be the sum function throughout the paper, i.e.,  
%---------------
\begin{align}
    f(s_1,\ldots,s_K) = \sum_{k=1}^{K}s_k. 
\end{align}
%---------------
An encoding function $\mathscr{E}_q(\cdot)$ with parameter $q\in \mathbb{Z}^{+}$, maps $s_k$ to a channel symbol $x_k$, i.e., $x_k = \mathscr{E}_q(s_k)$\footnote{Because the sum is permutation-invariant, an identical encoder is employed across all transmitters~\cite{saeed2023ChannelComp}. }. We assume that $x_k$ is restricted to a discrete set of complex numbers with cardinality $Q_k$. For the sake of simplicity, we assume $Q_k$ is divisible by the parameter $q$ for all $[K]$, i.e., $Q_k= n_k \times q$ for some positive integer $n_k$\footnote{This structure induces a rectangular grid in the constellation, facilitating a simple expression for the energy of the constellation points, i.e., \eqref{eq:energy_conste}, and yielding more interpretable analytical results.}. In particular, we enforce that $x_k\in C:=\{c_1,\ldots,c_{Q_k}\}$, where $c_m \in \mathbb{C}$ are complex numbers and $\sum_{m=1}^{Q_k} |c_{m}|^2/Q_k\leq P$ for all $k\in [K]$. This restricts the power of each transmitted channel symbol to be less than or equal to $P$. Subsequently, all nodes transmit their modulated signals through the shared communication channel. Then, the CP receives  
%---------------
\begin{align}
{r} = \sum_{k=1}^{K}{x}_k + {z},
\end{align}
%---------------
where  ${r}$ represents the superimposed signal under the assumption of synchronous transmission, and ${z}$ represents the circularly symmetric Gaussian noise term with zero mean and variance ${\sigma}^2$, i.e., $z \sim \mathcal{CN}(0,{\sigma}^2)$. In addition to the Gaussian model, other noise distributions, such as heavy-tailed (e.g., Cauchy) noise, are also considered in Section~\ref{sec:heavy-noise}. Note that imperfect synchronization among the transmitters at the receiver can be handled using existing techniques, e.g.~\cite{razavikia2022blind,Hellstrom2023Filter,sahin2025feasibility}. The CP then estimates the desired function value using a decoding function  $\mathscr{D}:\mathbb{C} \mapsto \mathcal{Y}_f$:
%---------------
\begin{align}
\hat{f} := \mathscr{D}(r),
\end{align}
%---------------
 where $\mathcal{Y}_f$ denotes the output alphabet for $f$.

\subsection{Problem Statement}

The goal of this paper is to design an optimal pair of encoding function  $\mathscr{E}_q(\cdot)$ and decoding function $\mathscr{D}(\cdot)$ that jointly minimize the MSE between the true function value $f$ and its estimate $\hat{f}$ at the CP. Specifically, let us define the MSE 
%-------------
\begin{align}
 \label{eq:MSEdefinition}
\text{MSE}(\hat{f}, f) = \mathbb{E}_{s_1,\ldots,s_K, z}\Bigg[\bigg(\underbrace{\mathscr{D}\Big( \sum\nolimits_{k=1}^{K} \mathscr{E}_q(s_k) + z \Big)}_{=\hat{f}} - f(s_1, \dots, s_K)\bigg)^2\Bigg],
\end{align}
%-------------
where the expectation is over the randomness of the input values $s_1,\ldots,s_K$ and the channel noise $z$. Hence, we aim to determine $(\mathscr{E}_q(\cdot), \mathscr{D}(\cdot))$ such that the expected squared error between $f(s_1, \dots, s_K)$ and $\hat{f}$ is minimized, i.e., 
%-------------
\begin{align}
\min_{\mathscr{E}_q, \mathscr{D}} ~ \text{MSE}(\hat{f}, f),~~~~\text{s.t.}, ~~  |\mathscr{E}_q(s_k)|^2 \leq P,\quad \forall s_k\in \mathbb{Z}_{Q_k}, ~ Q_k\in \mathbb{Z}^{+}. \label{eq:OptmialMSEgeneral}
\end{align}
%-------------
 The solution to this optimization problem depends on the structure of the encoder $\mathscr{E}_q$  and the decoder $\mathscr{D}$ at the CP, described in the subsequent subsections.

\subsection{Encoding}\label{sec:encoding}

Solving the general optimization problem in \eqref{eq:OptmialMSEgeneral} for all possible functions is challenging. Therefore, we focus on the sum function, which can be straightforwardly extended to the class of \emph{nomographic} functions by employing  pre- and post-processing functions~\cite{goldenbaum2014nomographic}\footnote{The Kolmogorov–Arnold representation theorem ensures that any continuous multivariate function can be expressed as a superposition of univariate functions and addition~\cite{kolmogorov1963representation}. However, the result is non-constructive and does not, in general, provide a decomposition in the strict nomographic form $\psi\!\left(\sum_k \phi_k(x_k)\right)$ for arbitrary functions.}.  We require that $\mathscr{E}_q(s)$ and its input value $s$ form an additive group for computing the sum function. In this regard, for any number $s\in \mathbb{Z}_{Q_k}$, consider the base-$q$ quotient–remainder decomposition
%------------
\begin{align}
    s = c_1+c_2q,
\end{align}
%------------
where $(c_1,c_2)\in \mathbb{Z}^{2}$, $c_1 =s - q\cdot \left\lfloor{s}/{{q}}\right\rfloor $ and $c_2 =\left\lfloor{s}/{{q}} \right\rfloor$.  We propose the encoding function $\mathscr{E}_{q}(\cdot)$ for node $k$ as follows:
%------------
\begin{align}
 \label{eq:encoding_qam}
\mathscr{E}_{q}(s_k) := c_1 d_1  +  c_2d_2 i + \chi_k,\quad d_1,d_2\in \mathbb{R}^{+}, ~\forall s_k\in \mathbb{Z}_{Q_k},
\end{align}
%------------
where $i$ is the imaginary unit, the parameters $d_1$ and $d_2$ are positive real values that specify the spacing between constellation points along the real and imaginary axes. Constant \( \chi_k \in \mathbb{C} \) shifts the constellation to the origin, thereby enforcing symmetry with respect to the origin and improving average power efficiency. At the receiver, the aggregate offset $\sum_k\chi_k$ is subtracted before decoding. 

The encoding function in \eqref{eq:encoding_qam} extends the SumComp code proposed in \cite{Razavikia2024Ring} to a general two-dimensional grid parameterized by $(d_1,d_2)$.  This encoding scheme maps a one-dimensional input value $s_k$ onto a two-dimensional complex plane (i.e., constellation points), where the real and imaginary parts correspond to the in-phase and quadrature components of the modulated signal, respectively. Figure~\ref{fig:systemconst} illustrates both the constellation diagram of the encoded symbols and the resulting superimposed constellation at the receiver, which directly enables computation of the sum function.

\begin{remark}
In \eqref{eq:encoding_qam}, the decomposition of $s$ into only two integer components is deliberate, corresponding to the in-phase and quadrature components in complex baseband modulation.  The proposed framework naturally extends to $N$-dimensional mappings, as discussed in Section~\ref{sec:higherN}. 
\end{remark}

\begin{remark}\label{rem:rational}
The formulation in \eqref{eq:encoding_qam} assumes an integer base \( q \) for simplicity.  However, the framework can be extended by expressing \( s \) as a linear combination of two bases, i.e., \( s = c_1 q_1 + c_2 q_2 \), where \( q_1 \) and \( q_2 \) are co-prime integers. Under this formulation, the base \( q \) can be interpreted as a rational number whose ratio is given by \( q = q_1/q_2 \).  In this case, the encoding structure remains the same—the transmitter still maps \( c_1 \) and \( c_2 \) onto the real and imaginary axes, respectively—but the coefficients \( (c_1, c_2) \) must be computed according to the generalized decomposition rule described in~\cite{Razavikia2024Ring}. 
\end{remark}

\begin{remark}\label{rem:addative}
    For every input data $s_k$ at node $k$,  $(s_k,\mathscr{E}_q(s_k))$ form an additive group~\cite{Razavikia2024Ring}. Accordingly, the summation $\sum_{k=1}^Ks_k$ with corresponding pair $\sum_{k=1}^K\mathscr{E}_q(s_k)$ can be uniquely decoded.  Specifically, using a proper isomorphic decoding scheme $\mathscr{D}$, we can assign $\sum_k\mathscr{E}_q(s_k)$ to  $f = \sum_ks_k$ and compute the desired sum function. 
\end{remark}

As a result of Remark~\ref{rem:addative}, the class of nomographic functions~\cite{goldenbaum2014nomographic} can be computed using $\mathscr{E}_{q}$ and $\mathscr{D}$, by adding pre-processing and post-processing functions to the system model. 

%---------------------
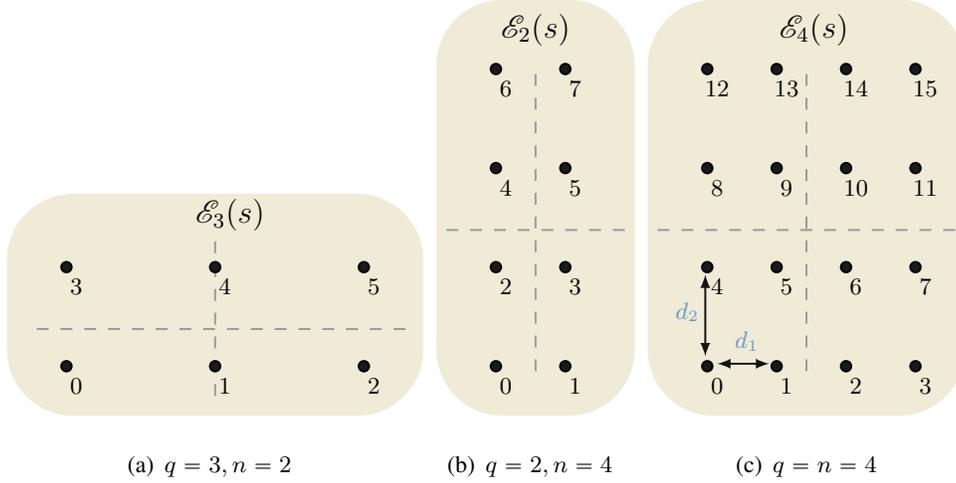
\begin{figure}[t]
    \centering

\definecolor{eggshell}{rgb}{0.94, 0.92, 0.84}
\definecolor{flavescent}{rgb}{0.97, 0.91, 0.56}
\definecolor{lightapricot}{rgb}{0.99, 0.84, 0.69}
\definecolor{peach-orange}{rgb}{1.0, 0.8, 0.6}
\definecolor{pearl}{rgb}{0.94, 0.92, 0.84}
\definecolor{peach-yellow}{rgb}{0.98, 0.87, 0.68}
\tikzset{every picture/.style={line width=0.75pt}} %set default line width to 0.75pt        
\subfigure[$q=3, n=2$]{\label{fig:GrayCode(a)}
    \begin{tikzpicture}[x=0.75pt,y=0.75pt,yscale=-1,scale=0.75]

\draw[draw opacity=0][fill=pearl, rounded corners=25pt] (210pt, 160pt) rectangle (420pt, 48pt) {};

 \draw [color={rgb, 255:red, 155; green, 155; blue, 155 } ] [dash pattern={on 4.5pt off 4.5pt}]  (300,155) -- (550,155) ;
% %Straight Lines 
\draw [color={rgb, 255:red, 155; green, 155; blue, 155 }] [dash pattern={on 4.5pt off 4.5pt}]  (420,200) -- (420,95) ;

      % draw 3 rows (j=1..3), 2 columns (i=1..2)
 \newcounter{za}\setcounter{za}{0}

\foreach \y in {110pt,60pt} {
            
         % \draw[-latex,color={rgb, 255:red, 74; green, 144; blue, 226 }] (375,\y+25pt) -- (530,\y+25pt);
           \foreach \x in {0pt,75pt,150pt} {
            
          \draw[fill=black!90] (240pt+\x,25pt+\y) node{}  circle  (3.5);
           
           \draw (245pt+\x,35pt+\y) node{\footnotesize $\theza$} ;
           \stepcounter{za};
          }
}
\draw (430,78) node   {$\mathscr{E}_3(s)$};
      % spacing arrows
      % \draw[<->] (1cm,0.5cm) -- (2cm,0.5cm) node[midway,below,font=\footnotesize] {$2d$};
      % \draw[<->] (0.5cm,1cm) -- (0.5cm,3cm) node[midway,left,font=\footnotesize] {$3d$};
    \end{tikzpicture}}\subfigure[$q=2, n=4$]{\label{fig:GrayCode(b)}
    \begin{tikzpicture}[x=0.75pt,y=0.75pt,yscale=-1,scale=0.75]

\draw[draw opacity=0][fill=pearl, rounded corners=25pt] (250pt, 210pt) rectangle (350pt, 0) {};

\draw [color={rgb, 255:red, 155; green, 155; blue, 155 } ] [dash pattern={on 4.5pt off 4.5pt}]  (340,155) -- (460,155) ;
%Straight Lines 
\draw [color={rgb, 255:red, 155; green, 155; blue, 155 }] [dash pattern={on 4.5pt off 4.5pt}]  (400,250) -- (400,45) ;

      % draw 3 rows (j=1..3), 2 columns (i=1..2)
 \newcounter{ca}\setcounter{ca}{0}

\foreach \y in {160pt,110pt,60pt,10pt} {
            
         % \draw[-latex,color={rgb, 255:red, 74; green, 144; blue, 226 }] (375,\y+25pt) -- (530,\y+25pt);
           \foreach \x in {40pt,75pt} {
            
          \draw[fill=black!90] (240pt+\x,25pt+\y) node{}  circle  (3.5);
           
           \draw (245pt+\x,35pt+\y) node{\footnotesize $\theca$} ;
           \stepcounter{ca};
          }
}
\draw (400,20) node   {$\mathscr{E}_2(s)$};
      % spacing arrows
      % \draw[<->] (1cm,0.5cm) -- (2cm,0.5cm) node[midway,below,font=\footnotesize] {$2d$};
      % \draw[<->] (0.5cm,1cm) -- (0.5cm,3cm) node[midway,left,font=\footnotesize] {$3d$};
    \end{tikzpicture}}\subfigure[$q=n=4$]{\label{fig:GrayCode(c)}
\begin{tikzpicture}[x=0.75pt,y=0.75pt,yscale=-1, scale=0.75]

\draw[draw opacity=0][fill=pearl, rounded corners=25pt] (250pt, 210pt) rectangle (410pt, 0) {};

\draw [color={rgb, 255:red, 155; green, 155; blue, 155 } ] [dash pattern={on 4.5pt off 4.5pt}]  (340,155) -- (540,155) ;
%Straight Lines 
\draw [color={rgb, 255:red, 155; green, 155; blue, 155 }] [dash pattern={on 4.5pt off 4.5pt}]  (440,250) -- (440,45) ;

% Text Node
\draw (445,20) node   {$\mathscr{E}_4(s)$};

\newcounter{wa}\setcounter{wa}{0}

\foreach \y in {160pt,110pt,60pt,10pt} {
            
         % \draw[-latex,color={rgb, 255:red, 74; green, 144; blue, 226 }] (375,\y+25pt) -- (530,\y+25pt);
           \foreach \x in {40pt,75pt,110pt,145pt} {
            
          \draw[fill=black!90] (240pt+\x,25pt+\y) node{}  circle  (3.5);
           
           \draw (245pt+\x,35pt+\y) node{\footnotesize $\thewa$} ;
           \stepcounter{wa};
          }
}

 \draw[latex-latex, color={rgb, 255:red, 20; green, 20; blue, 20 }] (380,245) -- (415,245);
\draw (400,230) node [font=\footnotesize]  {\color{bluegray} $d_1$};

\draw[latex-latex, color={rgb, 255:red, 20; green, 20; blue, 20 }] (372,240) -- (372,185);
\draw (360,210) node [font=\footnotesize]  {\color{bluegray}$d_2$};

\end{tikzpicture}
}

\caption{Examples of encoded constellation diagrams \( \mathscr{E}_q(s) \) for various combinations of modulation order \( Q \) and in-phase and quadrature levels $q$ and $n$, respectively. Each diagram represents the mapping of the input symbol \( s \in \{0,1,\ldots,q\times n-1\} \) to a multidimensional coordinate space, illustrating how constellation points are arranged. Subfigure \ref{fig:GrayCode(a)} shows \( q=3, n=2 \); \ref{fig:GrayCode(b)} depicts \( q=2, n=4 \); and \ref{fig:GrayCode(c)} illustrates \( q=4, n=4 \) using a 16-QAM constellation with labeled spacing parameters \( d_1 \) and \( d_2 \).}
    \label{fig:GrayCode}
\end{figure}
%---------------------

The encoding function divides the integer input into discrete segments and assigns each segment a corresponding complex value in the constellation plane.  The average power of such a two-dimensional grid can be expressed as \cite{goldsmith2005wireless}:
%----------
\begin{align}
\label{eq:energy_conste}
\mathbb{E}_{s_k}\big[|\mathscr{E}_q(s_k)|^2\big] = \frac{q^2-1}{12}d_1^2 + \frac{n_k^2-1}{12}d_2^2,
\end{align}
%----------
where $Q_k=q\times n_k$ for $k\in [K]$. Figure~\ref{fig:GrayCode} depicts the coded modulation diagram using the encoder in \eqref{eq:encoding_qam} for $q\in \{2,3,4\}$ with different values of $n$. 

\subsection{Decoding Using Maximum Likelihood (ML) Estimator}\label{sec:MLdecoder-Est}

To estimate the function value $\hat{f}$ from the received signal $r$,  we first need to map $r \in \mathbb{C}$ back to the superimposed symbol grid, which includes two-dimensional rectangular grid constellation points of size $N_{1, K} \times N_{2, K}$, where $N_{1, K} = (q-1)K+1, N_{2, K}=\sum_{k}n_k$. We define the set  $\mathcal{Y}$ containing all constellation points with cardinality $|\mathcal{Y}| = N_K$, where $N_K=N_{1,K} \times N_{2,K}$.
Then, we can use the maximum likelihood estimator, i.e., $\mathscr{D}: \mathbb{C} \mapsto \mathcal{Y}$, 
% -----------
\begin{align}
    \label{eq:Quntizer-ml}
    \mathscr{D}(r) =  \underset{y_j \in \mathcal{Y}}{\arg\max}~\mathrm{g}(r|y_j),
\end{align}
% -----------
where $\mathrm{g}(r|y_j)$ denotes the conditional channel transition probability, i.e.,  $$ \mathrm{g}(r|y) = { \exp{\Big(-\tfrac{\|r-y\|^2}{\sigma^2}}\Big)}/{\sigma\sqrt{\pi}},$$ due to the Gaussian distribution of the channel noise~\cite{saeed2023ChannelComp}. Since the real and imaginary components of the noise $z$ are independent, we can treat each component independently and decouple the estimator into two independent tasks. Consequently, the expression in \eqref{eq:Quntizer-ml} generates detection regions for the set
of all possible constellation points, $y_1, \ldots, y_{N_K}$ with the
corresponding regions $\{\mathcal{V}_1^{Re}, \ldots, \mathcal{V}^{Re}_{N_{1,K}}\}$ for the real components and $\{\mathcal{V}_1^{Im}, \ldots, \mathcal{V}^{Im}_{N_{2,K}}\}$, for the imaginary components,  where 
\begin{align}
     \nonumber
    \mathcal{V}_i^{Re}& := \{y \in \mathbb{R}|\mathrm{g}\big(y|\mathfrak{Re}(y_i)\big) \geq \mathrm{g}\big(y|\mathfrak{Re}(y_m)\big), \forall~y_m \in \mathcal{Y}\}, \\ \nonumber
    \mathcal{V}_j^{Im}& := \{y \in \mathbb{R}|\mathrm{g}\big(y|\mathfrak{Im}(y_j)\big) \geq \mathrm{g}\big(y|\mathfrak{Im}(y_m)\big), \forall~y_m \in \mathcal{Y}\},
\end{align}
for $i\in [N_{1,K}]$ and $j\in [N_{2,K}]$. Accordingly, the detection regions become 
% -----------
\begin{subequations}
\label{eq:Voronoi_RE}
    \begin{align}
    \mathcal{V}_1^{Re} = &\Big\{y\in \mathbb{R}| y \leq \frac{d_1}{2}+\mathfrak{Re}(y_1)\Big\}, \\\mathcal{V}_{N_{1,K}}^{Re} =& \Big\{y\in \mathbb{R}| y \geq \mathfrak{Re}(y_{N_{1,K}})-\frac{d_1}{2}\Big\}, \\
      \mathcal{V}_i^{Re} =& \Big\{y\in \mathbb{R}| |y - \mathfrak{Re}(y_i)|\leq \frac{d_1}{2}\Big\},
\end{align}
\end{subequations}
for $i\in \{2,3,\ldots,N_{1,K}-1\}$, and   
\begin{subequations}
\label{eq:Voronoi_Im}
    \begin{align}
    \mathcal{V}_1^{Im} = &\Big\{y\in \mathbb{R}| y \leq \frac{d_2}{2}+\mathfrak{Im}(y_1)\Big\}, \\\mathcal{V}_{N_{2,K}}^{Im} =& \Big\{y\in \mathbb{R}| y \geq \mathfrak{Im}(y_{N_{2,K}})-\frac{d_2}{2}\Big\}, \\
    \mathcal{V}_j^{Im} =& \Big\{y\in \mathbb{R}| |y - \mathfrak{Im}(y_j)|\leq \frac{d_2}{2}\Big\},
\end{align}
\end{subequations}
% -----------
for $j\in \{2,3,\ldots,N_{2,K}-1\}$. Then, the estimated value is given by 
% -----------
\begin{align}
    \hat{r} =  \sum_{i=1}^{N_{1,K}} \mathfrak{Re}(y_{i})\mathds{1}_{\mathcal{V}_i}(\mathfrak{Re}(r)) + \sum_{j=1}^{N_{2,K}} \mathfrak{Im}(y_j)\mathds{1}_{\mathcal{V}_j}(\mathfrak{Im}(r))\,i,
\end{align} 
% -----------
where $\mathds{1}_{\mathcal{V}_i}$ is the indicator function, i.e.,  
% -----------
\begin{align}
    \mathds{1}_{\mathcal{V}_i}(r):= \begin{cases}
            1, &    r \in \mathcal{V}_i, \\
            0, & \text{otherwise}.
        \end{cases}
\end{align}
% -----------
Hence, the desired function value is obtained by 
% -----------
\begin{align}
   \label{eq:deocder_qam}
    \hat{f} = \mathscr{D}(r) := \mathfrak{Re}(\hat{r}) + q\,\mathfrak{Im}(\hat{r}).
\end{align}
% -----------
We note that the factor $q$ appears because the imaginary component encodes $c_2$ in the decomposition $s=c_1+c_2q$, and must be scaled to recover $s$.

In the next subsection, we extend the decoding procedure to the case where prior information about the superimposed symbols is available.

\subsection{Decoding Using Maximum a Posteriori (MAP) Estimator}\label{sec:MAPDecoder}

Here, we use the MAP criterion to estimate the received signal $r$ for the decoder $\mathscr{D}$ because the superimposed symbols may not be equiprobable at the CP due to aggregation over the MAC. Accordingly,  we consider the estimator  $\mathscr{D}$ below:  
% -----------
\begin{align}
    \label{eq:Quntizer}
    \mathscr{D}(r) =  \underset{y_j \in\mathcal{Y}}{\arg\max}~p_j\mathrm{g}(r|y_j),
\end{align}
% -----------
where  $p_{j} = \Pr\{y=y_j\}$ is the prior probability that the transmitted constellation point over the MAC is $y_j\in \mathcal{Y}$, and $\mathrm{g}(r|y_j)$ denotes the conditional channel transition probability. Similar to the ML estimator, the expression in \eqref{eq:Quntizer} generates detection regions for the set of all possible constellation points, and $r$ is estimated by 
% -----------
\begin{align}
    \hat{r} =  \sum_{i=1}^{N_{1,K}} \mathfrak{Re}(y_{i})\mathds{1}_{\mathcal{V}_i^{Re}}\big(\mathfrak{Re}(r)\big) + \sum_{j=1}^{N_{2,K}} \mathfrak{Im}(y_j)\mathds{1}_{\mathcal{V}_j^{Im}}\big(\mathfrak{Im}(r)\big)\,i,
\end{align} 
% -----------
where $\mathds{1}_{\mathcal{V}_j}$ is the indicator function. We note that the boundaries of the detection regions depend on the prior distribution of aggregated symbols. For uniformly i.i.d. generated symbols, it has been shown that the distribution of aggregated symbols becomes a lattice distribution~\cite[Lemma 1]{liu2025digital}. For sufficiently large $K\gg 1$, the local limit theorem~\cite{gnedenko1968limit} suggests the following approximation by a normal distribution $\mathcal{N}(\mu_y, \sigma_y^2)$ with $\mu_y = (q-1)K/2$ and $\sigma_y^2 = (q^2-1)K/12$, i.e., 
% -----------
\begin{align}
    \label{eq:prior_dist}
    \Pr\{\mathfrak{Re}(y)=y_j\} \approx \begin{cases} \frac{\exp{\Big(-\frac{(y_j-\mu_y)^2}{2\sigma_y^2}\Big)}}{\sqrt{2\pi \sigma_y^2}}, ~~ \text{for}~y_j \in \mathcal{Y}, \\
      0, \quad   \quad  \text{otherwise}.
    \end{cases}
\end{align}
% -----------
Similarly, for the imaginary component, the prior distribution is approximated by a normal distribution with a mean $\sum_{k}(n_k-1)/2$ and variance $\sum_{k}(n_{k}^2-1)/12$. With this distribution, the detection regions $\mathcal{V}_i^{Re}$ and $\mathcal{V}_j^{Im}$ become identical to \eqref{eq:Voronoi_RE} and \eqref{eq:Voronoi_Im}, respectively, except that $d_1$ and $d_2$ are replaced by $d_1\eta$ and $d_2\eta$, where $\eta:= (1+\sigma^2/K)$~\cite[Proposition 1]{liu2025digital}. Notably, the prior distribution skews the decision boundaries by biasing them toward constellation points with higher prior probabilities. That is, the boundary shifts asymmetrically, resulting in a narrower region around low-probability points and a wider one around high-probability points. Also, for $\sigma^2/K \ll 1$—that is, either $K\gg 1$ or $\sigma^2\ll 1$—the decision boundaries converge to the midpoints between adjacent constellation points. 

Given this decoding procedure, we can now optimize the modulation diagram under such decision boundaries, which constitutes the main objective of this paper.

\section{Main Results}\label{sec:MainResults}

In this section, we summarize the main analytical results of this paper, which include the optimal values for parameters $d_1$ and $d_2$ such that the ML and MAP estimators can provide the minimum MSE solutions. Moreover, we obtain closed-form solutions for  $d_1$ and $d_2$  in terms of the generalized Lambert function in the high SNR region. Although the general communication model permits each node to employ a distinct modulation order $Q_{k} = q \times n_{k}$, for ease of exposition,  we assume a common modulation order across all nodes, namely $Q_{1} = \cdots = Q_{K} = Q = q \times n,$ for any integer $n$ greater than $2$, i.e.,  $ n \ge 2$.

\subsection{Optimal Modulation Diagram for ML Decoding}

The solution to \eqref{eq:OptmialMSEgeneral} under the ML estimator depends on the SNR level,  defined as $\xi:=P/\sigma^2$. We determine the exact optimal values $d_1^*$ and $d_2^*$,  corresponding to a given SNR $\xi$. These optimal parameters are characterized as the unique positive roots of specific polynomials, which are defined below.

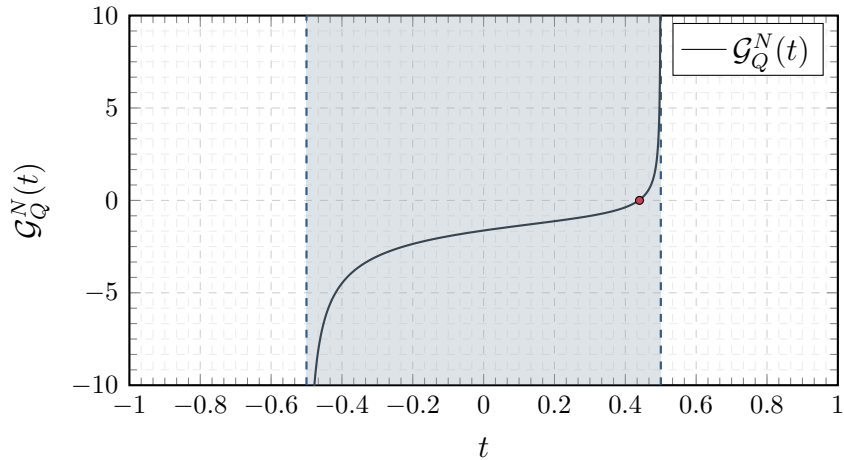
\begin{figure}[!t]
    \centering
    \pgfmathsetmacro{\K}{6}
\pgfmathsetmacro{\q}{4}
\pgfmathsetmacro{\n}{6}
\pgfmathsetmacro{\Mone}{\K*(\q-1)+1} % = 19
\pgfmathsetmacro{\Mtwo}{\K*(\n-1)+1} % = 91

% Set a and b (edit these):
\pgfmathsetmacro{\a}{0.42}
\pgfmathsetmacro{\b}{0.1}

\begin{tikzpicture}
  \begin{axis}[
    width=11cm, height=6.5cm,
    xlabel={$t$}, ylabel={$\mathcal{G}_{Q}^{N}(t)$},
    ymin =-10, ymax=10,
    xmin=-1, xmax=1,
     ymajorgrids=true,
            xmajorgrids=true,
            grid style=dashed,
            grid=both,
            minor tick num=5,
            grid style={line width=.1pt, draw=gray!15},
            major grid style={line width=.2pt,draw=gray!40},
            tick label style={font=\small},
            axis line style={thick},
    domain=-0.49:0.499, samples=2000,
  ]

  \draw[dashed, thick, color=darkcerulean]
    (axis cs:0.5,\pgfkeysvalueof{/pgfplots/ymin})
    -- (axis cs:0.5,\pgfkeysvalueof{/pgfplots/ymax});
  \draw[dashed, thick, color=darkcerulean]
    (axis cs:-0.5,\pgfkeysvalueof{/pgfplots/ymin})
    -- (axis cs:-0.5,\pgfkeysvalueof{/pgfplots/ymax});
      % gray background strip between x=-0.5 and x=0.5
    % shaded region between t=-0.5 and t=0.5
    \addplot [
      draw=none, fill=cadetgrey, fill opacity=0.3
    ] coordinates {(-0.5,-10) (-0.5,10) (0.5,10) (0.5,-10)} --cycle;
    \addplot+[mark=none, color=charcoal, thick, variable=\t]
      ({\t},
       { (\a*(1-1/\Mone))/sqrt(0.5-\t) * exp(-(\a*\a)*(0.5-\t)^2/8)
        -(\q*\q)*(\b*(1-1/\Mtwo))/sqrt(0.5+\t) * exp(-(\b*\b)*(0.5+\t)^2/8) }
      );
        \draw[fill=brickred](axis cs:0.44,0) circle[radius=1.5pt];
    \addlegendentry{$\mathcal{G}_{Q}^{N}(t)$}
  \end{axis}
\end{tikzpicture}
    \caption{Plot of $\mathcal{G}_{Q}^{N}(t)$ in \eqref{eq:MathcalG} with parameters $K=6$, $q=4$, and $n=16$, yielding $N_{1,K}=19$ and $N_{2, K}=91$. The interval $t\in[-0.5,0.5]$ is highlighted.}
    \label{fig:MathcalG}
\end{figure}

Recall that $N_{1, K}=K(q-1)+1$ and $N_{2, K}=K(n-1)+1$ denote the effective dimensions of the superimposed constellation grid, both strictly greater than three. We then define the following auxiliary polynomials used in the derivation of $d_1^*$ and $d_2^*$:
%-------------
\begin{subequations}
    \begin{align}
    P_{N_{i,K}}^{(1)}(x)& :=\sum_{m=1}^{N_{i,K}-1}\gamma_{i,m} {\rm e}^{-\theta_{m}x^{2}}(1+2\theta_{m}x^{2}), \quad i\in\{1,2\}, \\
     P^{(2)}_{N_{i,K}}(x) & := \frac{1}{x}\sum_{m=1}^{N_{i,K}-1}\gamma_{i,m}{\rm e}^{-\theta_{m}x^{2}},\quad i\in\{1,2\},
\end{align}
\end{subequations}
%-------------
where the coefficients are given by $\gamma_{i,m} = (2m-1)\big(2m-1+\frac{3m(1-m)-1}{N_{i,K}}\big)$ and $\theta_m= (2m-1)^2/4$. Then, we define the positive threshold $\xi_1$  as 
%-------------
\begin{align}
 \label{eq:r1r2def}
    \xi_1 := \begin{cases}  \frac{(q^2-1)x_1^2}{12} + \frac{(n^2-1)y_1^2}{12},&  N_{1, K}\geq 9, \\
        \frac{(q^2-1)x_2^2}{12} + \frac{(n^2-1)y_2^2}{12}, &  N_{1, K}\leq 8,~N_{2, K}\geq 10,
    \end{cases} \quad 
\end{align}
%-------------
where $(x_1, y_1)$ and $(x_2, y_2)$ are positive roots of the following polynomials: 
\begin{subequations}
    \begin{align}
      & P_{N_{1,K}}^{(1)}(x_1) = 0,  \quad q^2 P_{N_{2,K}}^{(2)}(y_1) = {P_{N_{1,K}}^{(2)}(x_1)}, \quad x_1,y_1 >0, \\ 
      & P_{N_{2,K}}^{(1)}(y_2) = 0,  \quad q^2 P_{N_{2,K}}^{(2)}(y_2) = {P_{N_{1,K}}^{(2)}(x_2)}, \quad x_2,y_2 >0.  
\end{align}
\end{subequations}
 Next, we define the auxiliary function $\mathcal{G}_{Q}^{N}(t)$ as 
%-------------
\begin{align}
     \label{eq:MathcalG}
    \mathcal{G}_{Q}^{N}(t) =  \sum_{m=1}^{N_{1,K}-1} {\gamma}_{1,m} \frac{{\rm e}^{\theta_{m}\Upsilon_1^2(-0.5+t)}}{\sqrt{0.5 - t}} -\sum_{m=1}^{N_{2,K}-1}{\gamma}_{2,m}\frac{\kappa q^2{\rm e}^{-\theta_{m}\Upsilon_2^2 (t+0.5)}}{\sqrt{0.5+t}},
\end{align}
%-------------
for $ t\in [0,0.5)$, $\kappa = \sqrt{(q^2-1)/(n^2-1)}$ and $\Upsilon_1 =\sqrt{12\xi/(q^2-1)}, \Upsilon_2 =\sqrt{12\xi/(n^2-1)}$.  The function $\mathcal{G}_{Q}^{N}(t)$ in \eqref{eq:MathcalG} captures the equilibrium between the real and imaginary components of the aggregated constellation points at the CP. It also explicitly incorporates key system parameters such as  $K$, $q$, and $n$, with the scaling factor $\kappa$ accounting for the asymmetry between the two-dimensional grid (see Figure~\ref{fig:MathcalG}). This function plays a key role in establishing the optimality conditions.

We are now ready to present the main optimality theorem.
%-----------------------------
%----------------------------

\begin{theorem}\label{th:Optmizaiton}
    For a network with $K$ transmitters using an encoding function $\mathscr{E}_{q}$ defined in \eqref{eq:energy_conste}, with  modulation order $Q=q\times n$, let $N_{1,K}:= K(q-1)+1, N_{2,K}:= K(n-1)+1$ and $\bar{N}_{1,K}=\lfloor 2N_{1,K}/3\rfloor, \bar{N}_{2,K}=\lfloor 2N_{2,K}/3\rfloor$.   Then, under an ML decoder, the optimal encoder parameters ${d}_1^*$ and ${d}_2^*$ are uniquely determined as follows. Let $\mathfrak{a}^{*}$ and $\mathfrak{b}^{*}$ be the unique positive roots of $\mathcal{G}_{Q}^{N}$ and $\mathcal{G}_{Q}^{\bar{N}}$, respectively, i.e., 
    %-------------
    \begin{align}
        \label{eq:uniqeg0}
         \mathcal{G}_{Q}^{N}(\mathfrak{a}^{*}) =0,\quad  \mathcal{G}_{Q}^{\bar{N}}(\mathfrak{b}^{*}) =0.
    \end{align}
    %-------------
    For any SNR $\xi$, when $3\leq N_{1,K}, N_{2,K} \leq8$,  the optimal parameters ${d}_1^*$ and ${d}_2^*$ are given by
    %-------------
    \begin{align}
        \label{eq:uniqegn}
        {d}_1^* = \sigma \Upsilon_1\sqrt{0.5- \mathfrak{a}^{*}},~~~~~{d}_2^* = \sigma \Upsilon_2\sqrt{0.5+\mathfrak{a}^{*}},
    \end{align}
    %-------------
    where $\sigma$ denotes the channel noise standard deviation. For $N_{1,K}\geq 9~\text{or}~ N_{1, K}\leq 8,~N_{2, K}\geq 10$, the optimal parameters are instead given by
    %-------------
        \begin{align}\label{eq:uniqeg(2)}
      {d}_1^* = \sigma \Upsilon_1\sqrt{0.5-\mathfrak{b}^{*}}, 
      \qquad 
  {d}_2^* = \sigma \Upsilon_2\sqrt{0.5+\mathfrak{b}^{*}}, \qquad  \xi \geq {\xi}_1,
    \end{align} 
    %-------------
where $\xi_1$ is the threshold defined in \eqref{eq:r1r2def}.  
\end{theorem}
\begin{proof}
    See Appendix~\ref{sec:Optimal}.
\end{proof}
We can approximate the boundaries for sufficiently large  $N_{1, K}$ and $N_{2, K}$, which leads to the following corollary.
\begin{cor}
    For $N_{1,K}, N_{2,K}\geq 30$, the SNR threshold $\xi_1$ can be approximated by $1.5n/K^2$. Accordingly, the optimal parameters ${d}_1^*$ and ${d}_2^*$  are given by
     %-------------
        \begin{align}\label{eq:uniqeg(2)-aprox}
      {d}_1^* = \sigma \Upsilon_1\sqrt{0.5-\mathfrak{b}^{*}}, %&  \xi \geq \frac{1.5n}{K^2},
     % \\0, &   \frac{1}{24K^4q^2}<\xi < \frac{1.5n}{K^2}, \\ 
      %    \Upsilon_1,  & \xi \leq    \frac{1}{24K^4q^2},
   \qquad
  {d}_2^* = \sigma \Upsilon_2\sqrt{0.5+\mathfrak{b}^{*}}, \qquad  \xi \geq \frac{1.5n}{K^2}.
   %   \\\Upsilon_2, &      \frac{1}{24K^4q^2}<\xi < \frac{1.5n}{K^2}, \\ 
    %     0, & \xi \leq     \frac{1}{24K^4q^2}.
    \end{align}
    %-------------
\end{cor}

Theorem~\ref{th:Optmizaiton} indicates that the optimal encoder parameters depend explicitly on the SNR.  For \(\xi \ge \xi_1\), the optimal pair \(({d}_1^*, {d}_2^*)\) is determined by the unique positive root of either \(\mathcal{G}_Q^N\) or \(\mathcal{G}_Q^{\bar{N}}\).  
In contrast, in the low-SNR regime (\(\xi \le \xi_1\)), the optimal in-phase component \(d_1\) tends toward zero, as the high noise variance dominates the received signal.   Moreover, for a large number of transmitters (\(K \gg 1\)), the threshold \(\xi_1\) approaches zero, i.e., \(\xi_1 \approx 0\).   In this asymptotic regime, the positive root of \(\mathcal{G}_Q^{\bar{N}}\) remains the unique optimal solution.

\begin{remark}\label{rem:smallroot}
 We note that the threshold \(\xi_1\) is very small and decreases rapidly as the number of transmitters \(K\) increases. Even for a small network with \(K=2\) and \(n=2\), we obtain \(\xi_1 \approx -1.25\,\mathrm{dB}\).
\end{remark}

Hitherto, we have derived the optimality conditions under the assumption of equiprobable induced constellation points over the MAC, without considering the prior distribution of the constellation diagram. 
In the following section, we extend this analysis to determine the optimal constellation design when the prior distribution of the constellation points is known.

\subsection{Optimal Modulation Diagram for MAP Decoding}

In this section, we incorporate the prior distribution of the induced constellation points to determine the optimal constellation diagram for each transmitter node. Assuming the normal approximation in \eqref{eq:prior_dist} for the prior distribution, we derive a closed-form expression for the MSE and identify the parameters that minimize it.  

Before presenting the main result, we define the auxiliary function $\mathcal{H}_Q(t)$ for any  $t \in [0, 0.5)$ as
%-------------
\begin{align}
    \label{eq:HQpoly}
    \mathcal{H}_Q(t) =  \sum_{m=1}^{2q}{\theta}_{m} \frac{{\rm e}^{-\theta_{m}  \Upsilon_1^2(0.5-t)}}{\sqrt{0.5-t}} -\sum_{m=1}^{2n}{\theta}_{m}\frac{\kappa q^2{\rm e}^{-\theta_{m} \Upsilon_2^2 (t+0.5)}}{\sqrt{0.5+t}},\quad  t\in [0,1/2),
\end{align}
where recall $\theta_m =(2m-1)^2/4$ for all $m$. Then, we provide the main result regarding the optimal constellation diagram under the MAP decoder.
%=========================
%=========================
\begin{theorem}\label{th:Optmizaiton-Map}
Let $d_1$ and $d_2$ denote the distances between constellation points along the in-phase and quadrature components, respectively, for a modulation order $Q = q \times n$.  
Assuming the normal approximation in \eqref{eq:prior_dist} for the prior distribution of the constellation points, the optimal constellation parameters that minimize the MSE under MAP decoding are given by
%-------------
\begin{align}
\label{eq:uniqeg-MAp-th}
{d}_1^* =  \frac{\sigma}{\eta} \Upsilon_1 \sqrt{0.5 - \mathfrak{a}^*}, 
\qquad 
{d}_2^* = \frac{\sigma}{\eta} \Upsilon_2 \sqrt{0.5 + \mathfrak{a}^*},
\end{align}
%-------------
where $\sigma$ is the standard deviation of the channel noise, $\eta = 1 + \sigma^2/K$,  and $\mathfrak{a}^*$ is the unique positive root of the polynomial $\mathcal{H}_Q(t)$ defined in \eqref{eq:HQpoly}, i.e.,
\[
\mathcal{H}_Q(\mathfrak{a}^*) = 0.
\]
\end{theorem}
\begin{proof}
    See Appendix~\ref{sec:prioir}.
\end{proof}
%=========================

\begin{remark}
An important insight from Theorem~\ref{th:Optmizaiton-Map} is that the MAP and ML solutions become identical as the number of transmitters increases, i.e., as \(K \to \infty\).  
In a massive-size network, we have \(\eta \approx 1\), \(\theta_m \approx \gamma_{1,m} \approx \gamma_{2,m}\), and \(\xi_1 \approx 0\).  
Consequently, \(\mathcal{G}_Q^N(t) \rightarrow \mathcal{H}_Q(t)\), implying that the optimal parameters \((d_1, d_2)\) obtained from \eqref{eq:uniqeg(2)} and \eqref{eq:uniqeg-MAp-th} coincide.   This result aligns with the classical Bernstein--von Mises phenomenon for Gaussian white noise~\cite{Haralambie2011Bernstein}.
\end{remark}

Thus far, we have established the conditions under which the optimal parameters can be uniquely determined by numerically solving the systems of equations in \eqref{eq:uniqeg0} and \eqref{eq:HQpoly}.  However, due to the strong nonlinearity in the low-SNR region, obtaining a closed-form analytical solution is challenging.  
In the next section, we demonstrate that a closed-form solution can indeed be derived in the high-SNR regime.
%====================
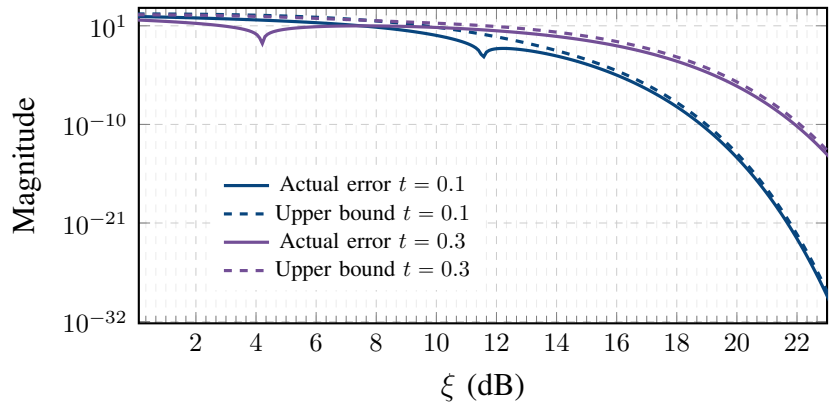
\begin{figure}
    \centering
    \begin{tikzpicture}
        \begin{axis}[
            width=0.65\linewidth,
            height=0.35\linewidth,
            xlabel={$\xi$ (dB)},
            ylabel={Magnitude},
            xmin=1e-1, xmax=23,
             ymax=1000,
            legend pos=north east,
            legend style={draw=none,font=\small},
            legend style={nodes={scale=0.85, transform shape}, at={(0.5,0.5)}}, 
            ymajorgrids=true,
            xmajorgrids=true,
            grid style=dashed,
            grid=both,
            ymode =log,
               minor tick num=5,
            grid style={line width=.1pt, draw=gray!15},
            major grid style={line width=.2pt,draw=gray!40},
            tick label style={font=\small},
            axis line style={thick},
        ]
        % Load data from file

    % Plot implicit curve
       \addplot[darkcerulean,  very thick] table [x=xi, y=actual_x1] {Data/error_vs_bound_upper.dat};

        \addplot[darkcerulean, dashed, very thick] table [x=xi, y=bound_x1] {Data/error_vs_bound_upper.dat};
        
       \addplot[darklavender,  very thick] table [x=xi, y=actual_x3] {Data/error_vs_bound_upper.dat};

        \addplot[darklavender, dashed, very thick] table [x=xi, y=bound_x3] {Data/error_vs_bound_upper.dat};

            \legend{ {Actual error $t=0.1$}, {Upper bound $t=0.1$}, {Actual error $t=0.3$}, {Upper bound $t=0.3$}};
        \end{axis}
    \end{tikzpicture}
    \caption{Comparison of the actual approximation error and the final tail‐based bound by Lemma~\ref{lem:apprximation_fg} as functions of $\xi$ for $t=0.1$ and $t=0.3$. Solid curves denote the computed error; dashed curves denote the analytical bound. The magnitude is plotted on a logarithmic scale. Parameters: $K=2$, and $q=n=4$.}

    \label{fig:upper_lambert}
\end{figure}
%====================

\subsection{Closed-form Solution for High SNR Region}

For sufficiently high SNR, the optimization problem in \eqref{eq:OptmialMSEgeneral} admits a closed-form solution, regardless of the prior distribution. 
In this regime, the exponential terms in both $\mathcal{G}_{Q}^{N}(t)$ and $\mathcal{H}_Q(t)$ vanish, leading to the approximation $\mathcal{G}_{Q}^{N}(t) \approx \mathcal{H}_Q(t)$. 
To obtain an analytical expression, we further consider the case where $\mathcal{G}_{Q}^{N}(t)$ is dominated by its first exponential term, and the contributions of higher-order terms are negligible.  Accordingly, for sufficiently large $\xi$, we can approximate
%-------------
\begin{align}
    \label{eq:approx_one}
     \mathcal{G}_Q^{N}(t) \approx \mathcal{F}_Q(t) :=  \frac{{\gamma}_{1,1}  {\rm e}^{-\theta_{1} \Upsilon_1^2 (0.5-t)}}{\sqrt{0.5-t}} -q^2\frac{\gamma_{2,1} \kappa {\rm e}^{-\theta_{1} \Upsilon_2^2 (t+0.5)}}{\sqrt{0.5+t}}. 
\end{align}
%-------------
Since $\Upsilon_1, \Upsilon_2 \propto \sqrt{\xi}$, it follows that $\Upsilon_1, \Upsilon_2 \to \infty$ as $\xi \to \infty$. Under the approximation in \eqref{eq:approx_one}, the optimal solution to \eqref{eq:uniqeg0} can be obtained in closed form for sufficiently high SNR, as stated in the following lemma.

\begin{lem}\label{lem:apprximation_fg}
    Let $\mathcal{F}_{Q}(t)$ be the approximation of $\mathcal{G}_{Q}^{N}(y)$ in \eqref{eq:approx_one},  the following bound holds
        \begin{align}
            |\mathcal{F}_Q(t)  - \mathcal{G}_Q^{N}(t)| \leq  36\sqrt{3\xi}  \bigg( \frac{{\rm e}^{-\frac{16\xi(0.5-t)}{3(q^2-1)}}}{\sqrt{(q^2-1){(0.5-t)}}} + \frac{\kappa q^2 {\rm e}^{-\frac{16\xi(0.5+t)}{3(n^2-1)}}}{\sqrt{(n^2-1){(0.5+t)}}}\bigg). 
    \end{align}
    Moreover, for $ \xi \geq {1}/{10} \max\Big\{{q^2}/{(0.5-t)}, {n^2}/{(0.5+t)} \Big\} $, the bound simplifies to
    %-------------
    \begin{align}
      |\mathcal{F}_Q(t)  - \mathcal{G}_Q^{N}(t)| \leq  18\,e^{-4/9}\,\biggl(\frac{1}{0.5-t} \;+\;\frac{\kappa q^2}{0.5+t}\biggr).
    \end{align}
    %-------------
    for $t\in [0,1/2)$.
\end{lem}
\begin{proof}
    See Appendix~\ref{ap:apprximation_FGQ}.
\end{proof}

Figure~\ref{fig:upper_lambert} illustrates the actual approximation error alongside the upper bounds derived in Lemma~\ref{lem:apprximation_fg}. 
The error decays as $\mathcal{O}\!\left(\sqrt{\xi}\,\min\{q,n\}^{-1} {\rm e}^{-\xi / \min\{q^2,n^2\}}\right)$, and thus vanishes rapidly in the high-SNR regime. 
This fast decay explains why the dashed curves in Figure~\ref{fig:upper_lambert} provide tight uniform upper bounds on the actual error (solid curves) even for moderate SNR values, i.e., $\xi \gtrsim 1$.

%---------------
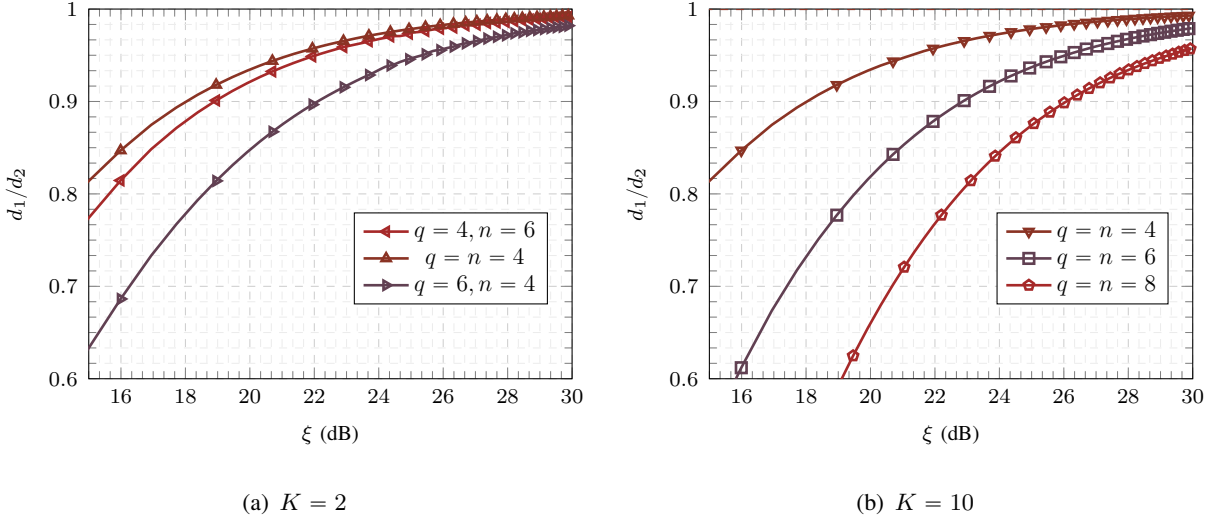
\begin{figure*}[!t]
\centering
  \subfigure[$K=2$]{\label{fig:Lambert_solution(a)}
    \begin{tikzpicture} 
    \begin{axis}[
        xlabel={${\xi}$ (dB)},
        ylabel={${d}_1/{d}_2$},
        label style={font=\scriptsize},
        % title = {\scriptsize $???$},
        %legend pos=,
        tick label style={font=\scriptsize}, 
        width=0.485\textwidth,
        height=6.5cm,
        xmin=15, xmax=30,
        ymin=0.6, ymax=1,
       legend style={nodes={scale=0.75, transform shape}, at={(0.95,0.45)}}, 
        minor tick num=5,
        %legend pos=east,
        % xmode =log,
        ymajorgrids=true,
        xmajorgrids=true,
        grid style=dashed,
        grid=both,
        grid style={line width=.1pt, draw=gray!15},
        major grid style={line width=.2pt,draw=gray!40},
    ]
% ==========
     \addplot[ mark=triangle,mark repeat=4,
     each nth point={5}, 
        color=brown(web),
        mark options = {rotate = 90},
        line width=1pt,
        mark size=2pt,
        ]
    table[x=xi,y=n6_q4]
    {Data/ratios_lambert.dat};
    \addplot[mark=triangle,mark repeat=4,
     each nth point={5}, 
        color=burntumber,
        line width=1pt,
        mark size=2pt,
        ]
    table[x=xi,y=n4_q4]
    {Data/ratios_lambert.dat};
% ============
     \addplot[mark=triangle,mark repeat=4,
     each nth point={5}, 
        color=eggplant,
        mark options = {rotate = 270},
        line width=1pt,
        mark size=2pt,
        ]
    table[x=xi,y=n4_q6]
    {Data/ratios_lambert.dat};
%    \addplot[dashed,
%      domain=5:48,
%      samples=2,            % only two points needed for a straight line
%      color=brown(web),
%      thick
%    ]{1.5};
%         \addplot[dashed,
%      domain=5:48,
%      samples=2,            % only two points needed for a straight line
%      color=eggplant,
%      thick
%    ]{0.66};
    \legend{ {$q=4, n=6$} ,{$q=n=4$} , {$q=6, n=4$}};
    \end{axis}
\end{tikzpicture}
  }\subfigure[$K=10$]{\label{fig:Lambert_solution(b)}
  \begin{tikzpicture} 
    \begin{axis}[
        xlabel={${\xi}$ (dB)},
        ylabel={${d}_1/{d}_2$},
        label style={font=\scriptsize},
        % title = {\scriptsize $???$},
        %legend pos=,
        tick label style={font=\scriptsize} , 
        width=0.485\textwidth,
        height=6.5cm,
        xmin=15, xmax=30,
        ymin=0.6, ymax=1,
       legend style={nodes={scale=0.75, transform shape}, at={(0.95,0.45)}}, 
        minor tick num=5,
        ymajorgrids=true,
        xmajorgrids=true,
        grid style=dashed,
        grid=both,
        grid style={line width=.1pt, draw=gray!15},
        major grid style={line width=.2pt,draw=gray!40},
    ]
% ==========
     \addplot[mark=triangle,
        color=  burntumber,
        mark repeat=4,
        each nth point={5}, 
        mark options = {rotate = 180},
        line width=1pt,
        mark size=2pt,
        ]
    table[x=xi,y=n4_q4]
    {Data/ratios_equal.dat};
    \addplot[mark=square,mark repeat=4,
     each nth point={5} ,
        color=eggplant,
        line width=1pt,
        mark size=2pt,
        ]
    table[x=xi,y=n6_q6]
    {Data/ratios_equal.dat};
% ============
     \addplot[mark=pentagon,mark repeat=4,
     each nth point={5} ,
        color=brown(web),
        line width=1pt,
        mark size=2pt,
        ]
    table[x=xi,y=n8_q8]
    {Data/ratios_equal.dat};
        \addplot[dashed,
      domain=5:48,
      samples=2,            % only two points needed for a straight line
      color=burntumber,
      thick
    ]{1};
    \legend{ {$q=n=4$} ,{$q=n=6$} , {$q=n=8$}};
    \end{axis}
    \end{tikzpicture}
  }
  \caption{ Ratio of the optimal scales $ d_1^*/ d_2^*$ versus the parameter $\xi$ for two array sizes.  
(a) $K=2$, showing curves for $(q,n)=(6,4),(4,4),(4,6)$ that asymptotically converge to $1$.  
(b) $K=10$, showing equal‐dimension cases $(q,n)=(4,4),(6,6),(8,8)$ with the unity asymptote.}\label{fig:Lambert_solution}
\end{figure*}
%---------------

\begin{prop}\label{Pr:Lambert}
     Let  $\mathfrak{a}^*$ denote the unique positive root of $ \mathcal{F}_Q$, i.e., 
    %-------------
    \begin{align}
       \label{eq:tildgforlambert}
      \mathcal{F}_Q(\mathfrak{a}^*)=0.  
    \end{align}
    %-------------
    Then, the approximated optimal parameters ${d}_1^* = \sigma\Upsilon_1 \sqrt{0.5 - \mathfrak{a}^*}$ and $ {d}_2^* =\sigma \Upsilon_2 \sqrt{0.5 + \mathfrak{a}^*}$  are expressed in closed form as
    %-------------
     \begin{subequations}
        \label{eq:d1d2optiamlLambewtr}
    \begin{align}
      d_1^*  = \sigma \Upsilon_1\sqrt{  \frac{2\mathcal{W}_{c}\Big(-c \frac{\Upsilon_1^2+\Upsilon_2^2}{2} \Big)}{\Upsilon_1^2+\Upsilon_2^2}    }, \quad  d_2^* = \sigma\Upsilon_2\sqrt{1-  \frac{2\mathcal{W}_{c}\Big(-c \frac{\Upsilon_1^2+\Upsilon_2^2}{2} \Big)}{\Upsilon_1^2+\Upsilon_2^2}   },
    \end{align}
     \end{subequations}
     %-------------
    where  $\tilde{\kappa} = {\rm e}^{\frac{\Upsilon_1^2-\Upsilon_2^2}{8}}\kappa q^2 (N_{1, K}N_{2 ,K}-N_{1 ,K})/(N_{1 ,K}N_{ 2,K}-N_{2 ,K})$, $c =  \tilde{\kappa}^2 {\rm e}^{-\frac{\Upsilon_1^2+\Upsilon_2^2}{4}}$,  and $\mathcal{W}_c(\cdot)$ denotes the generalized Lambert function, defined as the solution to
     %-------------
    \begin{align}
        {\rm e}^{-cx} = a_0(x-r_1)(x-r_2),
    \end{align}
     %-------------
     for some nonzero $r_1,r_2\neq 0$.
\end{prop}
\begin{proof}
    See Appendix~\ref{sec:PrLambert}. 
\end{proof}

The closed-form expressions in \eqref{eq:d1d2optiamlLambewtr} provide explicit high-SNR approximations for $d_1^*$ and $d_2^*$ in terms of the generalized Lambert function $\mathcal{W}_c(\cdot)$, which captures the dependence on both the SNR $\xi$ and the modulation parameter $Q$.  Practical computation of $\mathcal{W}_c(\cdot)$ can be performed via its series expansion~\cite{scott2014asymptotic,mezHo2017generalization}. 
Moreover, these closed-form expressions provide effective initialization points for numerical methods used to solve the general nonlinear system in~\eqref{eq:uniqeg0}.

To characterize the asymptotic behavior of the ratio $d_1^*/d_2^*$ derived from \eqref{eq:d1d2optiamlLambewtr}, we consider the high-SNR regime ($\xi \to \infty$). 
In this case, three asymptotic configurations of $(q, n)$ can be analyzed:
%-----------------
\begin{align}\label{eq:d_1d_2}
    \frac{d_1^*}{d_2^*}\Big(\xi\Big)=  \frac{\Upsilon_1}{\Upsilon_2} \sqrt{\frac{w'(\xi)}{\Upsilon_1^2+ \Upsilon_2^2-w'(\xi)}}\qquad  \lim_{\xi \rightarrow \infty} \frac{d_1^*}{d_2^*}\Big(\xi\Big)= \frac{\Upsilon_1}{\Upsilon_2}\times  \frac{\Upsilon_2}{\Upsilon_1} =1
\end{align}
%-----------------
where $w'(\xi)$ is a monotonic variant of the Lambert function parameterized by $(n, q, K)$, satisfying $\lim_{\xi \to \infty} w'(\xi) = \Upsilon_2$. 
The convergence rate depends on $(n, q, K)$. 
Equation~\eqref{eq:d_1d_2} shows that, as $\xi \to \infty$, the optimal constellation approaches a rectangle with an aspect ratio of $q/n$, since the total spans along the in-phase and quadrature axes are $d_1 q$ and $d_2 n$, respectively.

To illustrate this behavior, Figure~\ref{fig:Lambert_solution} depicts $(d_1^*, d_2^*)$ as functions of $\xi$ for three constellation configurations, $(q,n) = \{(6,4), (4,6), (4,4)\}$. 
As $\xi$ increases, the optimal ratio $d_1^*/d_2^*$ converges toward unity, consistent with~\eqref{eq:d_1d_2}.  In the high-SNR limit, the impact of Gaussian noise becomes negligible, and the symmetry in the constellation parameters naturally emerges.   Moreover, Figure~\ref{fig:Lambert_solution(b)} shows $d_1/d_2$ for equal-dimension constellations, $(q,n) = \{(4,4),(6,6), (8,8)\}$. 
These curves demonstrate that larger $(q,n)$ pairs delay convergence to unity, as higher-dimensional constellations exhibit stronger noise sensitivity at moderate SNR. 
Hence, an initially asymmetric configuration—featuring a larger quadrature span—is beneficial for mitigating computation errors in this regime.

\begin{remark}\label{rem:highsnreffect}
Similarly to the massive network regime, in the high-SNR region, the MAP and ML solutions converge to the closed-form result in Proposition~\ref{Pr:Lambert}.  As $\sigma \to 0$, it follows that $\eta \approx 1$, and the exponential terms in both $\mathcal{H}_Q(t)$ and $\mathcal{G}_Q^{N}(t)$ decay rapidly, yielding $\mathcal{H}_Q(t) \approx \mathcal{G}_Q^{N}(t) \approx \mathcal{F}_Q(t)$.  This equivalence implies that the ML and MAP decoders attain identical MSE performance at high SNR. Therefore, the closed-form solution for $(d_1, d_2)$ presented in Proposition~\ref{Pr:Lambert} coincides with the solutions obtained from the systems of equations in \eqref{eq:uniqeg(2)} and \eqref{eq:OptmialMSEgeneral}.
\end{remark}

%Henceforth,

%=================
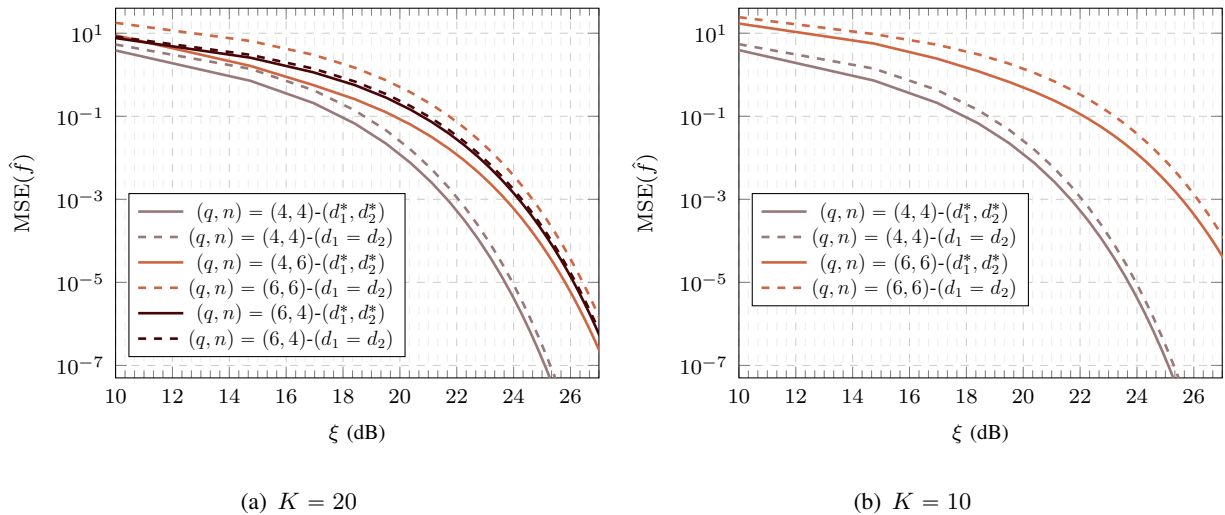
\begin{figure*}[!t]
\centering
\subfigure[$K=20$]{
    \label{fig:SUMQAM}
    \begin{tikzpicture} 
    \begin{axis}[
        xlabel={$\xi$ (dB)},
        ylabel={${\rm MSE}(\hat{f})$},
        label style={font=\scriptsize},
        % title = {\scriptsize $???$},
        %legend pos=,
        tick label style={font=\scriptsize} , 
        width=0.485\textwidth,
        height=6.5cm,
        xmin=10, xmax=27,
        ymin=5e-8, ymax=40,
        minor tick num=5,
         ymode = log,
      legend style={nodes={scale=0.65, transform shape}, at={(0.6,0.5)}}, 
        ymajorgrids=true,
        xmajorgrids=true,
        grid style=dashed,
        grid=both,
        grid style={line width=.1pt, draw=gray!15},
        major grid style={line width=.2pt,draw=gray!40},
    ]
% ==========
     \addplot[ color=bazaar,
        line width=1pt,
        each nth point={10}
        ]        
    table[x=xi_dB,y=mse_opt]
    {Data/Computation_exp/MSE_q4_n4_K10.dat};
% ==========    
     \addplot[ color=bazaar,
        line width=1pt,
        each nth point={10}, 
        dashed
        ]        
    table[x=xi_dB,y=mse_cf]
    {Data/Computation_exp/MSE_q4_n4_K10.dat};
% ==========    
    \addplot[color=burntsienna, each nth point={10}, 
        line width=1pt]
    table[x=xi_dB,y=mse_opt]
    {Data/Computation_exp/MSE_q6_n4_K10.dat};
% ==========    
    \addplot[color=burntsienna,dashed,
        line width=1pt,
        each nth point={10}]
    table[x=xi_dB,y=mse_cf]
    {Data/Computation_exp/MSE_q6_n4_K10.dat};
% ============
     \addplot[color=bulgarianrose,
        line width=1pt,
        each nth point={10}, ]
    table[x=xi_dB,y=mse_opt]
    {Data/Computation_exp/MSE_q4_n6_K10.dat};
% ==========    
     \addplot[color=bulgarianrose, dashed,
     each nth point={10}, 
        line width=1pt]
    table[x=xi_dB,y=mse_cf]
    {Data/Computation_exp/MSE_q4_n6_K10.dat};
% ============
    \legend{{$(q,n) =(4,4)$-$({d}_1^*,{d}_2^*)$},{$(q,n) =(4,4)$-$({d}_1={d}_2)$} ,{$(q,n) =(4,6)$-$({d}_1^*,{d}_2^*)$},{$(q,n) =(6,6)$-$({d}_1={d}_2)$}, {$(q,n) =(6,4)$-$({d}_1^*,{d}_2^*)$},{$(q,n) =(6,4)$-$({d}_1={d}_2)$}};
    \end{axis}
\end{tikzpicture}
}\subfigure[$K=10$ ]{
    \label{fig:SUMPAM}
    \begin{tikzpicture} 
    \begin{axis}[
        xlabel={$\xi$ (dB)},
        ylabel={${\rm MSE}(\hat{f})$},
        label style={font=\scriptsize},
        % title = {\scriptsize $???$},
        %legend pos=,
        tick label style={font=\scriptsize} , 
        width=0.485\textwidth,
        height=6.5cm,
        xmin=10, xmax=27,
        ymin=5e-8, ymax=40,
        minor tick num=5,
         ymode = log,
      legend style={nodes={scale=0.65, transform shape}, at={(0.6,0.5)}}, 
        ymajorgrids=true,
        xmajorgrids=true,
        grid style=dashed,
        grid=both,
        grid style={line width=.1pt, draw=gray!15},
        major grid style={line width=.2pt,draw=gray!40},
    ]
% ==========
     \addplot[ color=bazaar,
        line width=1pt,
        each nth point={10}
        ]        
    table[x=xi_dB,y=mse_opt]
    {Data/Computation_exp/MSE_q4_n4_K20.dat};
% ==========    
     \addplot[ color=bazaar,
        line width=1pt,
        each nth point={10}, 
        dashed
        ]        
    table[x=xi_dB,y=mse_cf]
    {Data/Computation_exp/MSE_q4_n4_K20.dat};
% ==========    
    \addplot[color=burntsienna, each nth point={10}, 
        line width=1pt]
    table[x=xi_dB,y=mse_opt]
    {Data/Computation_exp/MSE_q6_n6_K20.dat};
% ==========    
    \addplot[color=burntsienna,dashed,
        line width=1pt,
        each nth point={10}]
    table[x=xi_dB,y=mse_cf]
    {Data/Computation_exp/MSE_q6_n6_K20.dat};
% ============
    \legend{{$(q,n) =(4,4)$-$({d}_1^*,{d}_2^*)$},{$(q,n) =(4,4)$-$({d}_1={d}_2)$} ,{$(q,n) =(6,6)$-$({d}_1^*,{d}_2^*)$},{$(q,n) =(6,6)$-$({d}_1={d}_2)$}};
    \end{axis}
\end{tikzpicture}}
  \caption{Monte Carlo evaluation of the MSE for the summation function over $5\times10^4$ independent trials:  Figure~\ref{fig:SUMQAM} $K=20$ and Figure~\ref{fig:SUMPAM}  $K=10$. Solid curves denote the optimized distance parameters $({d}_1^*,{d}_2^*)$ obtained by Theorem~\ref{th:Optmizaiton}, whereas dashed curves correspond to the equal‐distance choice ${d}_1={d}_2=\sqrt{12/(n^2+q^2-2)}$.
  }
  \label{fig:SumSim}
\end{figure*}
%=================

%===================================================
\section{Computation Performance}\label{sec:computaion}
%===================================================

This section evaluates the computation performance of the proposed constellation design through numerical simulations. 
We analyze the MSE behavior across different network and modulation configurations to quantify the performance gains of the optimized parameters.  First, we consider the impact of the constellation shape and size on the MSE under varying SNR levels. Then, we compare the ML and MAP decoding schemes to highlight their relative performance and validate the analytical results derived in previous sections.

Figure~\ref{fig:SumSim} illustrates the resulting MSE as a function of $\xi \in \{10, \dots, 27\}\,\mathrm{dB}$ under two scenarios:   1) $K = 20$ nodes with $(q, n) \in \{(4,4), (6,4), (4,6)\}$, demonstrating the impact of the constellation aspect ratio $n/q$; and   2) $K = 10$ nodes with $(q, n) \in \{(4,4), (6,6)\}$, showing the effect of constellation size.  

In each configuration, the optimized solution $(d_1^*, d_2^*)$ is compared with the conventional equal-distance setting  $d_1 = d_2 = \sqrt{12 / (n^2 + q^2 - 2)}$, 
as used in the SumComp code~\cite{Razavikia2024Ring}.  Across a broad SNR range, the optimized design achieves a consistently lower MSE compared to the standard QAM-style grid. As $\xi$ increases, all curves converge, since Gaussian distortion becomes negligible and the ML and MAP optima coincide.  Overall, the optimized constellations yield approximately a $4$~dB improvement in MSE across a wide SNR range.  Furthermore, the configuration with $n = q = 4$ achieves the best performance among all cases, owing to its symmetric and balanced parameterization.

%=================
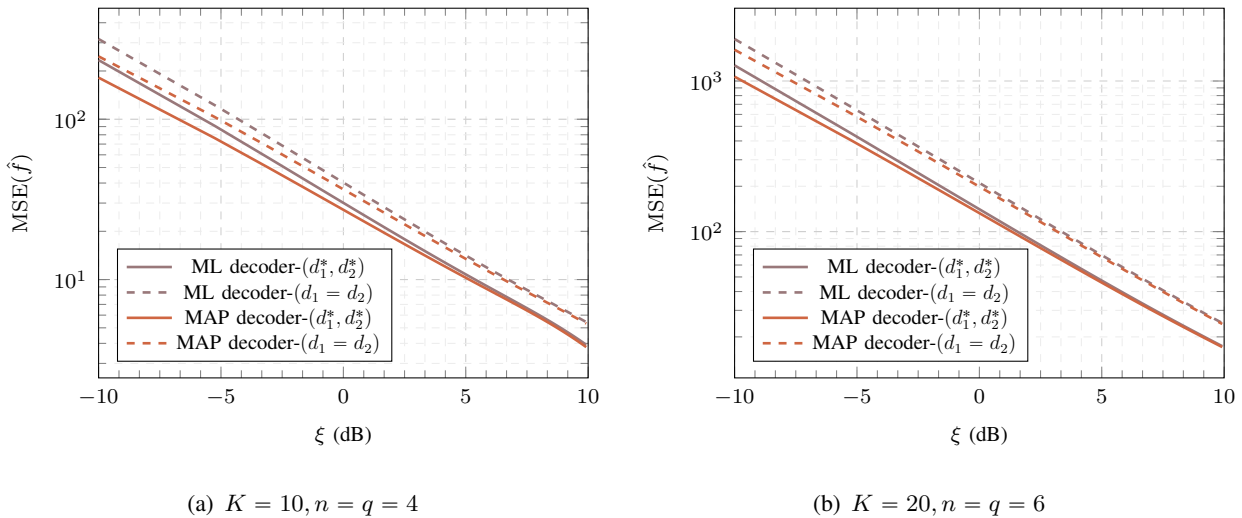
\begin{figure*}[!t]
\centering
\subfigure[$K=10, n=q=4$]{
    \label{fig:SimML4}
    \begin{tikzpicture} 
    \begin{axis}[
        xlabel={$\xi$ (dB)},
        ylabel={${\rm MSE}(\hat{f})$},
        label style={font=\scriptsize},
        % title = {\scriptsize $???$},
        %legend pos=,
        tick label style={font=\scriptsize} , 
        width=0.49\textwidth,
        height=6.5cm,
        xmin=-10, xmax=10,
        % ymin=5e-8, ymax=40,
        minor tick num=5,
         ymode = log,
      legend style={nodes={scale=0.65, transform shape}, at={(0.6,0.35)}}, 
        ymajorgrids=true,
        xmajorgrids=true,
        grid style=dashed,
        grid=both,
        grid style={line width=.1pt, draw=gray!15},
        major grid style={line width=.2pt,draw=gray!40},
    ]
% ==========
     \addplot[ color=bazaar,
        line width=1pt,
        each nth point={10}
        ]        
    table[x=xi_dB,y=mse_opt]
    {Data/Computation_exp/MLMSE_q4_n4_K10.dat};
% ==========    
     \addplot[ color=bazaar,
        line width=1pt,
        each nth point={10}, 
        dashed
        ]        
    table[x=xi_dB,y=mse_cf]
    {Data/Computation_exp/MLMSE_q4_n4_K10.dat};
% ==========    
    \addplot[color=burntsienna, each nth point={10}, 
        line width=1pt]
    table[x=xi_dB,y=mse_opt]
    {Data/Computation_exp/MAPMSE_q4_n4_K10.dat};
% ==========    
    \addplot[color=burntsienna,dashed,
        line width=1pt,
        each nth point={10}]
    table[x=xi_dB,y=mse_cf]
    {Data/Computation_exp/MAPMSE_q4_n4_K10.dat};
% ============
    \legend{{ML decoder-$({d}_1^*,{d}_2^*)$},{ML decoder-$({d}_1={d}_2)$} ,{MAP decoder-$({d}_1^*,{d}_2^*)$},{MAP decoder-$({d}_1={d}_2)$}};
    \end{axis}
\end{tikzpicture}
}\subfigure[$K=20, n=q=6$ ]{
    \label{fig:SimML6}
    \begin{tikzpicture} 
    \begin{axis}[
        xlabel={$\xi$ (dB)},
        ylabel={${\rm MSE}(\hat{f})$},
        label style={font=\scriptsize},
        tick label style={font=\scriptsize} , 
        width=0.49\textwidth,
        height=6.5cm,
         xmin=-10, xmax=10,
        % ymin=5e-8, ymax=40,
        minor tick num=5,
         ymode = log,
      legend style={nodes={scale=0.65, transform shape}, at={(0.6,0.35)}}, 
        ymajorgrids=true,
        xmajorgrids=true,
        grid style=dashed,
        grid=both,
        grid style={line width=.1pt, draw=gray!15},
        major grid style={line width=.2pt,draw=gray!40},
    ]
% ==========
     \addplot[ color=bazaar,
        line width=1pt,
        each nth point={10}
        ]        
    table[x=xi_dB,y=mse_opt]
    {Data/Computation_exp/MLMSE_q6_n6_K20.dat};
% ==========    
     \addplot[ color=bazaar,
        line width=1pt,
        each nth point={10}, 
        dashed
        ]        
    table[x=xi_dB,y=mse_cf]
    {Data/Computation_exp/MLMSE_q6_n6_K20.dat};
% ==========    
    \addplot[color=burntsienna, each nth point={10}, 
        line width=1pt]
    table[x=xi_dB,y=mse_opt]
    {Data/Computation_exp/MAPMSE_q6_n6_K20.dat};
% ==========    
    \addplot[color=burntsienna,dashed,
        line width=1pt,
        each nth point={10}]
    table[x=xi_dB,y=mse_cf]
    {Data/Computation_exp/MAPMSE_q6_n6_K20.dat};
%============
    \legend{{ML decoder-$({d}_1^*,{d}_2^*)$},{ML decoder-$({d}_1={d}_2)$} ,{MAP decoder-$({d}_1^*,{d}_2^*)$},{MAP decoder-$({d}_1={d}_2)$}};
    \end{axis}
 \end{tikzpicture}}
  \caption{Monte Carlo evaluation of the MSE of ML and MAP decoding: Figure~\ref{fig:SimML4} $K=10$, $q=n=4$; Figure~\ref{fig:SimML6}  $K=20$, $q=n=6$. Solid lines correspond to optimized distance parameters $({d}_1^*,{d}_2^*)$, and dashed lines to the equal‐distance choice ${d}_1={d}_2=\sqrt{12/(q^2+n^2-2)}$, for both the ML and MAP decoders. }
  \label{fig:SumSimML}
\end{figure*}
%=================

\subsection{ML vs. MAP Decoding}

In addition to optimizing the constellation distances, we compare the ML and MAP decoders under both the optimized and equal-distance designs defined by Theorems~\ref{th:Optmizaiton} and~\ref{th:Optmizaiton-Map}, respectively. 

Figure~\ref{fig:SumSimML} reports the MSE performance of both decoders over $5 \times 10^4$ Monte Carlo trials for two representative configurations:   (a) $K = 10$, $n = q = 4$; and   (b) $K = 20$, $n = q = 6$.  In both cases, the MAP decoder exhibits a slight MSE gain over ML at very low SNR due to its use of the prior distribution. At moderate SNR levels, the performances of both decoders converge, consistent with the observation in Remark~\ref{rem:highsnreffect}.  Notably, the gains provided by the optimized $(d_1^*, d_2^*)$ persist for both decoders, confirming the robustness and generality of the proposed constellation-shaping framework.

%===================================================
\section{Extensions}\label{sec:general}
%===================================================

\subsection{Heavy-tailed Noise Distribution}\label{sec:heavy-noise}

In practical wireless environments, the AWGN assumption may be overly idealistic, particularly in interference-limited scenarios where the noise exhibits heavy-tailed behavior~\cite{clavier2020experimental}. To better capture these conditions, we extend the framework by modeling the additive noise \( z \) as Cauchy distributed. Specifically, we model \( z \) as a centered Cauchy random variable characterized by a scale parameter \( \gamma \). We note that the centered Cauchy distribution does not possess a mean value.

Note that although the Cauchy distribution has heavy tails and lacks a finite second moment, the input data is digitally represented and undergoes hard decoding at the receiver. As a result, the MSE becomes finite~\cite{chen2023quantizing}. Hence, we consider the corresponding optimization problem 
%--------------
    \begin{align}
        \min_{d_1, d_2} \quad  J_q(d_1, d_2),~~~ \text{s.t.} \quad  \frac{d_1^2(q^2-1)}{12}+\frac{d_2^2(n^2-1)}{12} = P,
    \end{align}
%--------------
where $J_q := \mathbb{E}\big[|f-\hat{f}|^2\big]$,  $P$ denotes power budget and $(q,n)$ are the number of in-phase and quadrature levels. The probability density function of a Cauchy stable distribution around its location is given by
%--------------
\begin{equation}
    f(x) = \frac{\gamma}{\pi (\gamma^2 + x^2)},\quad \gamma >0, 
\end{equation}
%--------------
where $\gamma$ is the scale parameter. While this replacement modifies the analytical expressions for error probability, it does not alter the ML decision regions due to the symmetry of the Cauchy distribution about its location.

Following a procedure analogous to that used for the Gaussian case (Appendix~\ref{sec:Optimal}), we derive modified optimality conditions for the constellation design under Cauchy noise.  In the high-SNR regime, a series expansion of the Cauchy characteristic function for small arguments yields a nonlinear system that determines the optimal inter-point distances.  Let \( \mathcal{G}_{Q,\gamma}^{N}(d_1, d_2) \) denote the modified optimality function under Cauchy noise.  Then, for sufficiently large SNR (\( \xi > \bar{\xi}_1 \)), the optimal constellation parameters \( d_1^* \) and \( d_2^* \) for the ML decoder are obtained as the unique solution to $\mathcal{G}_{Q,\gamma}^{N}(d_1, d_2) = 0,$ where \( \mathcal{G}_{Q,\gamma}^{N}(t) \) is defined as
%-------------
\begin{align}
\label{eq:MathcalCaucy}
\mathcal{G}_{Q,\gamma}^{N}(t) 
=  
\sum_{m=1}^{N_{1,K}-1} 
\frac{\gamma_{1,m}}
{\sqrt{0.5 - t}\left(1 + \theta_{m}\Upsilon_1^2(0.5 - t)\right)}
-
\sum_{m=1}^{N_{2,K}-1} 
\frac{\kappa q^2 \gamma_{2,m}}
{\sqrt{0.5 + t}\left(1 + \theta_{m}\Upsilon_2^2(0.5 + t)\right)},
\end{align}
%-------------
where the parameters $\gamma_{1,m}$, $\gamma_{2,m}$, $\theta_{m}$, $\Upsilon_1$, and $\Upsilon_2$ are defined identically to the Gaussian case except for replacing $\sigma^2$ by $\gamma$. For general scenarios, the optimal parameters follow the same piecewise structure as in \eqref{eq:uniqeg(2)}, but with modified SNR thresholds denoted by $\bar{\xi}_1$. 
Due to the heavier tails of the Cauchy distribution, these thresholds are lower than in the Gaussian case, meaning that nonlinear effects appear at smaller SNR values.

In particular, the first threshold $\bar{\xi}_1$ can be obtained by solving
\begin{align}
    \bar{\xi}_1 =  \frac{(q^2-1)\bar{x}_1^2}{12} + \frac{(n^2-1)\bar{y}_1^2}{12}, \quad   P_{N_{1,K}}^{(3)}(\bar{x}_1) = 0,  \quad q^2 P_{N_{2,K}}^{(4)}(\bar{y}_1) = P_{N_{1,K}}^{(4)}(\bar{x}_1), 
\end{align}
%----------------
where, 
\begin{align*}
    P_{N_{1,K}}^{(3)}(x) &:=\sum_{m=1}^{N_{1,K}-1}\frac{\gamma_{1,m}}{\left(1+\frac{(2m-1)^{2}}{4}x^{2}\right)^{2}}\left(1+\frac{3(2m-1)^{2}}{4}x^{2}\right), \\
    P_{N_{i,K}}^{(4)}(x) &= \sum_{m=1}^{N_{i,K}-1}\frac{\gamma_{i,m}}{\left(1+\frac{(2m-1)^{2}}{4}x^{2}\right)^{2}}, \quad i \in \{1,2\}.
\end{align*}
%----------------
This extension broadens the applicability of the proposed encoding framework to more realistic wireless settings where heavy-tailed interference dominates.  Future research will focus on quantifying the performance trade-offs and potential enhancements offered by this approach under various classes of heavy-tailed noise distributions.

\subsection{Computation Over the Real Domain}\label{sec:hybrid}

Here, we consider the case where the input variable \( s_k \) lies in a subset of the real domain, i.e., \( s_k \in \mathbb{G} \subseteq \mathbb{R} \) for all \( k \in [K] \). In this setting, the previously proposed encoder operator \( \mathscr{E}(\cdot) \) is no longer directly applicable, since \( s_k \) is continuous rather than discrete.  This limitation arises because \( s_k \) can take infinitely many values in \( \mathbb{R} \), whereas conventional digital modulation schemes in the complex domain rely on a finite set of constellation points.  To overcome this challenge, we introduce a hybrid digital–analog modulation scheme that maps real-valued inputs onto a continuous curve in the complex plane \cite{skoglund2006hybrid}. 

\begin{figure}[t]
    \centering

\definecolor{eggshell}{rgb}{0.94, 0.92, 0.84}
\definecolor{flavescent}{rgb}{0.97, 0.91, 0.56}
\definecolor{lightapricot}{rgb}{0.99, 0.84, 0.69}
\definecolor{peach-orange}{rgb}{1.0, 0.8, 0.6}
\definecolor{pearl}{rgb}{0.94, 0.92, 0.84}
\definecolor{peach-yellow}{rgb}{0.98, 0.87, 0.68}
\tikzset{every picture/.style={line width=0.75pt}} %set default line width to 0.75pt        
\subfigure[$q=3$]{\label{fig:systemconstHybrid(a)}
    \begin{tikzpicture}[x=0.75pt,y=0.75pt,yscale=-1,scale=0.75]

\draw[draw opacity=0][fill=pearl, rounded corners=25pt] (210pt, 160pt) rectangle (420pt, -10pt) {};

 \draw [color={rgb, 255:red, 155; green, 155; blue, 155 } ] [dash pattern={on 4.5pt off 4.5pt}]  (300,113) -- (550,113) ;
% %Straight Lines 
\draw [color={rgb, 255:red, 155; green, 155; blue, 155 }] [dash pattern={on 4.5pt off 4.5pt}]  (420,200) -- (420,25) ;

\foreach \y in {110pt,60pt,10pt} {
            
         % \draw[-latex,color={rgb, 255:red, 74; green, 144; blue, 226 }] (375,\y+25pt) -- (530,\y+25pt);
    \draw[fill=black!90] (240pt,25pt+\y) -- (390pt,25pt+\y);

}

\draw (240pt,125pt) node   {\footnotesize $0$};
\draw (390pt,125pt) node   {\footnotesize $3$};
\draw (240pt,75pt) node   {\footnotesize $3$};
\draw (390pt,75pt) node   {\footnotesize $6$};
\draw (240pt,25pt) node   {\footnotesize $6$};
\draw (390pt,25pt) node   {\footnotesize $9$};

\draw (430,5) node   {$\mathscr{E}_3(s)$};
      % spacing arrows
      % \draw[<->] (1cm,0.5cm) -- (2cm,0.5cm) node[midway,below,font=\footnotesize] {$2d$};
      % \draw[<->] (0.5cm,1cm) -- (0.5cm,3cm) node[midway,left,font=\footnotesize] {$3d$};
    \end{tikzpicture}}\subfigure[$q=4$]{\label{fig:systemconstHybrid(c)}
\begin{tikzpicture}[x=0.75pt,y=0.75pt,yscale=-1, scale=0.75]

\draw[draw opacity=0][fill=pearl, rounded corners=25pt] (250pt, 210pt) rectangle (410pt, 0) {};

\draw [color={rgb, 255:red, 155; green, 155; blue, 155 } ] [dash pattern={on 4.5pt off 4.5pt}]  (340,155) -- (540,155) ;
%Straight Lines 
\draw [color={rgb, 255:red, 155; green, 155; blue, 155 }] [dash pattern={on 4.5pt off 4.5pt}]  (440,250) -- (440,45) ;

% Text Node
\draw (445,20) node   {$\mathscr{E}_4(s)$};

\draw (280pt,175pt) node   {\footnotesize $0$};
\draw (385pt,175pt) node   {\footnotesize $4$};
\draw (280pt,125pt) node   {\footnotesize $4$};
\draw (385pt,125pt) node   {\footnotesize $8$};
\draw (280pt,75pt) node   {\footnotesize $8$};
\draw (385pt,75pt) node   {\footnotesize $12$};
\draw (280pt,25pt) node   {\footnotesize $12$};
\draw (385pt,25pt) node   {\footnotesize $16$};

% \newcounter{wa}\setcounter{wa}{0}

\foreach \y in {160pt,110pt,60pt,10pt} {
            
         % \draw[-latex,color={rgb, 255:red, 74; green, 144; blue, 226 }] (375,\y+25pt) -- (530,\y+25pt);
         \draw[fill=black!90] (280pt,25pt+\y) -- (385pt,25pt+\y);
}

 \draw[latex-latex, color={rgb, 255:red, 20; green, 20; blue, 20 }] (380,255) -- (510,255);
\draw (430,265) node [font=\footnotesize]  {\color{bluegray} $d_1$};

\draw[latex-latex, color={rgb, 255:red, 20; green, 20; blue, 20 }] (362,240) -- (362,185);
\draw (352,210) node [font=\footnotesize]  {\color{bluegray}$d_2$};

\end{tikzpicture}
}
\caption{Examples of hybrid digital–analog modulation constellations \( \mathscr{E}_q(s) \) for \( (q, n) \in \{(3,3), (4,4)\} \), with \( \mathbb{G} = [0,9] \) and \( \mathbb{G} = [0,16] \).}
    \label{fig:systemconstHybrid}
\end{figure}
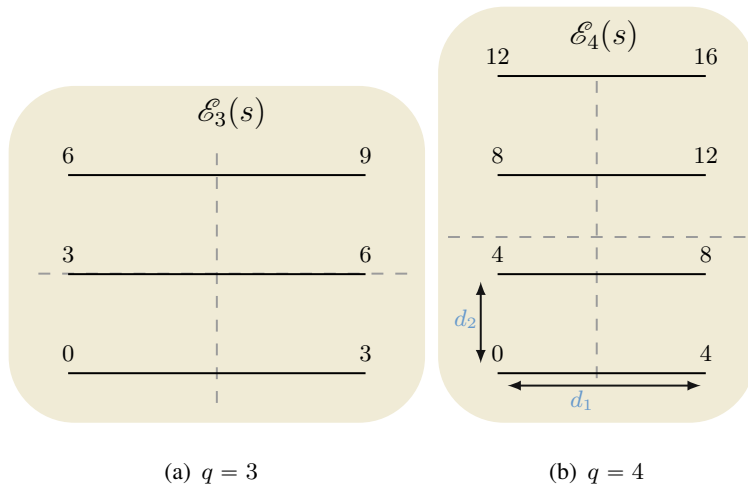

Let \( \mathbb{G} = [0, a] \), where \( a = n \times q \) for two positive parameters \( (q, n) \). The interval \( \mathbb{G} \) is partitioned into \( n \) subintervals of length \( q \), i.e., \( [q(i-1), q i) \) for \( i \in \{1,2,\ldots,n\} \). Each subinterval corresponds to a discrete level represented along the imaginary axis, while the value within each subinterval is transmitted by analog amplitude modulation along the real axis. In this way, the scheme combines a digital pulse-amplitude modulation (PAM) to select the subinterval index \( i = \lfloor s/q \rfloor \) with an analog modulation to transmit the residual component \( s - q\lfloor s/q \rfloor \). More precisely, for any \( s \in \mathbb{G} \), the proposed hybrid encoder is defined as
%-------------
\begin{align}
\label{eq:hybrid}
\mathscr{E}_q(s)
:= \big(s - q \lfloor s/q \rfloor \big)d_1
+ \lfloor s/q \rfloor d_2\,i + \chi_k, 
\qquad s \in [0, a],
\end{align}
%-------------
where \( d_1 \) represents the amplitude scaling for the analog component, and \( d_2 \) determines the spacing between the digital constellation points along the imaginary axis. 
This formulation enables continuous-valued inputs to be transmitted efficiently while preserving the digital constellation structure.  Figure~\ref{fig:systemconstHybrid} illustrates the structure of the proposed hybrid encoding scheme for \( (q,n) \in \{(3,3),(4,4)\} \) and $\mathbb{G}=[0,9]$ and $\mathbb{G}=[0,16]$.

%--------------------
%--------------------

The average transmit power of this constellation is given by
%-------------
\begin{equation}
\label{eq:power-hybrid}
\mathbb{E}\!\left[|\mathscr{E}_q(s)|^2\right]
= |d_1|^2\,\frac{q^2}{3}
+ |d_2|^2\,\frac{(n - 1)(2n - 1)}{6}.
\end{equation}
%-------------
As shown in Sections~\ref{sec:MLdecoder-Est} and~\ref{sec:MAPDecoder}, the real and imaginary components decouple and are detected independently along orthogonal axes.  The receiver output estimate is given by
%---------------
\begin{align}
    \hat{f} = \mathscr{D}(r) := \mathfrak{Re}(\hat{r}) + q\,\mathfrak{Im}(\hat{r}),
\end{align}
%---------------
where the ML and MAP decoding rules are as defined in \eqref{eq:Quntizer-ml}–\eqref{eq:deocder_qam}. 
Along the analog (real) axis, the ML decision rule behaves almost linearly except for truncation at the boundaries, i.e.,
%-------------
\begin{align}
\mathfrak{Re}(\hat{r}) = 
\begin{cases}
0, & \mathfrak{Re}(r) < -\chi,\\
\frac{\mathfrak{Re}(r)-K\chi}{d_1} , 
& -\chi \le \mathfrak{Re}(r) \le {K q d_1}-\chi,\\
K q, & \mathfrak{Re}(r) > {q K d_1}-\chi,
\end{cases}
\end{align}
%-------------
where $\chi = \sum_k\chi_k$,  On the digital (imaginary) axis, the ML decision corresponds to a nearest-neighbor slicing on a one-dimensional PAM grid with spacing \( d_2 \) and \( N_{2,K} = K(n - 1) + 1 \) constellation points, as described in Section~\ref{sec:MLdecoder-Est}.  For the imaginary component, both the decoding procedure and the corresponding MSE expressions remain identical to those derived previously. Moreover, we note that the sum is still decodeable as the real part axis is just a linear operator. 

To determine the optimal modulation parameters for the hybrid scheme, the same analytical framework used in Appendix~\ref{sec:Optimal} can be applied. 
The key difference lies in the treatment of the analog component, where the derivation must explicitly account for the truncation boundaries in the real domain. The proposed hybrid modulation framework can thus be regarded as a generalization of analog OAC schemes found in the literature.  Notably, given the optimal parameter pair \( (d_1, d_2) \), the proposed hybrid approach can outperform purely analog modulation, which can be viewed as a special case of this generalized scheme.

\subsection{Generalization to Higher Dimensions}\label{sec:higherN}

Thus far, we have derived the optimal parameters for a two-dimensional constellation grid, corresponding to the in-phase and quadrature components of digital modulation used for sum computation over the MAC.  However, the proposed framework naturally extends to higher dimensions, allowing computation over multiple communication resources. 
Indeed, the optimization method is not restricted to a two-dimensional grid and can be generalized to an $N$-dimensional grid, where each dimension represents a communication resource such as a time slot or a frequency band.

Consider the computation of a function $f(s_1, \ldots, s_K)$ over $N$ communication resources. 
Let a number $c \in [0,1]$ be represented as
%-------------
\begin{align}
    c = \sum_{i=1}^{N} c_i q^{i-1} = \langle \bm{c}, \bm{q} \rangle,
\end{align}
%-------------
where $\bm{c} = [c_1, \ldots, c_N]^{\mathsf{T}} \in \mathbb{R}^{N}$ and $\bm{q} = [1, q, q^2, \ldots, q^{N-1}]^{\mathsf{T}} \in \mathbb{R}^{N}$. 
The encoder $\mathscr{E}_q^N : \mathbb{Z} \to \mathbb{Z}^N$ is defined for an input vector $\bm{c} \in \mathbb{Z}_q^N$ and a parameter vector $\bm{d} := [d_1,\ldots, d_N]^{\mathsf{T}}\in \mathbb{R}^N$ as
%-------------
\begin{align}
\mathscr{E}_q^N(c) := \bm{d} \odot \bm{c} = [d_1 c_1,\, d_2 c_2,\, \ldots,\, d_N c_N],
\end{align}
%-------------
which maps one-dimensional data onto an $N$-dimensional grid. 
For simplicity, we assume square constellations such that $Q_k = Q = q \times q$ for all $k \in [K]$.  The average power of the resulting $N$-dimensional constellation is then given by
%-------------
\begin{align}
  \mathbb{E}_{c}\big[  \|\mathscr{E}_q^N(c)\|_2^2] = \frac{q^2 - 1}{12} \|\bm{d}\|_2^2.
\end{align}
%-------------
Since each dimension corresponds to an independent channel use, the noise variance may differ across dimensions. Let \( z_i \) denote the noise component on dimension \( i \), with \( z_i \sim \mathcal{N}(0, \sigma_i) \) for \( i \in [N] \), where $\sigma_i^2$ is the variance of noise along dimension $i$. Then,  the corresponding MSE of the $N$-dimensional constellation is expressed as
%-------------
\begin{align}
\label{eq:MSEboundN}
    \text{MSE}(\hat{f}, \bm{d}) = \sum_{i=1}^{N} q^{i-1} \mu_i(d_i), 
\end{align}
%-------------
where $\mu_i(x) = \sum_{m=1}^{N-1} \alpha_m Q((2m-1)x/ \sigma_i)$ with $\alpha_m := 2m -1 + \frac{3m(1-m)-1}{K(q - 1) + 1}$.  The optimal design problem is thus formulated as
%-------------
\begin{align}
\label{eq:OptmialMSEgeneralN}
    \min_{\bm{d}} \sum_{i=1}^{N} q^{i-1} \mu_i(d_i), \quad \|\bm{d}\|_2^2 = \frac{12P}{q^2 - 1}.
\end{align}
%-------------
where $P$ is the total power budget. For simplicity, we assume that the power is uniform across all dimensions. When \( N = 2 \),\( q = n \), and $\sigma_1=\cdots=\sigma_N$, the above formulation reduces to the two-dimensional constellation design derived in Section~\ref{sec:system}. 

%----------------
% --- where you want the figure ---
\begin{figure}[t]
  \centering
  \begin{tikzpicture}
    \begin{axis}[
      width=0.8\linewidth,
      height=6cm,
      xlabel={\scriptsize  SNR $\xi$ (dB)},
      ylabel={\scriptsize Value},
      xmin=0, xmax=22,
      grid=both,
    minor tick num=5,
    legend style={nodes={scale=0.75, transform shape}, at={(0.2,0.75)}}, 
        ymajorgrids=true,
        xmajorgrids=true,
        grid style=dashed,
        grid=both,
        grid style={line width=.1pt, draw=gray!15},
        major grid style={line width=.2pt,draw=gray!40},
      % grid style={densely dotted},
      % legend cell align=left,
      % legend pos=north west,
      % tick align=outside,
      % tick style={black},
      % line width=0.9pt,
    ]
      % File: Data/OPTd_unscaled_q4_K10_P1.dat
      % Columns: [0]=xi_dB, [1]=d1, [2]=d2, [3]=d3
      \addplot+[ mark=triangle, mark options = {rotate = 90}, color=eggplant, thick]
        table[x=xi_dB,y=d1] {Data/Noptimal.dat};
      \addplot+[mark=triangle,  mark options = {rotate = 90}, color= brown(web), thick, dashed]
        table[x=xi_dB,y=d2] {Data/Noptimal.dat};
      % \addlegendentry{$d_2^\ast$}
      \addplot+[ mark= triangle, mark options = {rotate = 90}, color= burntumber, thick, dotted]
       table[x=xi_dB,y=d3]
       {Data/Noptimal.dat};
      % \addlegendentry{$d_3^\ast$}
       \legend{$d_1$,$d_2$,$d_3$};
    \end{axis}
  \end{tikzpicture}
  \caption{Optimal spacings $d_1,d_2,d_3$ versus different SNR $\xi$ for $N=3$, $K=10$, $q=4$, and $P=1$, obtained via \eqref{eq:optimal_nonlinear_N}.}
  \label{fig:opt-d-vs-snr}
\end{figure}
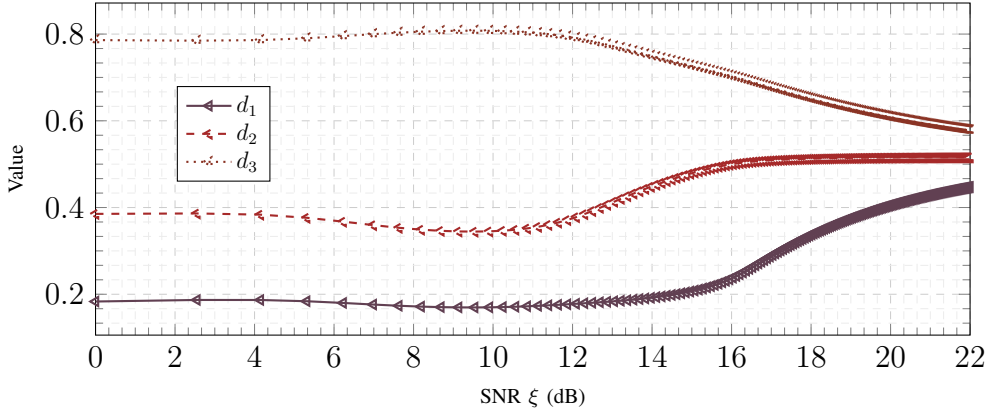
%----------------

Applying the Lagrangian formulation to \eqref{eq:OptmialMSEgeneralN} and enforcing the first-order optimality conditions yields the following nonlinear system:
%-------------
\begin{align}
\label{eq:optimal_nonlinear_N}
    {g}_{q}\Big(\frac{{d}_i}{\sigma_i},\frac{{d}_{i+1}}{\sigma_{i+1}}\Big) = 0, \quad \forall~i \in \{1,\ldots, N-1\}, \quad \|\bm{d}\|_2^2 = r^2,
\end{align}
%-------------
where $g_q(x,y)$ is defined in \eqref{eq:gq(x,y)} for $N_{1,K} = N_{2,K} = K(q - 1) + 1$, 
$r = \sqrt{12P / (q^2 - 1)}$ with $d_1 \le {d}_2 \le \cdots \le {d}_N$. 
%and $\bm{{\Delta}} = [{\Delta}_1, \ldots, {\Delta}_N]^{\mathsf{T}}$ 
%The parameters are related through $\bm{d} = \bm{\sigma} \odot \bm{{\Delta}}$, where $\bm{\sigma} := [\sigma_1,\ldots,\sigma_N]^{\mathsf{T}}$ and $\odot$ stands for Hadamard product between two vectors. 

Solving the nonlinear system in \eqref{eq:optimal_nonlinear_N} may yield multiple optimal or suboptimal solutions.  To ensure uniqueness, the implicit function $g_q(x,y)$ can be replaced by its convexified approximation $\tilde{g}_q(x,y)$, defined in \eqref{eq:gq2tild}, resulting in the modified system
%-------------
\begin{align}
\label{eq:optimal_nonlinear_approx_N}
    \tilde{g}_q\Big(\frac{{d}_i}{\sigma_i},\frac{{d}_{i+1}}{\sigma_{i+1}}\Big) = 0, \quad \forall~i \in \{1,\ldots, N-1\}, \quad \|\bm{{d}}\|_2^2 = r^2.
\end{align}
%-------------
The solution to \eqref{eq:optimal_nonlinear_approx_N} is guaranteed to be unique for sufficiently large $r$, corresponding to the high-SNR regime.  Here, $\tilde{g}_q$ is derived by removing the negative coefficients from $g_q$, which can be interpreted as a convex relaxation of the original problem.  However, characterizing the exact SNR conditions that ensure uniqueness in the low-SNR regime remains an open problem for future investigation. The simulation results in Figure~\ref{fig:opt-d-vs-snr} illustrate that, as $\xi$ increases, the optimal spacings $d_1^{*}, d_2^{*}, d_3^{*}$ converge to equal values (each $\approx 1/3$ of the total power), consistent with the observed trend in Figure~\ref{fig:Lambert_solution}.

%===================================================
\section{Conclusion}\label{sec:conclusion}
%===================================================

We studied the design of optimal digital constellations for OAC over MACs, focusing particularly on the summation function.  Using a two-dimensional grid representation, we formulated a power-constrained optimization problem that minimizes the MSE of function computation.  We derived the corresponding optimality conditions as a system of nonlinear equations and established the uniqueness of their solutions through geometric analysis. 
In the high SNR regime, we obtained closed-form approximations based on the generalized Lambert function, which provide valuable analytical insight into the system behavior. Furthermore, we extended the framework to accommodate MAP decoding and derived the associated optimal modulation parameters.

The proposed approach also generalizes naturally to multidimensional constellations and non-Gaussian noise environments. The solution can also be easily extended to other nomographic functions.  Numerical results demonstrate that the optimized constellations significantly outperform conventional QAM-shaped schemes, achieving lower computation error and enhanced robustness. A key takeaway is that the SNR thresholds required for attaining the optimal MSE are relatively low—owing to the inherent signal superposition property of OAC—and the KKT conditions uniquely determine the optimal constellation parameters once these thresholds are met.

Future research will explore hybrid analog–digital constellations with continuous or mixed alphabets for OAC and will derive the MMSE-optimal encoder parameters for continuous-valued function estimation.  Another direction of interest is the analytical characterization of the optimal modulation pair $(q, n)$ as a function of SNR. 
Preliminary results indicate a transition from a pulse-amplitude-modulation–like configuration ($n=1$, $q=Q$) at very low SNR to a square-QAM structure ($q=n=\sqrt{Q}$) at high SNR.  Determining the exact SNR threshold for this transition remains an open problem for future investigation.

\appendix

\subsection{Proof of Theorem~\ref{th:Optmizaiton}}\label{sec:Optimal}

\begin{figure}[t]
\centering
\usetikzlibrary{arrows.meta, positioning, shapes.multipart}
\tikzset{
  block/.style={
    rectangle split, rectangle split parts=2,
    rectangle split part fill={black!10,white},
    draw=black!70, rounded corners=2pt,
    text width=42mm, align=center,
    font=\footnotesize
  },
  arrow/.style={-Latex, thick}
}
\scalebox{0.9}{
\begin{tikzpicture}[node distance=15mm and 20mm]

  % Nodes
  \node[block] (mse)
    {Proposition~\ref{Pr:MSE(1)}
     \nodepart{second}  Derivation of the exact MSE as a function of $(d_1,d_2)$};

  \node[block, right=of mse] (kkt)
    {KKT Conditions
     \nodepart{second} Reformulation as a nonlinear system~\eqref{eq:optimal_nonlinear}};

  \node[block, right=of kkt] (stationary)
    {From \eqref{eq:optimal_nonlinear}
     \nodepart{second} Identifying all possible solutions};

  \node[block, below=of stationary] (hessian)
    {Lemmas~\ref{lem:Boundries} and  \ref{lem:Boundries2}
     \nodepart{second} Bordered Hessian test to \\ identify the saddle points and SNR threshold~$\xi_1$};

  % \node[block, left=of hessian] (global)
  %   { Lemma \ref{lem:Boundries2}
  %    \nodepart{second}   };

  \node[block, left=of hessian] (encoder)
    {Proposition~\ref{Pr:Optmizaiton}
     \nodepart{second} 
  Ensuring the uniqueness of the central intersection};

  % Arrows
  \draw[-latex] (mse) -- (kkt);
  \draw[-latex] (kkt) -- (stationary);
  \draw[-latex] (stationary) -- (hessian);
  \draw[-latex] (hessian) -- (encoder);
  % \draw[arrow] (global) -- (encoder);

\end{tikzpicture}
}
\caption{Proof outline for Theorem~\ref{th:Optmizaiton}. 
Proposition~\ref{Pr:MSE(1)} provides a closed-form MSE approximation in terms of the encoder parameters.  The KKT conditions lead to the nonlinear system~\eqref{eq:optimal_nonlinear}. 
Applying Lemmas~\ref{lem:Boundries} and~\ref{lem:Boundries2} with the bordered Hessian identifies the saddle points and establishes the SNR threshold~$\xi_1$. 
Finally, Proposition~\ref{Pr:Optmizaiton} proves the uniqueness of the global minimizer and the corresponding optimal encoder parameters~$(d_1^*,d_2^*)$}
\label{fig:proof_theorem1}
\end{figure}
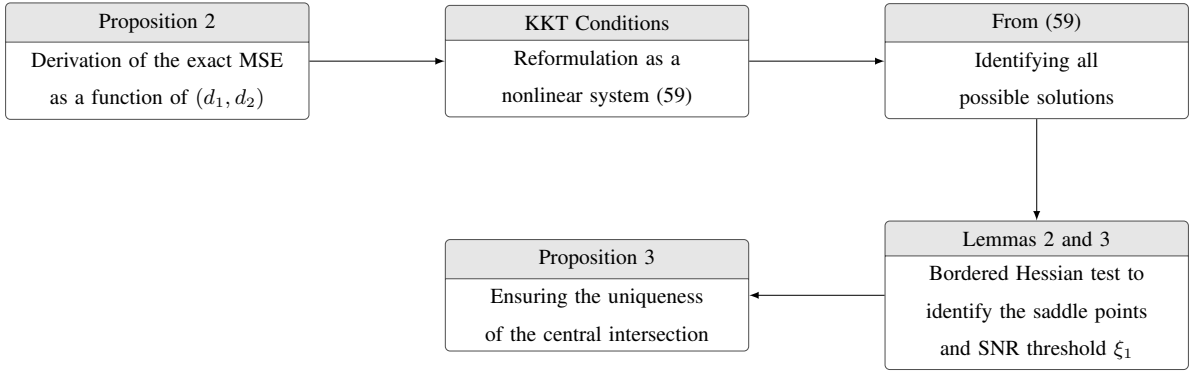

Note that the distances among the constellation points of the superimposed signal, $\sum_{k}x_k$, are the same as the distances of the constellation points of each node, i.e., $x_k$ for all $k$. Hence,  the task of designing the optimal constellation by solving \eqref{eq:OptmialMSEgeneral} for the encoding scheme in Section \ref{sec:encoding} is equivalent to selecting the optimal values of the parameters $d_1$ and  $d_2$.  To proceed, we first derive a closed-form expression for the MSE of the decoder and encoding in terms of $d_1$ and  $d_2$. Thereafter, we formulate and solve a constrained optimization problem that minimizes this MSE under an average power budget for the constellation. Figure~\ref{fig:proof_theorem1} summarizes the overall proof procedure.    Specifically, for transmission over an AWGN channel, the resulting MSE can be expressed in terms of the Q-function, as detailed below.

%------------
\begin{prop}\label{Pr:MSE(1)}\cite[Proposition 1]{Razavikia2024Ring}
    Consider a communication network with $K$ nodes where the nodes use the encoder $\mathscr{E}_q$ in \eqref{eq:encoding_qam} and ML decoder $\mathscr{D}$ in \eqref{eq:deocder_qam} to compute the sum $f = \sum_{k}s_k$, where $s_k\in \mathbb{Z}$ over the noisy MAC with $\mathcal{CN}(0,\sigma^2)$. Assume that the constellation points induced by $\sum_{k}s_k$ have a uniform distribution over $\mathcal{Y}$. Then, the MSE of the function $f$ in Eq.~\eqref{eq:MSEdefinition} becomes  
    \begin{align}
        \label{eq:MSEd1q2d2}
        \text{MSE}(f,\hat{f})  = \mu_1(d_1) + q^2\mu_2(d_2),
    \end{align}
    where
    \begin{align}
\label{eq:mu_definiation}
         \mu_1(x) =  2    \sum_{m=1}^{N_{1,K}-1}\alpha_{1,m}  Q\Big(\frac{(2m-1)x}{\sqrt{2}\sigma}\Big), \quad \mu_2(x) =  2    \sum_{m=1}^{N_{2,K}-1}\alpha_{2,m}  Q\Big(\frac{(2m-1)x}{\sqrt{2}\sigma}\Big),
    \end{align}
      and $\alpha_{1,m} := 2m -1 + \frac{3m(1-m)-1}{N_{1,K}}$, and $\alpha_{2,m} := 2m -1 + \frac{3m(1-m)-1}{N_{2,K}}$ with $N_{1,K}=K \times (q-1)+1$, $N_{2,K}=K \times (n-1)+1$, and $Q(x)$ is the Gaussian $Q$ function, i.e., 
% -----------
\begin{align}
    Q(x)  = \frac{1}{\sqrt{2\pi}}\int_{x}^{\infty}{\rm e}^{-\frac{t^2}{2}}dt.
\end{align}
% -----------
\end{prop}
% -----------
Proposition~\ref{Pr:MSE(1)} reformulates the MSE in Eq.~\eqref{eq:MSEdefinition} as a weighted sum of Q-functions, where the arguments of the Q-function are determined by $d_1$ and $d_2$. By invoking the closed-form expression of the MSE in \eqref{eq:MSEd1q2d2}, we can recast the general optimization problem in \eqref{eq:OptmialMSEgeneral} to a minimization over parameters $d_1$ and $d_2$ of the encoder $\mathscr{E}_q$. i.e.,   
%-------------
\begin{align}
\min_{d_1,d_2} \mu_1(d_1)+ q^2\mu_2(d_2),~~\text{s.t.}~ \frac{d_1^2}{\tilde{\Upsilon}_1^2} + \frac{d_2^2}{\tilde{\Upsilon}_2^2} = 1,  \label{eq:msed1d2}
\end{align}
%-------------
where $\tilde{\Upsilon}_1 = \sqrt{12P/(q^2-1)}$ and $\tilde{\Upsilon}_2 = \sqrt{12P/(n^2-1)}$. Hence, given the order of the modulation $Q= q\times n$, the power budget $P$, and the variance of the noise $\sigma^2$, solving the optimization in~\eqref{eq:msed1d2} gives us the optimal $d_1$ and $d_2$  to achieve minimum MSE. Note that $P$ and $\sigma^2$ do not act as independent factors; in the sequel, we show that both can be encapsulated by a single parameter—the available signal-to-noise ratio (SNR).

%========================

%========================

To solve the constrained minimization in \eqref{eq:msed1d2}, we form the Lagrangian
%-------------
\begin{align}
 \hspace{-2pt}\mathcal{L}(d_1, d_2, \lambda) =  \mu_1(d_1)+ q^2\mu_2(d_2) + \lambda \left( \frac{1}{\tilde{\Upsilon}_1^2}d_1^2 + \frac{1}{\tilde{\Upsilon}_2^2}d_2^2 - 1 \right),  \label{eq:Lagrangian_function}
\end{align}
%-------------
where $\lambda$ is the Lagrange multiplier, enforcing the power constraint.  The key idea is to apply the Karush–Kuhn–Tucker conditions of the constrained optimization in \eqref{eq:OptmialMSEgeneral} to locate all the critical points of the Lagrangian function  $\mathcal{L}(d_1, d_2, \lambda)$. Subsequently, we employ the bordered Hessian theorem~\cite{d2001bordered} to determine which of these stationary points correspond to the global minimizer of the constrained MSE. Concretely,  to obtain these stationary points, we take the partial derivatives of the Lagrangian with respect to $d_1$, $d_2$, and $\lambda$:
%------------------
\begin{subequations}
\label{eq:Lagderivations}
\begin{align}
\frac{\partial \mathcal{L}}{\partial d_1} &= \frac{ \partial \mu_1{(d_1)}}{\partial d_1} + 2\lambda \frac{d_1}{\tilde{\Upsilon}_1^2} = 0,  \Rightarrow \frac{\tilde{\Upsilon}_1^2}{d_1} \frac{ \partial \mu_1{(d_1)}}{\partial d_1} =-2\lambda,  \\
\frac{\partial \mathcal{L}}{\partial d_2} &= \frac{q^2\partial \mu_2{(d_2)}}{ \partial d_2} + 2\lambda \frac{d_2}{\tilde{\Upsilon}_2^2} = 0,  \Rightarrow \frac{\tilde{\Upsilon}_2^2q^2}{d_2} \frac{\partial \mu_2{(d_2)}}{\partial d_2} =-2\lambda, \\
\frac{\partial \mathcal{L}}{\partial \lambda} &=\frac{d_1^2}{\tilde{\Upsilon}_1^2} + \frac{d_2^2}{\tilde{\Upsilon}_2^2} - 1= 0.
\end{align}
\end{subequations}
%------------------
Eliminating the Lagrange multiplier $\lambda$  by equating the first two expressions in  \eqref{eq:Lagderivations} yields the relationship
%-------------
\begin{align}
\label{eq:MSEequald1d2}
\frac{\partial \mu_1{(d_1)}}{d_1 \partial d_1}  = \frac{\tilde{\Upsilon}_2^2q^2\partial \mu_2{(d_2)}}{\tilde{\Upsilon}_1^2d_2 \partial d_2}, \qquad 
\bigg(\frac{d_1}{\tilde{\Upsilon}_1}\bigg)^2 + \bigg(\frac{d_2}{\tilde{\Upsilon}_2}\bigg)^2 = 1.
\end{align}
%-------------
Since both $\mu_1$ and $\mu_2$ involve Gaussian $Q$-function, their derivatives are computed using
%-------------
\begin{align}
\label{eq:derivative_q}
\frac{d}{dx} Q(x) = -\frac{1}{\sqrt{2\pi}}{\rm e}^{-\frac{x^2}{2}},
\end{align}
%-------------
which facilitates the evaluation of ${\partial \mu_1(d_1)}/{\partial d_1}$ and ${\partial \mu_2(d_2)}/{\partial d_2}$ required in \eqref{eq:MSEequald1d2}. After substitution, we rewrite \eqref{eq:MSEequald1d2} as the following system of equations. 
%-------------
\begin{align*}
\sum_{m=1}^{N_{1,K}-1} \alpha_{1,m}(2m-1)  {\rm e}^{-\theta_{m}  \frac{d_1^2}{\sigma^2}} = 
\frac{d_1\tilde{\Upsilon}_2^2q^2}{\tilde{\Upsilon}_1^2d_2}\sum_{m=1}^{N_{2,K}-1} \alpha_{2,m}(2m-1)  {\rm e}^{-\theta_{m} \frac{d_2^2}{\sigma^2}}, ~
\bigg(\frac{d_1}{\tilde{\Upsilon}_1}\bigg)^2 + \bigg(\frac{d_2}{\tilde{\Upsilon}_2}\bigg)^2  = 1.
\end{align*}
%-------------
where $\theta_{m} = \frac{(2m-1)^2}{4}$, $\forall~m$. To eliminate the dependence on the noise variance $\sigma^2$, we rewrite the equations in terms of SNR, i.e.,
%-------------
     \begin{align}
     \label{eq:exppoly-snr}
 \sum_{m=1}^{N_{1,K}-1} \frac{\gamma_{1,m}}{\Delta_1} {\rm e}^{-\theta_{m}  \Delta_1^2} = 
q^2\sum_{m=1}^{N_{2,K}-1} \frac{\gamma_{2,m}}{\Delta_2} {\rm e}^{-\theta_{m} \Delta_2^2}, \qquad
\frac{\Delta_1^2}{\Upsilon_1^2} + \frac{\Delta_2^2}{\Upsilon_2^2} = 1. 
\end{align}
%-------------
where $\gamma_{1,m} :=\alpha_{1,m}(2m-1)\Upsilon_1^2$ and $\gamma_{2,m} :=\alpha_{2,m}(2m-1)\Upsilon_2^2$. Also, $\Delta_1 :={d}_1/\sigma$ and $\Delta_2 := {d}_2/\sigma$, $\Upsilon_1 =\sqrt{12\xi/(q^2-1)}, \Upsilon_2 =\sqrt{12\xi/(n^2-1)}$  and $\xi = {P}/\sigma^2$ denotes the SNR. Here, \eqref{eq:exppoly-snr} consists of the equality of two nonlinear functions in terms of $\Delta_1$ and $\Delta_2$, which are highly nonlinear.

%-------------
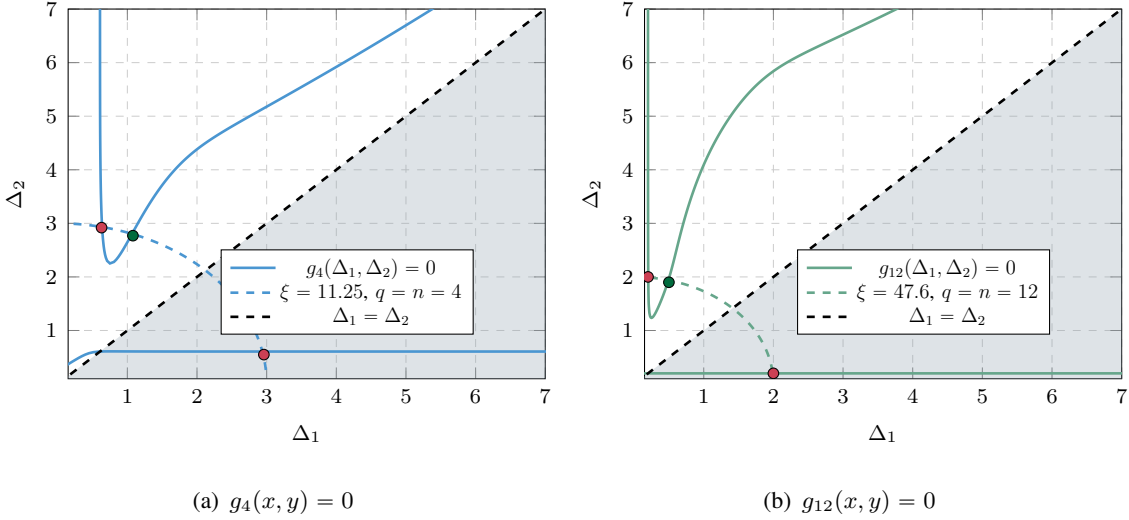
\begin{figure}[!t]
\subfigure[$g_{4}(x,y)=0$]{
\centering
    \begin{tikzpicture} 
    \begin{axis}[
        xlabel={${\Delta_1}$},
        ylabel={$\Delta_2$},
        label style={font=\scriptsize},
        % title = {\scriptsize $???$},
        %legend pos=,
        tick label style={font=\scriptsize} , 
        width=0.48\textwidth,
        height=6.5cm,
        xmin=0.15, xmax=7,
        ymin=0.1, ymax=7,
        xtick={0,1,2,3,4,5,6,7},
        ytick={0,1,2,3,4,5,6,7},
        %ytick={0, 20, 40, 60, 80, 100},
         % ymode = log,
       legend style={nodes={scale=0.65, transform shape}, at={(0.85,0.35)}}, 
        ymajorgrids=true,
        xmajorgrids=true,
        grid style=dashed,
        grid=both,
        grid style={line width=.1pt, draw=gray!15},
        major grid style={line width=.2pt,draw=gray!40},
    ]
    % ============
  % \addplot[
  %      color=cadmiumgreen,
  %       line width=1pt,
  %       % mark size=2pt,
  %       ]
  %   table[x=d1,y=d2]
  %   {data/Contour-12.dat};
    \addplot[
    color=celestialblue,
        line width=1pt,
        mark size=2pt,
        ]
    table[x=d1,y=d2]
    {Data/Contour-41.dat};
    % \draw[dashed] (axis cs:0,0) circle [color=bazaar, radius=2];
    \addplot[mark=none,dashed,
    domain = 0:3,
    line width=1pt,
    color= celestialblue,
    samples=400,
    unbounded coords=jump
    ] { sqrt(9-x^2)};
    % \addplot[mark=none,dashed,
    % domain = 0:2,
    % line width=1pt,
    % color= cadmiumgreen,
    % samples=400,
    % unbounded coords=jump
    % ] { sqrt(4-x^2)};
     \addplot[mark=none,dashed,
    domain = 0:9,
    line width=1pt,
    color= black,
    samples=400,
    unbounded coords=jump
    ] {x};
    \addplot[name path=line, draw=none, domain=0:9] {x};
   \addplot[name path=vertical, draw=none] coordinates {(0,0) (9,0)};
   \addplot[fill=cadetgrey, fill opacity=0.3, draw=none] fill between[of=line and vertical];
    \draw[fill=cadmiumgreen](axis cs:1.08,2.77) circle[radius=2pt];
    \draw[fill=brickred](axis cs:2.96,0.55) circle[radius=2pt];
    \draw[fill=brickred](axis cs:0.63,2.92) circle[radius=2pt];
    \legend{ {$g_{4}(\Delta_1,\Delta_2)=0$}, {$\xi=11.25$, $q=n=4$},{$\Delta_1=\Delta_2$}};
    \end{axis}
\end{tikzpicture}}\subfigure[$g_{12}(x,y)=0$]{
\centering
    \begin{tikzpicture} 
    \begin{axis}[
        xlabel={${\Delta_1}$},
        ylabel={$\Delta_2$},
        label style={font=\scriptsize},
        % title = {\scriptsize $???$},
        %legend pos=,
        tick label style={font=\scriptsize} , 
        width=0.48\textwidth,
        height=6.5cm,
        xmin=0.15, xmax=7,
        ymin=0.1, ymax=7,
        xtick={0,1,2,3,4,5,6,7,8,9},
        ytick={0,1,2,3,4,5,6,7,8,9},
        %ytick={0, 20, 40, 60, 80, 100},
         % ymode = log,
       legend style={nodes={scale=0.65, transform shape}, at={(0.85,0.35)}}, 
        ymajorgrids=true,
        xmajorgrids=true,
        grid style=dashed,
        grid=both,
        grid style={line width=.1pt, draw=gray!15},
        major grid style={line width=.2pt,draw=gray!40},
    ]
    % ============
    % \addplot3[surf, shader=flat, colormap name=hot] file {Data/surface_data.dat};
  \addplot[
       color=cadmiumgreen!60,
        line width=1pt,
        ]
    table[x=d1,y=d2]
    {Data/Contour-12.dat};
    % \draw[dashed] (axis cs:0,0) circle [color=bazaar, radius=2];
    \addplot[mark=none,dashed,
    domain = 0:2,
    line width=1pt,
    color= cadmiumgreen!60,
    samples=400,
    unbounded coords=jump
    ] { sqrt(4-x^2)};
     \addplot[mark=none,dashed,
    domain = 0:9,
    line width=1pt,
    color= black,
    samples=400,
    unbounded coords=jump
    ] {x};
    \addplot[name path=line, draw=none, domain=0:9] {x};
   \addplot[name path=vertical, draw=none] coordinates {(0,0) (9,0)};
   \addplot[fill=cadetgrey, fill opacity=0.3, draw=none] fill between[of=line and vertical];
    \draw[fill=cadmiumgreen](axis cs:0.5,1.9) circle[radius=2pt];
    \draw[fill=brickred](axis cs:2,0.2) circle[radius=2pt];
    \draw[fill=brickred](axis cs:0.2,2) circle[radius=2pt];
    \legend{{$g_{12}(\Delta_1,\Delta_2)=0$}, {$\xi=47.6$, $q=n=12$},{$\Delta_1=\Delta_2$}};
    \end{axis}
\end{tikzpicture}}
  \caption{The implicit function ${g}_{q}(\Delta_1,\Delta_2) =0$, with power constraint for $n=q$, which is a circle, i.e.,  $\Delta_1^2 + \Delta_2^2= \frac{6\xi}{q^2-1}$.   }
  \label{fig:ExpG_qF_q}
\end{figure}
%-------------

Therefore, solving the optimization problem~\eqref{eq:OptmialMSEgeneral}, that is, finding the optimal parameters of the constellation diagram, reduces to finding the solutions of the following nonlinear system:
%-------------
\begin{align}
    \label{eq:optimal_nonlinear}
     g_q(\Delta_1,\Delta_2)= 0, ~ \frac{\Delta_1^2}{\Upsilon_1^2} + \frac{\Delta_2^2}{\Upsilon_2^2} = 1,
\end{align}
%-------------
where, 
%-------------
\begin{align}
     \label{eq:gq(x,y)}
    g_q(x,y) :=  \sum_{m=1}^{N_{1,K}-1} \gamma_{1,m} \frac{{\rm e}^{-\theta_{m}  x^2}}{x} - q^2\sum_{m=1}^{N_{2,K}-1}\gamma_{2,m}\frac{{\rm e}^{-\theta_{m}  y^2}}{y},\quad x,y>0.
\end{align}
%-------------
It is worth noting that $\mathcal{G}_Q^{N}(t)$ defined in \eqref{eq:MathcalG} corresponds exactly to the above system. In particular, by introducing the parametric representation  $x = \Upsilon_1\sqrt{0.5-t}$ and $y = \Upsilon_2\sqrt{0.5+t}$ for $t\in [0,1/2)$, we can express
%-------------
\begin{align}
   \label{eq:bivariaterep}
    g_q( \Upsilon_1\sqrt{0.5-t}, \Upsilon_2\sqrt{0.5+t} ) = \mathcal{G}_Q^{N}(t). 
\end{align}
%-------------
Therefore, for clarity of exposition, we employ $g_q(x,y)$ in place of $\mathcal{G}_Q^{N}$.

The solutions to \eqref{eq:optimal_nonlinear} depend solely on three parameters: the modulation levels $(q,n)$, SNR $\xi$, and the number of nodes $K$. Determining the optimal values of $\Delta_1$ and $\Delta_2$ thus requires analyzing the geometry of the implicit curve defined by $g_q(x,y)=0$.

To illustrate this relationship, we plot the implicit function $ g_q(\Delta_1,\Delta_2) = 0$ for various values of $q$ and $\xi$\footnote{Note that the number of nodes $K$ also plays a role in the function $g_q$. However, because $\theta_m$ increases quadratically, the effect of exponential terms in the sum $g_q(x,y)\approx 0$ for $m\geq q+1$ is almost negligible, and the effect of $K\geq2$ is mainly shown in the low SNR region.} in Figure~\ref{fig:ExpG_qF_q}. For sufficiently large SNR values $\xi$,  i.e.,  when the ellipse in the power constraint becomes wider, system~\eqref{eq:optimal_nonlinear} typically yields three intersection points between the power constraint ellipse and the $g_q(x,y) = 0 $ contour, corresponding to the critical points of the Lagrangian function.

%-------------
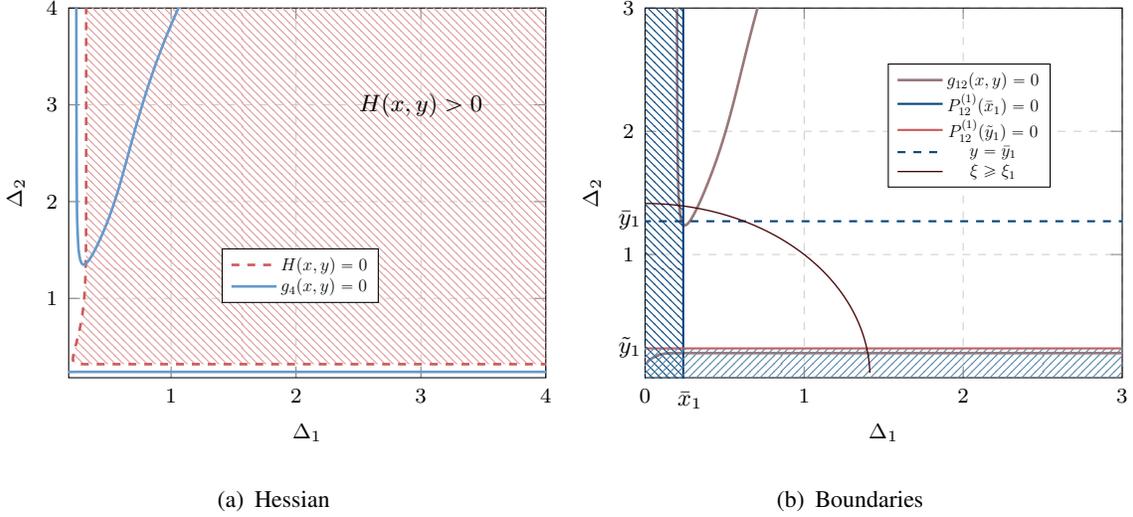
\begin{figure*}[!t]
\centering
\subfigure[Hessian]{  \label{fig:Hessianzero}
    \begin{tikzpicture} 
    \begin{axis}[
        xlabel={${\Delta_1}$},
        ylabel={$\Delta_2$},
        label style={font=\scriptsize},
        % title = {\scriptsize $???$},
        %legend pos=,
        tick label style={font=\scriptsize} , 
        width=0.48\textwidth,
         height=6.5cm,
        xmin=0.18, xmax=4,
        ymin=0.18, ymax=4,
        xtick={0,1,2,3,4},
        ytick={0,1,2,3,4},
        %ytick={0, 20, 40, 60, 80, 100},
         % ymode = log,
       legend style={nodes={scale=0.55, transform shape}, at={(0.65,0.35)}}, 
        ymajorgrids=true,
        xmajorgrids=true,
        grid style=dashed,
        grid=both,
        grid style={line width=.1pt, draw=gray!15},
        major grid style={line width=.2pt,draw=gray!40},
    ]
     \addplot[name path=curve, color=chestnut, dashed,
        line width=1pt,
        mark size=2pt,
        ]
    table[x=d1,y=d2]
    {Data/Hessian_o.dat};
    \addplot[color=bluegray,
        line width=1pt,
        ]
    table[x=d1,y=d2]
    {Data/Hessian_g.dat};
% axis edges (relative coords are in [0,1])
\path[name path=rightedge] (rel axis cs:1,0) -- (rel axis cs:1,1);
\path[name path=topedge]   (rel axis cs:0,1) -- (rel axis cs:1,1);
% fill RIGHT of the curve (toward x = xmax)
\addplot[
  on layer=axis background,
  pattern=north west lines, % or: fill=blue!12
  pattern color=chestnut!60,
  draw=none,
  forget plot
] fill between[
    of=curve and rightedge,
    split
];
  \node at (3,3) {\footnotesize $H(x,y)>0$};
    % ============
    \legend{{$H(x,y) =0$}, {$g_4(x,y) = 0$}};
    \end{axis}
\end{tikzpicture}}\subfigure[Boundaries]{\label{fig:g_tilde(b)bound}
    \begin{tikzpicture} 
    \begin{axis}[
        xlabel={${\Delta_1}$},
        ylabel={$\Delta_2$},
        label style={font=\scriptsize},
        % title = {\scriptsize $???$},
        %legend pos=,
        tick label style={font=\scriptsize} , 
        width=0.48\textwidth,
        height=6.5cm,
        xmin=0, xmax=3,
        ymin=0, ymax=3,
        xtick={0,1,2,3,4,5,6,7,8,9},
        ytick={1,2,3,4,5,6,7,8,9},
        %ytick={0, 20, 40, 60, 80, 100},
         % ymode = log,
       legend style={nodes={scale=0.55, transform shape}, at={(0.85,0.85)}}, 
        ymajorgrids=true,
        xmajorgrids=true,
        grid style=dashed,
        grid=both,
        grid style={line width=.1pt, draw=gray!15},
        major grid style={line width=.2pt,draw=gray!40},
    ]
    % ============
  \addplot[
       color=bazaar,
        line width=1pt,
        ]
    table[x=d1,y=d2]
    {Data/Contour-12.dat};
    %------ q=4
    \addplot[thick, color=darkcerulean, mark=none] (0.24,x);
    \addplot[thick, color=chestnut, mark=none] {0.238};
     \addplot[thick, color=darkcerulean, dashed, mark=none] {1.27};
    \draw[pattern=north west lines, pattern color=darkcerulean, draw=none] (0,0) rectangle (0.24,3);
      \draw[pattern=north east lines, pattern color=airforceblue, draw=none] (0,0) rectangle (3,0.237);
        \addplot[
    domain = 0:1.42,
    line width=0.5pt,
    mark size = 1pt,
    color= bulgarianrose,
    samples=450,
    unbounded coords=jump
    ] { sqrt(2-x^2)};
    \legend{ {$g_{12}(x,y)=0$}, {$P^{(1)}_{12}(\bar{x}_1)=0$},  {$P^{(1)}_{12}(\tilde{y}_1)=0$}, {$y=\bar{y}_1$}, {$\xi\geq \xi_1$}};
    \end{axis}
    \node at (0.6,-0.3) {\footnotesize $\bar{x}_1$};
    \node at (-0.2,2.1) {\footnotesize $\bar{y}_1$};
      \node at (-0.2,0.4) {\footnotesize $\tilde{y}_1$};
\end{tikzpicture}}
\caption{The determinant $H(x,y)$ corresponds to the Hessian of the Lagrangian cost function defined in~\eqref{eq:Lagrangian_function}. The plot illustrates the zero-level set of $H(x,y)$, which delineates the local maxima and minima of the optimization problem in~\eqref{eq:OptmialMSEgeneral}.}
\end{figure*}
%-------------

\subsubsection{Optimal Versus Sub-optimal Solutions}\label{sec:optimalverssub}

Here, we derive the conditions that characterize the optimal solution of the nonlinear system in~\eqref{eq:optimal_nonlinear}. Among the three intersection points between the power-constraint ellipse and the implicit curve \(g_q(x,y)=0\), it can be shown that the two outer intersections correspond to saddle points. Consequently, by exclusion, the middle intersection represents the global minimizer of the Lagrangian function. 

To substantiate this, Figure~\ref{fig:Hessianzero} illustrates the sub-level set where the determinant of the Hessian of the Lagrangian, denoted by \(\mathcal{L}\), equals zero. The region located below and to the left of the red boundary in Figure~\ref{fig:Hessianzero} corresponds to points for which the Hessian determinant is negative—indicating the absence of a local minimum. 

The subsequent lemmas analytically determine the boundary demarcating these regions by deriving and evaluating the roots of the characteristic polynomials associated with the Hessian determinant, thereby providing precise criteria for distinguishing between saddle points and the global minimum.

%-------------
\begin{lem}\label{lem:Boundries}
    Consider the following polynomials:
   %--------------
    \begin{subequations}
       \label{eq:Polydefinition}
       \begin{align}
   P^{(1)}_{N}(x)& := \sum_{m=1}^{N-1}\gamma_{m}{\rm e}^{-\theta_{m}x^{2}}\left(2\theta_{m}x^{2}+1\right), \\
      P^{(2)}_{N}(x) & := \frac{1}{x}\sum_{m=1}^{N-1}\gamma_{m}{\rm e}^{-\theta_{m}x^{2}}, 
   \end{align}
    \end{subequations}
    %--------------
    where we recall $\gamma_m = (2m-1)(2m-1+\frac{3m(1-m)-1}{N}), \theta_m=(2m-1)^2/4$ for all $m$.  Then, we have the following cases depending on the value of $N$: 
    \begin{enumerate}
       \item For $3 \leq N\leq 8$, the polynomial $P^{(1)}_{N}(x)$ has no positive real root. 
        \item  For $N = 9$, the polynomial $P^{(1)}_{N}(x)$ has two positive real roots at $\tilde{x}_1 = 9.6369\times 10^{-3}$ and $\bar{x}_1 = 0.2484$. 
        \item For $N\geq 10$,   both $P^{(1)}_{N}(x)$ and $P^{(2)}_{N}(x)$ each have a single positive real root.
    \end{enumerate}
\end{lem} 
%-------------
\begin{proof}
    See Appendix~\ref{ap:Boundries}. 
\end{proof}

\begin{lem}\label{lem:Boundries2}
     Let $(x^*,y^*)$ denote a solution to \eqref{eq:optimal_nonlinear}. Then, the following statements hold:
    \begin{itemize}
        \item For $3 \leq N_{1,K}, N_{2,K} \leq 8$, there exists a single intersection between $g_q(x,y)$ and the ellipse constraint. 
        \item For $N_{1,K}~\text{or}~N_{2,K}\geq 9$, we define 
        %-------------
    \begin{align}
    \label{eq:r1r2def2}
    \xi_1 := \begin{cases}  \frac{(q^2-1)\bar{x}_1^2}{12} + \frac{(n^2-1)\bar{y}_1^2}{12},&  N_{1, K}\geq 9, \\
        \frac{(q^2-1)\bar{x}_2^2}{12} + \frac{(n^2-1)\bar{y}_2^2}{12}, &  N_{1, K}\leq 8,~N_{2, K}\geq 10,
    \end{cases} \quad 
    \end{align}
    %-------------
    where $(x_1, y_1)$ and $(x_2, y_2)$ are positive roots of the following polynomials: 
\begin{subequations}
    \begin{align}
      & P_{N_{1,K}}^{(1)}(x_1) = 0,  \quad g_q(\bar{x}_1, \bar{y}_1) = 0, \\ 
      & P_{N_{2,K}}^{(1)}(y_2) = 0,  \quad g_q(\bar{x}_2, \bar{y}_2) = 0. 
\end{align}
\end{subequations}
                For $N_{1,K} = 9$ or $N_{2,K} = 9$, $\bar{x}_1$ and $\bar{y}_2$ are the largest positive root of $P^{(1)}_{9}(x)$ and $P^{(1)}_{9}(y)$, respectively. Then:
        \begin{itemize}
            \item For $\xi \geq \xi_1$, system~\eqref{eq:optimal_nonlinear} admits multiple solutions, among which only the intersection corresponding to the middle branch of $g_q(x,y)$ yields the optimal solution.
            \item If $\xi \ll \xi_1$, all solutions to~\eqref{eq:optimal_nonlinear} correspond to saddle points, and the optimal solutions lie on the boundary of the feasible region, i.e., at the intersections of the ellipse (power constraint) with the $x$- or $y$-axis.
        \end{itemize}
    \end{itemize}
   %-------------
\end{lem}
\begin{proof}
 See Appendix~\ref{qp:Boundries2}. 
\end{proof}

Thus, based on Lemma~\ref{lem:Boundries2}, three distinct regions of optimality can be identified:  
\begin{itemize}
    \item \textbf{Main region:} When $\xi$ is sufficiently large, the optimal solution to~\eqref{eq:OptmialMSEgeneral} corresponds to the intersection of the ellipse constraint 
    \[
    \frac{\Delta_1^2}{\Upsilon_1^2} + \frac{\Delta_2^2}{\Upsilon_2^2} = 1
    \]
    and the middle branch of the implicit curve $g_q(x,y)=0$. 

    \item \textbf{Low-SNR region:} As $\xi$ decreases, the ellipse intersects the coordinate axes at $(\Delta_1, \Delta_2) = (\Upsilon_1, 0)$ and $(0, \Upsilon_2)$. Direct substitution of these points into the MSE expression shows that the intersection on the $y$-axis yields a smaller MSE than that on the $x$-axis, due to the scaling factor $q$ applied to the quadrature component. Consequently, in the low-SNR regime, the optimal configuration satisfies $(\Delta_1, \Delta_2) = (\epsilon, \Upsilon_2 - \epsilon)$ for a small $\epsilon \ll 1$. Alternatively, the modulation order can be reduced to $q$ (instead of $q^2$), with input symbols quantized to the set $[0, q, 2q, \ldots, q^2 - q]$. In practical terms, this implies that a PAM-type modulation scheme transmitted over two time slots is preferable; otherwise, transmission errors become dominant.

    \item \textbf{Extremely low-SNR region:} When $\xi \approx 0$, the intersection of the ellipse constraint with the $x$-axis yields a smaller MSE than that with the $y$-axis. In this regime, the noise variance is so large that the term $q^2\Upsilon_2^2/\Upsilon_1^2$ becomes negligible. The SNR threshold at which this behavior appears corresponds to extremely low SNR levels (e.g., below $-60$~dB), and its exact value is therefore omitted.
\end{itemize}

Although Lemmas~\ref{lem:Boundries} and \ref{lem:Boundries2} provide exact analytical boundaries for these regions, the algebraic complexity of the polynomial $P_{N_{i,K}}^{(1)}(x)$ hinders direct interpretation. Therefore, in the following lemma, we present an approximate characterization of these critical values to facilitate analytical insight.

%-------------
\begin{lem}\label{lem:crictical}
Let $N_{1,K}, N_{2,K} \geq 30$, and let $\xi_1$ denote the threshold separating different feasible regions as defined in Lemma~\ref{lem:Boundries}. Then, the following lower bound holds:
%-------------
\begin{align}
   \bar{\Delta}_1 &  \geq \frac{1.3}{N_{1, K}}, 
   \qquad \bar{\Delta}_2   \geq \frac{4.2}{\sqrt{n}K}, \\
    \xi_1 & \geq \sqrt{\frac{1.3^2}{12N_{1,K}^2}(q^2 - 1) + \frac{4.2^2}{12nK^2}(n^2 - 1)}
    \approx \frac{1.5n}{K^2}.
\end{align}
%-------------
\end{lem}
\begin{proof}
    See Appendix~\ref{ap:crictical}.
\end{proof}

Figure~\ref{fig:g_tilde} illustrates the implicit curve \(g_q(x, y)\) together with the corresponding points \((\Bar{\Delta}_1, \Bar{\Delta}_2)\) derived from Lemma~\ref{lem:crictical} for the cases \(q = n = 4\) and \(q = n = 12\).

From Lemma~\ref{lem:Boundries2}, the optimal solution can generally be obtained by solving the nonlinear system in~\eqref{eq:optimal_nonlinear}, depending on the operating regime. For the case where \(N_{1,K}, N_{2,K} \geq 10\) and \(\xi \geq \xi_1\), it is essential to verify that solving~\eqref{eq:optimal_nonlinear} indeed converges to the global optimum (the middle branch intersection), regardless of the initial conditions. To this end, we approximate the implicit curve \(g_q(\Delta_1, \Delta_2) = 0\) by a smooth function that preserves only its middle branch.  Specifically, recalling the definition of \(g_q(x,y)\),
%-------------
\begin{align*}
    g_q(x,y) =  \sum_{m=1}^{N_{1,K}-1} \gamma_{1,m} \frac{e^{-\theta_{m}x^2}}{x}
    - q^2 \sum_{m=1}^{N_{2,K}-1} \gamma_{2,m} \frac{e^{-\theta_{m}y^2}}{y},
    \qquad x,y>0,
\end{align*}
%-------------
we observe that for indices \(m \geq \lceil \frac{2N_{1,K}}{3} \rceil\), the coefficients \(\gamma_{1,m}\) and \(\gamma_{2,m}\) become negative, i.e., \(\gamma_{1,m}, \gamma_{2,m} < 0\). For small \(x \ll 1\) or \(y \ll 1\), the exponential terms can be approximated as \(\exp(-\theta_m x^2) \approx 1\) and \(\exp(-\theta_m y^2) \approx 1\), respectively. In this regime, the factor \(1/x\) tends to infinity, causing \(g_q(x,y)\) to diverge to \(-\infty\). Consequently, to satisfy the equality \(g_q(x,y) = 0\), the corresponding variable \(y\) must approach \(+\infty\).  Therefore, to suppress the nonlinear behavior, we omit the higher-order terms corresponding to indices greater than $\bar{N}_{1,K} = \lfloor \frac{2N_{1,K}}{3} \rfloor$ and $\bar{N}_{2,K} = \lfloor \frac{2N_{2,K}}{3} \rfloor$, leading to the following approximation:
%-------------
\begin{align}\label{eq:gq2tild}
    \tilde{g}_q(x,y) =
    \sum_{m=1}^{\bar{N}_{1,K}} \gamma_{1,m} \frac{e^{-\theta_{m}x^2}}{x}
    - q^2 \sum_{m=1}^{\bar{N}_{2,K}} \gamma_{2,m} \frac{e^{-\theta_{m}y^2}}{y},
    \qquad x, y > 0.
\end{align}
%-------------
Next, we plot $\tilde{g}_q(x,y) = 0$ and $g_q(x,y) = 0$ together in Figure~\ref{fig:g_tilde}. As shown, unlike the zero-level set of $g_q(x,y)$, the zero-level set of $\tilde{g}_q(x,y)$ yields a single intersection with the power-constraint ellipse, thereby ensuring the uniqueness of the solution for $x, y > 0$.

%--------------
\definecolor{airforceblue}{rgb}{0.36, 0.54, 0.66}
	\definecolor{babyblue}{rgb}{0.54, 0.81, 0.94}
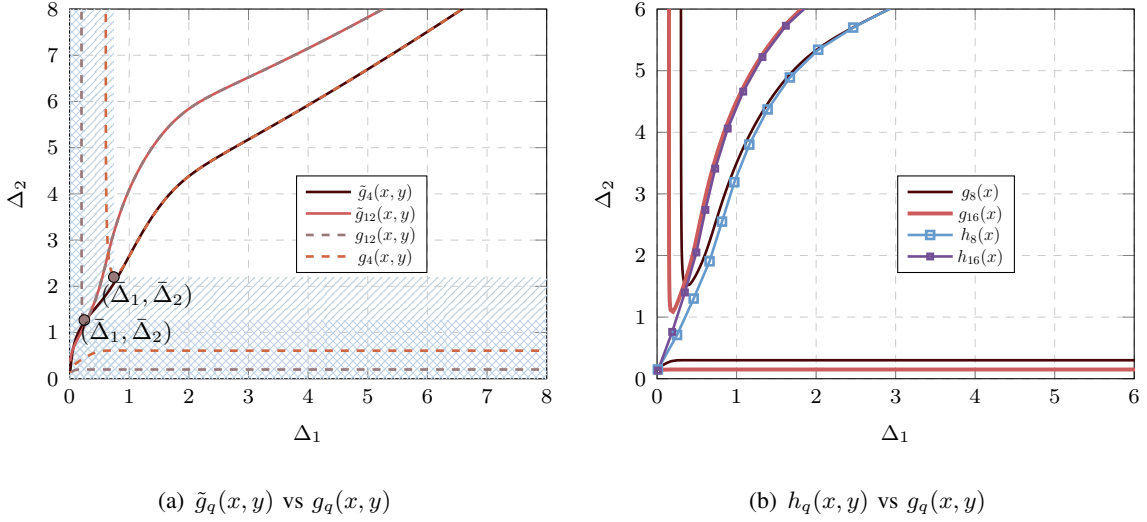
\begin{figure*}[!t]
\centering
\subfigure[$\tilde{g}_q(x,y)$ vs $g_q(x,y)$]{  \label{fig:g_tilde}
    \begin{tikzpicture} 
    \begin{axis}[
        xlabel={${\Delta_1}$},
        ylabel={$\Delta_2$},
        label style={font=\scriptsize},
        % title = {\scriptsize $???$},
        %legend pos=,
        tick label style={font=\scriptsize} , 
        width=0.48\textwidth,
        height=6.5cm,
        xmin=0, xmax=8,
        ymin=0, ymax=8,
        xtick={0,1,2,3,4,5,6,7,8,9},
        ytick={0,1,2,3,4,5,6,7,8,9},
        %ytick={0, 20, 40, 60, 80, 100},
         % ymode = log,
       legend style={nodes={scale=0.55, transform shape}, at={(0.75,0.55)}}, 
        ymajorgrids=true,
        xmajorgrids=true,
        grid style=dashed,
        grid=both,
        grid style={line width=.1pt, draw=gray!15},
        major grid style={line width=.2pt,draw=gray!40},
    ]
    \fill[pattern=north east lines, pattern color=airforceblue!40] (0,0) rectangle (8,2.2);
    \fill[pattern=north east lines, pattern color=airforceblue!40] (0,0) rectangle (0.74175,8);
    %----------
    \fill[pattern=north west lines, pattern color=bluegray!40] (0,0) rectangle (8,1.27);
    \fill[pattern=north west lines, pattern color=bluegray!40] (0,0) rectangle (0.24725,8);
    %----------
     \addplot[color=bulgarianrose,
        line width=1pt,
        mark size=2pt,
        ]
    table[x=d1,y=d2]
    {Data/Countor-solid4.dat};
    \addplot[color=chestnut,
        line width=1pt,
        ]
    table[x=d1,y=d2]
    {Data/Countor-solid12.dat};
    % ============
  \addplot[dashed,
       color=bazaar,
        line width=1pt,
        ]
    table[x=d1,y=d2]
    {Data/Contour-12.dat};
    \addplot[dashed,
    color=burntsienna,
        line width=1pt,
        mark size=2pt,
        ]
    table[x=d1,y=d2]
    {Data/Contour-4.dat};
    %------ q=4
    % \addplot[name path=lower, draw=none, domain=0.74175:9] {2.2};
    % \addplot[name path=upper, draw=none, domain=0.74175:9] {9};
    % \addplot[fill=bluegray, opacity=0.5] fill between[of=lower and upper];
    \draw[fill=bazaar](axis cs:0.74175,2.2) circle[radius=2pt];
    \node at (axis cs:1.3,1.8) {\footnotesize $(\Bar{{\Delta}}_1,\Bar{{\Delta}}_2)$};
    %------ q=12
    % \addplot[name path=lower, draw=none, domain=0.24725:9] {1.27};
    % \addplot[name path=upper, draw=none, domain=0.24725:9] {9};
    % \addplot[fill=airforceblue, opacity=0.2] fill between[of=lower and upper];
    \draw[fill=bazaar](axis cs:0.24725,1.27) circle[radius=2pt];
    \node at (axis cs:0.95,1) {\footnotesize $(\Bar{{\Delta}}_1,\Bar{{\Delta}}_2)$};
    \legend{{$\tilde{g}_{4}(x,y)$},{$\tilde{g}_{12}(x,y)$}, {$g_{12}(x,y)$}, {$g_{4}(x,y)$}};
    \end{axis}
\end{tikzpicture}}
\subfigure[$h_q(x,y)$ vs $g_q(x,y)$]{
    \label{fig:H_function}
\begin{tikzpicture} 
    \begin{axis}[
        xlabel={${\Delta_1}$},
        ylabel={$\Delta_2$},
        label style={font=\scriptsize},
        % title = {\scriptsize $???$},
        %legend pos=,
        tick label style={font=\scriptsize} , 
        width=0.48\textwidth,
        height=6.5cm,
        xmin=0, xmax=6,
        ymin=0, ymax=6,
        xtick={0,1,2,3,4,5,6},
        ytick={0,1,2,3,4,5,6},
        %ytick={0, 20, 40, 60, 80, 100},
         % ymode = log,
       legend style={nodes={scale=0.55, transform shape}, at={(0.75,0.55)}}, 
        ymajorgrids=true,
        xmajorgrids=true,
        grid style=dashed,
        grid=both,
        grid style={line width=.1pt, draw=gray!15},
        major grid style={line width=.2pt,draw=gray!40},
    ]
     \addplot[ color=bulgarianrose,
        line width=1pt, 
        ]
    table[x=d1,y=d2]
    {Data/contour_data_g8.dat};
    \addplot[
    color=chestnut,
        line width=1.5pt,
        ]
    table[x=d1,y=d2]
    {Data/contour_data_g16.dat};
    % ============
  \addplot[mark= square, mark size=1.5pt,
       color=bluegray,each nth point=10,
        line width=1pt,
        ]
    table[x=d1,y=d2]
    {Data/contour_data_h8.dat};
    \addplot[mark= square,  mark size=1pt,
    color=darklavender,
        line width=1pt,each nth point=10,
        ]
    table[x=d1,y=d2]
    {Data/contour_data_h16.dat};
    % \addplot[
    % domain = 0:5,
    % line width=0.5pt,
    % mark size = 1pt,
    % color= bulgarianrose,
    % samples=400,
    % unbounded coords=jump
    % ] { sqrt(12-x^2)};
    % \addplot[
    % mark size = 1pt,
    % domain = 0:4,
    % line width=0.5pt,
    % color= burntsienna,
    % samples=400,
    % unbounded coords=jump
    % ] { sqrt(5.17-x^2)};
    \legend{{$g_{8}(x)$},{$g_{16}(x)$}, {$h_{8}(x)$}, {$h_{16}(x)$}};
    \end{axis}
\end{tikzpicture}
}
  \caption{ Comparison between ${g}_{q}(\Delta_1,\Delta_2)$ with $\tilde{g}_{q}(\Delta_1,\Delta_2) $ and $h_q(\Delta_1,\Delta_2)$.  Figure~\ref{fig:g_tilde} compares between the implicit functions ${g}_{q}(\Delta_1,\Delta_2) =0$ with their approximation  $\tilde{g}_{q}(\Delta_1,\Delta_2)=0$ for $q=\{4,12\}$ with $K=3$ and $n=q$. Figure\ref{fig:H_function} illustrate the implicit functions $h_q(\Delta_1,\Delta_2)$ and $g_q(\Delta_1,\Delta_2)$ for $q=\{8,16\}$.}
\end{figure*}
%--------------

In the following proposition, we derive closed-form expressions for the solution in each SNR regime.

%-------------
\begin{prop}\label{Pr:Optmizaiton}
    Let $K$, $q$, and $n$ denote the number of nodes, and the modulation levels for the in-phase and quadrature components, respectively. 
    Let $P^{(1)}_{N_{i,K}}(x)$ be the polynomial defined in~\eqref{eq:Polydefinition}, and let $\xi_1$ denote its corresponding root. 
    Then, regardless of the radius $\xi$, for $3 \leq N_{1,K}, N_{2,K} \leq 8$, the optimal pair $(\Delta_1^*, \Delta_2^*)$ is uniquely obtained from
    %-------------
    \begin{align}
        \label{eq:uniqeg0-pr}
        g_q(\Delta_1, \Delta_2) = 0, 
        \qquad 
        \frac{\Delta_1^2}{\Upsilon_1^2} + \frac{\Delta_2^2}{\Upsilon_2^2} = 1.
    \end{align}
    %-------------
    For $N_{1,K}~\text{or}~N_{2,K} \geq 9$, the optimal $\Delta_1^*$ and $\Delta_2^*$ are given by
    %-------------
    \begin{align}
        \label{eq:uniqeg}
        \tilde{g}_q(\Delta_1, \Delta_2) = 0, 
        \qquad 
        \frac{\Delta_1^2}{\Upsilon_1^2} + \frac{\Delta_2^2}{\Upsilon_2^2} = 1,
        \quad \text{if}~~\xi \geq \xi_1,
    \end{align}
    %-------------
    otherwise, the optimal solution lies on the coordinate axes, i.e., 
    \((\Delta_1^*, \Delta_2^*) = (\Upsilon_1, 0)\) or 
    \((\Delta_1^*, \Delta_2^*) = (0, \Upsilon_2)\).
\end{prop}
%-------------
\begin{proof}
    See Appendix~\ref{Ap:Optmizaiton}.
\end{proof}

Finally, by invoking~\eqref{eq:bivariaterep} and Proposition~\ref{Pr:Optmizaiton}, the proof is completed.

\subsection{Proof of Theorem~\ref{th:Optmizaiton-Map}}\label{sec:prioir}

We first derive the MSE expression for the MAP decoder. The approximation of the prior distribution by a normal distribution, as introduced in Section~\ref{sec:MAPDecoder}, enables a tight closed-form approximation of the MSE. 
%------------
\begin{prop}
    \label{Pr:MSE}
    Consider a communication network with $K$ nodes, each employing the encoder $\mathscr{E}_q$ in~\eqref{eq:encoding_qam} and the MAP decoder $\mathscr{D}$ in~\eqref{eq:deocder_qam} to compute the function $f = \sum_{k}s_k$, where $s_k \in \mathbb{Z}$, over a noisy MAC with noise $\mathcal{CN}(0,\sigma^2)$. 
    Assume that the constellation symbols $s_k$ are uniformly distributed over $[0, q^2 - 1]$. 
    Then, the MSE of $f$, as defined in~\eqref{eq:MSEdefinition}, can be approximated as
    \begin{align}
     \label{eq:MSEbound2}
        {\rm MSE}(d_1,d_2)  = \omega_1(d_1) + q^2\omega_2(d_2) +\delta,
    \end{align}
    where 
    \begin{align}
        \nonumber
         \omega_{1}(x) =  2   \sum_{m=1}^{2q}\beta_{m} Q\Big(\frac{\eta \beta_{m}x}{\sqrt{2}\sigma}\Big), \quad \omega_{2}(x) =  2   \sum_{m=1}^{2n}\beta_{m} Q\Big(\frac{\eta \beta_{m}x}{\sqrt{2}\sigma}\Big), 
    \end{align}
    with $\beta_m := 2m - 1$, $\eta = (1 + \sigma/K)$, and $Q(\cdot)$ denoting the Gaussian $Q$-function. The approximation error in~\eqref{eq:MSEbound2} relative to~\eqref{eq:MSEdefinition}, denoted by $\delta$, is upper-bounded as
    %------------
    \begin{align*}
        |\delta| \leq A_1 e^{-c_1 \left(\frac{d_1 q \eta}{\sigma}\right)^2} 
        + q^2 A_2 e^{-c_2 \left(\frac{d_2 n \eta}{\sigma}\right)^2},
    \end{align*}
    %------------
    where $A_1 = \mathcal{O}\left(\frac{d_1 q \eta}{\sigma}\right)$ and 
    $A_2 = \mathcal{O}\left(\frac{d_2 n \eta}{\sigma}\right)$, while $c_1$ and $c_2$ are positive constants.
% -----------
\end{prop}
% -----------
\begin{proof}
    See Appendix~\ref{Ap:proof-MSE}.
\end{proof}
%------------
Using the MSE approximation established in Proposition~\ref{Pr:MSE} and following the procedure outlined in Appendix~\ref{sec:Optimal}, the optimization problem in~\eqref{eq:OptmialMSEgeneral} reduces to solving the following system of equations:
%-------------
\begin{align}
    \label{eq:optimal_nonlinear_MAp}
     h_q(\Delta_1, \Delta_2) = 0, 
     \qquad 
     \frac{\Delta_1^2}{\Upsilon_1^2} + \frac{\Delta_2^2}{\Upsilon_2^2} = 1,
\end{align}
%-------------
where $\Upsilon_1 = \sqrt{12\xi \eta / (q^2 - 1)}$ and $\Upsilon_2 = \sqrt{12\xi \eta / (n^2 - 1)}$. The function $h_q(x,y)$ is defined as
%-------------
\begin{align}
    \nonumber
    h_q(x,y) =  
    \sum_{m=1}^{2q} \Upsilon_1^2 \theta_m \frac{e^{-\theta_m x^2}}{x}
    - q^2 \sum_{m=1}^{2n} \Upsilon_2^2\theta_m \frac{e^{-\theta_m y^2}}{y},
    \qquad x, y > 0.
\end{align}
%-------------
% where $\kappa = \sqrt{(n^2 - 1)/(q^2 - 1)}$. 

Figure~\ref{fig:H_function} depicts the zero-level set of $h_q(x,y)$ for $q = 8$ and $q = 16$. The function $h_q(x,y)$ is monotonically increasing, and its intersection with the ellipse representing the power constraint occurs at a single point. Consequently, the system of equations in~\eqref{eq:optimal_nonlinear_MAp} admits a unique solution. 

Furthermore, since the MSE in Proposition~\ref{Pr:MSE} is expressed as a weighted sum of $Q$-functions with strictly positive weights, the objective function is convex. Therefore, the uniqueness of the solution is guaranteed, as formally stated in the following proposition.

%---------------
% \input{Figs/Fig_H_function}
%---------------

%-------------
\begin{prop}\label{Pr:Optmizaiton-Map}
    Let $\sigma^2$ denote the noise variance and $\eta = 1 + \sigma^2 / K$. 
    Then, the optimal constellation points that solve~\eqref{eq:optimal_nonlinear_MAp} are uniquely determined by the following system of equations:
    %-------------
    \begin{align}
        \label{eq:uniqeg-MAp}
        h_q(\Delta_1, \Delta_2) = 0, 
        \qquad 
        \frac{\Delta_1^2}{\Upsilon_1^2} + \frac{\Delta_2^2}{\Upsilon_2^2} = 1.
    \end{align}
    %-------------
\end{prop}

\begin{proof}
Define the function $\hat{h}_q(t) = h_q\!\left(\Upsilon_1\sqrt{0.5 - t},\, \Upsilon_2\sqrt{0.5 + t}\right)$, which can be explicitly written as
 %-------------
\begin{align}
    h_q(t) = 
    \sum_{m=1}^{2q} \Upsilon_1 \theta_m 
    \frac{e^{-\theta_m \Upsilon_1^2 (0.5 - t)}}{\sqrt{0.5 - t}}
    - q^2 \sum_{m=1}^{2n} \Upsilon_2 \theta_m 
    \frac{e^{-\theta_m \Upsilon_2^2 (0.5 + t)}}{\sqrt{0.5 + t}}.
\end{align}
 %-------------
Taking the derivative of $h_q(t)$ yields a sum in which each term contains exponential functions multiplied by positive coefficients $\theta_m > 0$. Hence, $h_q(t)$ is continuous and strictly increasing on $(0, \Upsilon_1)$.  Consequently, the function $\hat{h}_q(t)$ is also continuous and strictly increasing. Moreover,
 %-------------
\begin{align*}
    \lim_{t \to 0^+} \hat{h}_q(t) = -\infty,
\qquad 
\lim_{t \to \Upsilon_1^-} \hat{h}_q(t) = +\infty.
\end{align*}
 %-------------
By the intermediate value theorem, there exists exactly one $t \in (0, \Upsilon_1)$ such that $\hat{h}_q(t) = 0$. 
Therefore, the system~\eqref{eq:uniqeg-MAp} admits a unique solution, completing the proof.
\end{proof}
%-------------

\subsection{Proof of Lemma~\ref{lem:Boundries}}\label{ap:Boundries}

%----------------------
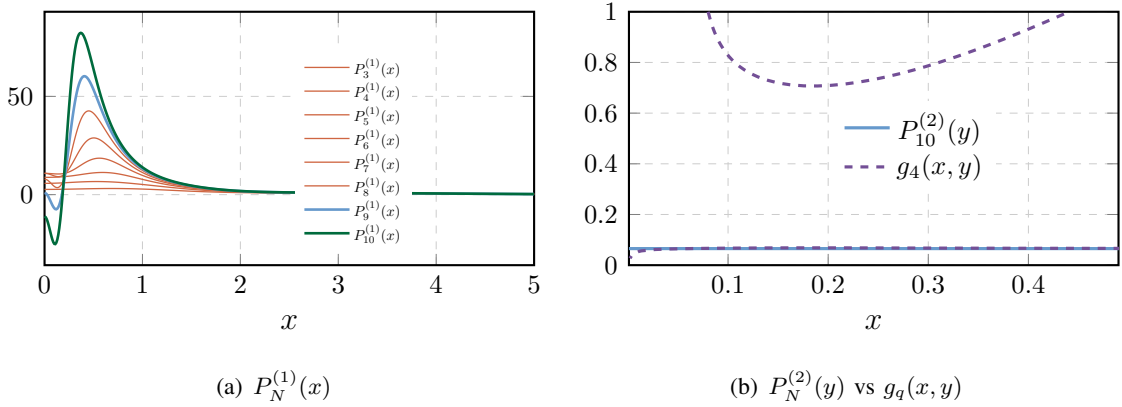
\begin{figure*}[t]
    \centering
    \subfigure[$P_{N}^{(1)}(x)$]{
     \label{fig:polynomial_roots}
    \begin{tikzpicture}
        \begin{axis}[
            width=0.49\textwidth ,
            height=0.3\textwidth,
            xlabel={$x$},
            xmin=0, xmax=5,
            legend pos=north east,
            legend style={draw=none,font=\small},
            legend style={nodes={scale=0.55, transform shape}, at={(0.75,0.85)}}, 
            ymajorgrids=true,
            xmajorgrids=true,
            grid style=dashed,
            grid=both,
            grid style={line width=.1pt, draw=gray!15},
            major grid style={line width=.2pt,draw=gray!40},
            tick label style={font=\small},
            axis line style={thick},
        ]
        % Explicitly adding each M to avoid \foreach issue
        \addplot[mark size = 0.5pt,line width=0.5pt,color=burntsienna] table[x=x, y=P_M3] {Data/P_x_all_M.dat}; \addlegendentry{$P_3^{(1)}(x)$};
        \addplot[mark size = 0.5pt,line width=0.5pt,color=burntsienna] table[x=x, y=P_M4] {Data/P_x_all_M.dat}; \addlegendentry{$P_4^{(1)}(x)$};
        \addplot[mark size = 0.5pt,line width=0.5pt,color=burntsienna] table[x=x, y=P_M5] {Data/P_x_all_M.dat}; \addlegendentry{$P_5^{(1)}(x)$};
        \addplot[mark size = 0.5pt,line width=0.5pt,color=burntsienna] table[x=x, y=P_M6] {Data/P_x_all_M.dat}; \addlegendentry{$P_6^{(1)}(x)$};
        \addplot[mark size = 0.5pt,line width=0.5pt,color=burntsienna] table[x=x, y=P_M7] {Data/P_x_all_M.dat}; \addlegendentry{$P_7^{(1)}(x)$};
        \addplot[mark size = 0.5pt,line width=0.5pt,color=burntsienna] table[x=x, y=P_M8] {Data/P_x_all_M.dat}; \addlegendentry{$P_8^{(1)}(x)$};
        \addplot[mark size = 0.5pt,line width=1pt, color=bluegray] table[x=x, y=P_M9] {Data/P_x_all_M.dat}; \addlegendentry{$P_9^{(1)}(x)$};
        \addplot[mark size = 0.5pt, line width=1pt,color=cadmiumgreen] table[x=x, y=P_M10] {Data/P_x_all_M.dat}; \addlegendentry{$P_{10}^{(1)}(x)$};
        \end{axis}
    \end{tikzpicture} }\subfigure[$P_{N}^{(2)}(y)$ vs $g_q(x,y)$]{\label{fig:PM_plot_M10}
    \begin{tikzpicture}
        \begin{axis}[
            width=0.49\linewidth,
            height=0.3\linewidth,
            xlabel={$x$},
            xmin=1e-3, xmax=0.49,
            ymin=0, ymax=1,
            legend pos=north east,
            legend style={draw=none,font=\small},
            legend style={nodes={scale=1, transform shape}, at={(0.75,0.65)}}, 
            ymajorgrids=true,
            xmajorgrids=true,
            grid style=dashed,
            grid=both,
            grid style={line width=.1pt, draw=gray!15},
            major grid style={line width=.2pt,draw=gray!40},
            tick label style={font=\small},
            axis line style={thick},
        ]
        % Load data from file
        \addplot[bluegray,very thick,  domain=1e-3:0.5] {0.0654}; 
       \addlegendentry{$P_{10}^{(2)}(y)$}
    % Plot implicit curve
       \addplot[darklavender, dashed, very thick] table [x=x, y=y] {Data/Imlicit_curve_P_m2.dat};
       \addlegendentry{$g_4(x,y)$}
        \end{axis}
    \end{tikzpicture}
    }
    \caption{Behavior of the polynomials \( P_N^{(1)}(x) \)  and $ P_N^{(2)}(y)$ for different values of \( N \). In Figure~\ref{fig:polynomial_roots}, the number of positive real roots varies as \( N \) changes: for \( N \leq 8 \), there are no positive real roots; for \( N=9 \), two positive roots exist; for \( N \geq 10 \), only one positive real root remains. This transition highlights the fundamental structural shift in the polynomial's root distribution. In Figure\ref{fig:PM_plot_M10}, we depict $P_{10}^{(2)}(y)=0$ versus $g_4(x,y)$ for $N=10$.}
\end{figure*}
%----------------------

We split the domain into an analytically certified region and a compact remainder. Specifically, we show \(P_{N}^{(1)}(x)>0\) for all \(x\ge 1.62\) (uniformly in \(N\ge 3\)), and then validate the behavior on \([0,1.62]\) numerically.  Let \(\bar N=\big\lceil2N/3\big\rceil\). Then \(\gamma_m\ge 0\) for \(m\le \bar N\) and \(\gamma_m\le 0\) for \(m\ge \bar N\). Since \(\theta_m\) is strictly increasing (\(\theta_m\le \theta_{m+1}\)) and \(\theta_{\bar N}-\theta_{\bar N-1}= 2\bar N-2\), we obtain
\begin{align}
P_N^{(1)}(x)
=\sum_{m=1}^{N-1}\gamma_m \, \mathrm{e}^{-\theta_m x^2}\!\left(2\theta_m x^2+1\right)
\;\ge\;
2x^2\!\left(A_{+}\,\mathrm{e}^{-\theta_{\bar N-1}x^2}-A_{-}\,\mathrm{e}^{-\theta_{\bar N}x^2}\right),
\end{align}
where $A_{+}:=\sum_{m=1}^{\bar N-1}\theta_m\gamma_m$
and $A_{-}:=\sum_{m=\bar N}^{N-1}\theta_m|\gamma_m|$. A sufficient condition for the right-hand side (and hence $P_N^{(1)}(x)$) to be positive is
\begin{align}
x^2 \;\ge\; \frac{\ln\!\big(A_{-}/A_{+}\big)}{\theta_{\bar N}-\theta_{\bar N-1}}
\;=\; \frac{\ln\!\big(A_{-}/A_{+}\big)}{2\bar N-2}.
\end{align}
Elementary calculations yield \(A_{-}/A_{+}\le 14\). Consequently, we get 
\begin{align}\label{eq:rootlimit}
x
\;\ge\; \sqrt{\frac{\ln 14}{2\bar N-2}}
\;\le\; \sqrt{\frac{3\ln 14}{4N-6}}
\;=\; \sqrt{\frac{3.96}{2N-3}}.
\end{align}
Thus, \(P_N^{(1)}(x)>0\) for all \(x\ge 1.15\) and \(N\ge 3\) so any positive real root must lie in \([0,1.15]\).  Figure~\ref{fig:polynomial_roots} plots \(P_{N}^{(1)}(x)\) on \([0,5]\) for \(N\in\{3,\ldots,10\}\). For \(N=3,\ldots,8\) no positive real root is observed; for \(N=9\) two positive real roots appear at $\tilde{x}_1 = 9.6369\times 10^{-3}$ and $\bar{x}_1 = 0.2484$; for \(N=10\) a single positive real root remains.

 To argue that this single-root pattern persists for all \(N\ge 10\), note that \(\mathrm{e}^{-\theta_m\cdot 0}=1\) implies
\begin{equation}\label{eq:f0}
P_{N}^{(1)}(0)=\sum_{m=1}^{N-1}\gamma_m \;<\; 0,\qquad N\ge 10.
\end{equation}
Moreover, \(P_{N}^{(1)}(x)>0\) for sufficiently large \(x\) (e.g., \(x\ge 1.62\)). Hence, by intermediate value theorem, $P_{N}^{(1)}(x)$ admits at least a single positive root in $(0,\infty)$. Moreover, in the limit \(N\to\infty\), we have \(\gamma_m \approx (2m-1)^2>0\), so for any fixed \(x>0\) the polynomial is positive. By continuity on \((0,\infty)\) and the monotone growth of \(\gamma_m\) toward its positive limit together with the upper bound in \eqref{eq:rootlimit}, the unique positive root \(x^\ast(N^{-0.5})\) moves toward \(0\) as \(N\) increases. Hence, for all \(N\ge 10\), \(P_{N}^{(1)}(x)\) admits exactly one positive real root, located in \((0,\mathcal{O}(N^{-0.5})]\).

Note that $P^{(2)}_{N}(x)$ has the same positive roots as $xP^{(2)}_{N}(x)$. Then, because $xP^{(2)}_{N}(x)$  in contrast to $P_{N}^{(1)}(x)$  can be written in the format of an algebraic polynomial and there is only a single sign change in the sequence of its coefficients $\gamma_m$,  based on Descartes’ Rule~\cite{anderson1998descartes}, $xP^{(2)}_{N}(x)$ can have at most one positive root.  Therefore, $P^{(2)}_{N}(x)$ has at most one positive root. In Figure~\ref{fig:PM_plot_M10}, we depict the polynomial  $P^{(2)}_{N}$ and the implicit $g_4(x,y)$ for $N=10$ and $q=4$.

%----------
% \input{Figs/FigP_2m7to8}
%----------

%============================
\subsection{Proof of Lemma~\ref{lem:Boundries2}}\label{qp:Boundries2}
%============================
The proof follows directly from applying the bordered Hessian theorem~\cite{boothby2003introduction,d2001bordered}.  We begin by computing the second-order derivative of the Lagrangian $\mathcal{L}$ as
%------------
\begin{align}
    \label{eq:lxx}
    \mathcal{L}_{x,x} := \frac{\partial^2 \mathcal{L}}{\partial x^2} 
    = \frac{\partial^2 \mu(x)}{\partial x^2} + 2\lambda.
\end{align}
%------------
From~\eqref{eq:Lagderivations}, we know that $2\lambda = -{x}^{-1}{\partial \mu(x)}/{\partial x} = {h_q^{1}(x)}/{\sqrt{2\pi}}$,
where $h_q^{1}(x) := \sum_{m=1}^{N_{1,K}-1} \gamma_m {e^{-\theta_m x^2}}/x$. Substituting this expression for $\lambda$ into~\eqref{eq:lxx} yields
%------------
\begin{align}
\nonumber
    \mathcal{L}_{x,x} 
    &= \frac{\partial^2 \mu(x)}{\partial x^2} 
    - \frac{1}{x}\frac{\partial \mu(x)}{\partial x} 
    = \frac{1}{\sqrt{2\pi}} \left( \frac{\partial}{\partial x}\big(-x h_q^{1}(x)\big) + h_q^{1}(x) \right), \\
    &= \frac{1}{\sqrt{2\pi}} \left( -x\frac{\partial h_q^{1}(x)}{\partial x} - h_q^{1}(x) + h_q^{1}(x) \right)
    = \frac{-x}{\sqrt{2\pi}} \frac{\partial h_q^{1}(x)}{\partial x}.
    \label{eq:partialgq(x)}
\end{align}
%------------
Similarly, we obtain
%------------
\begin{align}
\nonumber
    \mathcal{L}_{y,y} 
    = \frac{-y\Upsilon_2^2 q^2}{\Upsilon_1^2 \sqrt{2\pi}} 
      \frac{\partial h_q^{2}(y)}{\partial y}, 
    \qquad 
    h_q^{2}(y) := \sum_{m=1}^{N_{2,K}-1} \gamma_m \frac{e^{-\theta_m y^2}}{y}.
\end{align}
%------------
Next, consider the bordered Hessian matrix:
%------------
\begin{align}
    H(x,y) 
    &= 
    \begin{vmatrix} 
        0 & 2x & 2y \\[3pt]
        2x &  \mathcal{L}_{x,x} & 0 \\[3pt]
        2y & 0 & \mathcal{L}_{y,y}
    \end{vmatrix}
    = -\mathcal{L}_{y,y}(2x)^2 - \mathcal{L}_{x,x}(2y)^2
    = -4\big(x^2 \mathcal{L}_{y,y} + y^2 \mathcal{L}_{x,x}\big).
    \label{eq:Hessian_matrix}
\end{align}
%------------
Substituting~\eqref{eq:partialgq(x)} into~\eqref{eq:Hessian_matrix}, and noting that 
\[
\frac{\partial g_q(x,y)}{\partial x} = -\frac{\partial h_q(x)}{\partial x}, 
\qquad 
\frac{\partial g_q(x,y)}{\partial y} = -\frac{q^2 \Upsilon_2^2}{\Upsilon_1^2}\frac{\partial h_q(y)}{\partial y},
\]
we obtain
%------------
\begin{align}
    \label{eq:Hessian_borded}
    H(x,y) 
    = -\frac{4xy}{\sqrt{2\pi}} 
      \left( x \frac{\partial g_q(x,y)}{\partial y} 
      + y \frac{\partial g_q(x,y)}{\partial x} \right).
\end{align}
%------------
Next, consider the intersections along the side branches of the stationary set, i.e., where 
$\frac{\partial g_q(x,y)}{\partial y} = 0$ or $\frac{\partial g_q(x,y)}{\partial x} = 0$. 
On these branches,~\eqref{eq:Hessian_borded} simplifies to
%------------
\begin{align}
    \label{eq:Hessian_borded2}
    H(x,y) = -\frac{4xy^2}{\sqrt{2\pi}} \frac{\partial g_q(x,y)}{\partial x},
    \qquad \text{or} \qquad
    H(x,y) = -\frac{4yx^2}{\sqrt{2\pi}} \frac{\partial g_q(x,y)}{\partial y},
\end{align}
%------------
respectively. Since we assume $x > 0$ and $y > 0$ throughout, the multiplicative factors in~\eqref{eq:Hessian_borded2} are strictly positive. Next, consider the \emph{vertical branch}, where $\big(\partial g_q / \partial y = 0\big)$. 
Along this branch, the sign of $H(x,y)$ coincides with the sign of $-\partial g_q / \partial x$. A direct computation yields
%------------
\begin{align}
    -\frac{\partial g_q(x,y)}{\partial x} = 0
    \quad \Longleftrightarrow \quad
    \sum_{m=1}^{N_{1,K}-1} \gamma_m e^{-\theta_m x^2} 
    \left( 2\theta_m + \frac{1}{x^2} \right) = 0,
\end{align}
%------------
which, after multiplying by $x^2 > 0$, becomes the polynomial equation
%------------
\begin{align}
    P^{(1)}_{N_{1,K}}(x) 
    := \sum_{m=1}^{N_{1,K}-1} 
    \gamma_m e^{-\theta_m x^2} (2\theta_m x^2 + 1) = 0.
\end{align}
%------------
Hence, for $x > 0$, the sign of $H(x,y)$ coincides with that of $P^{(1)}_{N_{1,K}}(x)$. 
For the \emph{horizontal branch}, where $\big(\partial g_q / \partial x = 0\big)$, an identical argument holds upon interchanging $x$ and $y$. 
Based on this correspondence, the remainder of the proof follows by applying Lemma~\ref{lem:Boundries} to $H(x,y)$, which yields the following distinct cases.
 
\begin{enumerate}
\item {$3 \le N_{1,K}, N_{2,K} \le 8$.}  
Both polynomials $P^{(1)}_{N_{1,K}}(x)$ and $P^{(1)}_{N_{2,K}}(x)$ have no positive real roots. 
Hence, $\frac{\partial g_q(x,y)}{\partial x} > 0$ and $\frac{\partial g_q(x,y)}{\partial y} > 0$ for all $x, y > 0$. 
From~\eqref{eq:Hessian_borded2}, it follows that $H(x,y) > 0$. 
Therefore, there is no side branch, and the only intersection corresponds to a (bordered-Hessian) locally minimizing point, which is also the global minimizer.

\item {$N_{1,K} \ge 10$.}  
In this case, the polynomials $P^{(1)}_{N_{1,K}}$ admits a unique positive root. 
Let $\bar{x}_1 > 0$ denote the root satisfying $P^{(1)}_{N_{1,K}}(\bar{x}_1) = 0$.   
Since $P^{(1)}_{N_{1,K}}(0) < 0$ and $P^{(1)}_{N_{1,K}}(x) > 0$ for large $x$, we have
\[
P^{(1)}_{N_{1,K}}(x) < 0 \quad \text{for } 0 < x < \bar{x}_1, 
\qquad 
P^{(1)}_{N_{1,K}}(x) > 0 \quad \text{for } x > \bar{x}_1.
\]
Accordingly, along the vertical branch in \eqref{eq:Hessian_borded2}, $H(x,y) < 0$ for $0 < x < \bar{x}_1$ (corresponding to a saddle point or local maximum) and $H(x,y) > 0$ for $x > \bar{x}_1$ (local minimum).  A similar argument holds along the horizontal branch with $y\leq \tilde{y}_1 $, where $\tilde{y}_1 > 0$ denotes the positive root of $P^{(1)}_{N_{2,K}}(\tilde{y}_1) = 0$.  Hence, to ensure $H(x,y) > 0$, it suffices to require $\{(x,y)~|~x \ge \bar{x}_1,\, y \ge \tilde{y}_1\}$, which yields a unique minimizer.   Moreover, let $(\bar{x}_1, \bar{y}_1)$ denote the pair satisfying $P^{(1)}_{N_{1,K}}(\bar{x}_1) = 0$ and $g_q(\bar{x}_1, \bar{y}_1) = 0$.  
Since $\bar{y}_1 \ge \tilde{y}_1$\footnote{By monotonicity, 
$\bar{y}_1(N_{2,K}) \le \tilde{y}_1(9)$ and $\bar{y}_1(N_{2,K}) \ge \bar{y}_1(\infty)$.  
Therefore, it is sufficient to numerically verify the worst case $\tilde{y}_1(9) \approx0.09 < \bar{y}_1(\infty)\approx 0.095$, which confirms the stated inequality.}, 
the feasibility region can be restricted to $\{(x,y)~|~x \ge \bar{x}_1,\, y \ge \bar{y}_1\}$. To illustrate these regions, we depict them with the the implicit curve $g_q(x,y) =0$ in Figure~\ref{fig:g_tilde(b)bound} for $N_{1,k}=N_{2,k}=12$. 

\item {$N_{1,K} =9$ and $N_{2,K} \leq 9$.}  For $N_{2,K} \le 8$, the polynomial $P^{(1)}_{N{2,K}}(y)$ admits no positive root, and hence, no horizontal branch exists.   Polynomial $P^{(1)}_{N_{1,K}}(x)$ has exactly two positive roots, $0 < \tilde{x}_1 < \bar{x}_1$. Consequently, 
\begin{align*}
    & P^{(1)}_{N_{1,K}}(x) > 0 \quad \text{for } x \in (0, 9.6369\times 10^{-3}) \cup (0.2484, \infty), \\
    \qquad 
    & P^{(1)}_{N_{1,K}}(x) < 0 \quad \text{for } x \in (9.6369\times 10^{-3}, 0.2484).
\end{align*}
 The side branches of the locus $g_q(x,y)=0$ lie within $x\in (9.6369\times 10^{-3}, 0.2484)$ (numerically can be checked).   Thus, along either side branches, $H(x,y)$ is positive before the first root, negative between the two roots (saddle or local maximum), and positive again beyond the larger root.  

For $\{(x,y)~|~x \le 9.6369\times 10^{-3},\, y \ge \tilde{y}_1\}$, we have $H(x,y) > 0$, and the corresponding solution, which is the intersection with $y$-axis,  is optimal.   
However, since $\tilde{x}_1 = 9.6369\times 10^{-3}\approx 0$, meaning that the solution lies near the boundary (the $y$-axis), which is not of practical interest, as noted in Remark~\ref{rem:smallroot}.  
Therefore, in this case, we consider the same feasible region as before, i.e., $\{(x,y)~|~x \ge \bar{x}_1,\, y \ge \bar{y}_1\}$, where $\bar{x}_1 = 0.2484$ and the $\bar{y}_1$ comes from $g_q(0.2484,\bar{y}_1)=0$. 
For $N_{2,K}=9$, although $P^{(1)}_{N{2,K}}(y)$ possesses two positive roots, substitution into $g_q(x,y)=0$ yields no positive solutions.
Consequently, no horizontal branch appears, and the preceding arguments remain valid in this case.

% \item {$N_{1,K} \leq 8$ and $N_{2,K} = 9$.}  
% In this case, 

\item {$N_{2,K} \ge 10$~\text{and}~$N_{1,K} \leq 8$.}  In this case, two intersections occur—one along the main branch and another along the horizontal branch.   Clearly, for all $(x,y)$ satisfying $\{x \ge 0,\, y \ge \tilde{y}_1\}$, we have $H(x,y) > 0$.

\end{enumerate}
As long as $(x,y)$ satisfies $\{x \ge \bar{x}_1,\, y \ge \bar{y}_1\}$, we have $H(x,y) > 0$.  
Therefore, although not necessary, it is sufficient to require that  $\xi \ge \xi_1 =  \frac{\sqrt{(q^2 - 1)\bar{x}_1^2 + (n^2 - 1)\bar{y}_1^2}}{2\sqrt{3}},$ which guarantees that the ellipse constraint fully contains the region $\{x \ge \bar{x}_1,\, y \ge \bar{y}_1\}$.   Consequently, the optimal solution must lie within this region, and the proof is complete.

%==========================
\subsection{Proof of Lemma~\ref{lem:crictical}}\label{ap:crictical}
%==========================

Let $x_1$ be the positive root of the polynomial $P_N^{(1)}(x)$, i.e., 
 %-------------
\begin{align}
P_N^{(1)}(\bar{x}_1)= \sum_{m=1}^{N_{1, K}-1} \gamma_{m} {\rm e}^{-\theta_{m}\bar{x}_1^{2}}\left( 1+2\theta_{m}\bar{x}_1^2\right)=0. \label{eq:roundgxpe}
\end{align}
 %-------------
Obtaining a closed form expression for $x_1$ in \eqref{eq:roundgxpe} is challenging due to the combination of exponential and algebraic terms,so we derive a lower bound.  By Lemma~\ref{lem:Boundries},  the root $\bar{x}_1$ is small, specifically  $x_1 \leq 1/\sqrt{N_{1, K}}$. Hence, the exponential terms can be approximated by  ${\rm e}^{-\theta_{m}\bar{x}_1^{2}} \approx 1$. Accordingly, the terms with larger $m$ have smaller (negative) coefficients, so replacing ${\rm e}^{-\theta_{m}\bar{x}_1^{2}}$ by $1$, and dropping the last term yields a smaller polynomial. Therefore,
%-------------
\begin{align}
\sum_{m=1}^{N_{1, K}-1} \gamma_{m} {\rm e}^{-\theta_{m}\bar{x}_1^{2}}\left( 1+2\theta_{m}\bar{x}_1^2\right) \geq  \sum_{m=1}^{N_{1, K}-1} \gamma_{m} \left( 1+2\theta_{m}\bar{x}_1^2\right).
\end{align}
 %-------------
 The right-hand side is quadratic in $\bar{x}_1$.   Let $x_0$ be the positive root of the right hand side expression. With simple algebraic steps, we obtain 
 %--------------------
 \begin{align}
     x_0 = \sqrt{-\frac{\sum_{m=1}^{N_{1, K}-1}\gamma_m}{2\sum_{m=1}^{N_{1, K}-1}\theta_m\gamma_m}} \approx \frac{1.3}{N_{1, K}}. 
 \end{align}
%-----------------
We note that $\sum_{m=1}^{N_{1,K}-1}\gamma_m <0 $ for $N_{1, K} \geq 10$,  so the quantity under the square root is positive. Consequently, the lower bound for  $\bar{x}_1$ is given by
 %-------------
\begin{align}
   \label{eq:simplexbar}
    \bar{x}_1 \geq x_0 = \frac{1.3}{N_{1, K}}. 
\end{align}
 %-------------
Next,  obtain $\bar{y}_0$ by substituting $\bar{x}_0$ into
the equation $g(\bar{x}_0,\bar{y}_0)= 0$. Using  $P_{N_{1, K}}^{(3)}(x)$, we write  
 %-------------
\begin{align}
    \nonumber
    P_{N_{2, K}}^{(3)}(\bar{y}_0)  = \frac{P_{N_{1, K}}^{(3)}(\bar{x}_0)}{q^2}\rightarrow
    \bar{y}_0  = \frac{q^2}{P_{N_{1, K}}^{(3)}(\bar{x}_0)} \sum_{m=1}^{N_{2, K}-1}\gamma_{m}{\rm e}^{-\theta_{m}\bar{y}_0^{2}}.
\end{align}
 %-------------
To sharpen this approximation,  we linearize the exponential terms via a first-order Taylor expansion around $\bar{y}_0=4/\sqrt{n}K$ (guided by a numerical estimate), which yields 
 %-------------
\begin{align}
    {y}_0  = \frac{q^2}{P_{N_{1, K}}^{(3)}({x}_0)} \sum_{m=1}^{N_{2, K}-1}\gamma_{m} {\rm e}^{-\theta_{m} \frac{16}{q K^2}}\times \Big(1-2\theta_{m}\sqrt{\frac{16} {q}}\big({y}_0-\sqrt{\frac{16}{q}}\big)\Big).  \nonumber
\end{align}
 %-------------
By solving for $\bar{y}_0$, we obtain 
 %-------------
\begin{align}
    {y}_0 = \frac{\sum_{m=1}^{N_{2, K}-1} (1+2 \theta_m \frac{16}{K^2n})\gamma_m {\rm e}^{-\theta_m \frac{16}{K^2n}}  }{\frac{P_{N_{1, K}}^{(3)}({x}_0)}{q^2} +  \sum_{m=1}^{N_{2, K}-1} 2 \gamma_m \theta_m \sqrt{\frac{16}{K^2n}} {\rm e}^{-\theta_m \frac{16}{K^2n}}}.
\end{align}
 %-------------
By bounding the exponential terms and approximating the sums by integrals, one shows that this fraction exceeds the crude estimate, leading to
 %-------------
\begin{align}
    \bar{y}_1 \geq {y}_0\geq \frac{4.2}{\sqrt{n}K}, \quad N_{1,K}, N_{2,K} \geq 30.  
\end{align}
 %-------------
Hence, the radius $\xi_1$ is be bounded below by $\sqrt{\bar{x}_1^2+\bar{y}_1^2}$.

\subsection{Proof of Proposition~\ref{Pr:Optmizaiton}}\label{Ap:Optmizaiton}

Throughout Appendix~\ref{sec:optimalverssub}, we observed that depending on the value of $\xi$, different regions arise for the solutions to the system of equations in~\eqref{eq:optimal_nonlinear}. 
For $\xi \leq \xi_1$, the proof directly follows from Lemmas~\ref{lem:Boundries} and~\ref{lem:Boundries2}. 
Here, we focus on the main region corresponding to $\xi \geq \xi_1$, where $\xi_1$ is defined in Lemma~\ref{lem:Boundries}. 
In this regime, three intersections always exist, and according to Lemmas~\ref{lem:Boundries} and~\ref{lem:Boundries2}, the middle branch of $g_q(x,y)$ provides the global optimum. 

The remaining task is to establish that solving the system of equations in~\eqref{eq:uniqeg} using a numerical solver yields a unique solution. 
To this end, it suffices to show that, for $x, y \in \mathbb{R}^{+}$, the zero-level set of the function $\tilde{g}_q(x,y)$ and the ellipse $x^2/\Upsilon_1^2 + y^2/\Upsilon_2^2 - 1 = 0$ intersect at a single point. 
Equivalently, the function $\tilde{g}_q(\Upsilon_1\sqrt{0.5 - t}, \Upsilon_2\sqrt{0.5 + t})$ must be invertible for $t \in [0, 1/2)$. 
To verify this, consider
%-------------
\begin{align*} 
\tilde{g}_q\big(\Upsilon_1\sqrt{0.5 - t}, \Upsilon_2\sqrt{0.5 + t}\big) 
= \sum_{m=1}^{\bar{N}_{1,K}} \gamma_{1,m} 
  \frac{e^{-\theta_m \Upsilon_1^2 (0.5 - t)}}{\Upsilon_1 \sqrt{0.5 - t}} 
  - q^2 \sum_{m=1}^{\bar{N}_{2,K}} \gamma_{2,m} 
  \frac{e^{-\theta_m \Upsilon_2^2 (0.5 + t)}}{\Upsilon_2 \sqrt{0.5 + t}}.
\end{align*}
%-------------
We now demonstrate that this function is bijective.  Taking the derivative with respect to $t$ yields
%-------------
\begin{align}
  \nonumber
   \frac{\partial \tilde{g}_q}{\partial t} 
   &= \sum_{m=1}^{\bar{N}_{1,K}} \gamma_{1,m} 
   \frac{e^{-\theta_m \Upsilon_1^2 (0.5 - t)}}{2\Upsilon_1 \sqrt{0.5 - t}}
   \Big(2\theta_m \Upsilon_1^2 + (0.5 - t)^{-1}\Big) \\ 
   &\quad + q^2 \sum_{m=1}^{\bar{N}_{2,K}} \gamma_{2,m} 
   \frac{e^{-\theta_m \Upsilon_2^2 (0.5 + t)}}{2\Upsilon_2 \sqrt{0.5 + t}}
   \Big(2\theta_m \Upsilon_2^2 + (0.5 + t)^{-1}\Big) 
   \ge 0.
   \label{eq:Derivative}
\end{align}
%-------------
Since $\gamma_{1,m}, \gamma_{2,m} > 0$ and $\theta_m$ are positive for all $m \leq \bar{N}_{1,K}, \bar{N}_{2,K}$, it follows that $\frac{\partial \tilde{g}_q}{\partial t} > 0$ for all $t \in [0, 1/2)$. 
Hence, $\tilde{g}_q$ is strictly increasing and therefore injective. 
Furthermore, by the intermediate value theorem~\cite{huang1992intermediate}, and noting that $\lim_{t \to +\infty} \tilde{g}_q(t) = +\infty$ and $\tilde{g}_q(t)$ is continuous, the function is also surjective. Consequently, $\tilde{g}_q(t)$ is bijective, implying that the system admits a unique solution.

\subsection{Proof of Lemma~\ref{lem:apprximation_fg}}\label{ap:apprximation_FGQ}

We derive an upper bound on the approximation error between $\mathcal{F}_Q$ and $\mathcal{G}_Q^{N}$ as follows:
%-------------
\begin{align}
 \nonumber
    |\mathcal{F}_Q -\mathcal{G}_Q^{N}|& = \Big|\sum_{m=2}^{N_{1,K}-1}\gamma_{1,m} \frac{{\rm e}^{-\theta_m \Upsilon_1^2(0.5-t)}}{\Upsilon_1\sqrt{0.5-t}}-\sum_{m=2}^{N_{2,K}-1}\gamma_{2,m}\frac{\kappa q^2{\rm e}^{-\theta_m\Upsilon_2^2(0.5+t)}}{\Upsilon_2\sqrt{0.5+t}}\Big|,\\ \nonumber
    & \leq\Big|\sum_{m=2}^{N_{1,K}-1}\gamma_{1,m}\frac{{\rm e}^{-\theta_m \Upsilon_1^2(0.5-t)}}{\Upsilon_1\sqrt{0.5-t}}\Big|+\Big|\sum_{m=2}^{N_{2,K}-1}\gamma_{2,m}\frac{\kappa q^2{\rm e}^{-\theta_m\Upsilon_2^2(0.5+t)}}{\Upsilon_2\sqrt{0.5+t}}\Big|, \nonumber \\
    & = \frac{\Upsilon_1}{\sqrt{0.5-t}} \Big|\sum_{m=2}^{N_{1,K}-1}\frac{\gamma_{1,m}}{\Upsilon_1^2} {\rm e}^{-\theta_m \Upsilon_1^2(0.5-t)}\Big|+\frac{\Upsilon_2\kappa q^2}{\sqrt{0.5+t}}\Big|\sum_{m=2}^{N_{2,K}-1}\frac{\gamma_{2,m}}{\Upsilon_2^2}{\rm e}^{-\theta_m\Upsilon_2^2(0.5+t)}\Big|,\label{eq:upperalaki}
\end{align}
%-------------
where the first inequality follows from the triangle inequality.  
For each summation term in~\eqref{eq:upperalaki}, we apply an integral upper bound. 
Since $\lim_{N_{i,K} \to \infty} \gamma_{i,m}/\Upsilon_i^2 = (2m - 1)^2 > 0$, the coefficients become positive, and the series is monotonically decreasing. Thus,
%-----------
\begin{align}
    \nonumber
    \frac{1}{\Upsilon_i^2}\sum_{m=2}^{N_{i,K}-1}\gamma_{i,m}{\rm e}^{-\theta_m a}  & \leq \sum_{m=2}^{\infty }\gamma_{i,m}{\rm e}^{-\theta_m a} \leq \int_{2}^{\infty}(2t-1)^2{\rm e}^{-a\frac{(2t-1)^2}{4}} dt + {\rm e}^{-\tfrac{9a}{4}},\\&= \frac{\sqrt{\pi}}{a^{1.5}} {\rm erfc}\Big(\frac{3 \sqrt{a}}{2}\Big) + \Big(\frac{3}{a}+9\Big){\rm e}^{-\frac{9a}{4}} \leq \Big(\frac{2\sqrt{\pi}}{3a^2}+\frac{3}{a} + 9\Big){\rm e}^{-\frac{9a}{4}},\nonumber \\
    & \leq 18 {\rm e}^{-\frac{9a}{4}},
     \label{eq:gammaint}
\end{align}
%----------
where the final inequality holds for $a \ge 0.566$. 
Substituting~\eqref{eq:gammaint} into~\eqref{eq:upperalaki} yields
%-----------
\begin{align}
\nonumber
    |\mathcal{F}_Q(t) -\mathcal{G}_Q^{N}(t)| & \leq \frac{\Upsilon_118{\rm e}^{-\frac{4\Upsilon_1^2(0.5-t)}{9}}}{\sqrt{0.5-t}} + \frac{18\kappa q^2\Upsilon_2{\rm e}^{-\frac{4\Upsilon_2^2(0.5+t)}{9}}}{\sqrt{0.5+t}},\\
    & = 36\sqrt{3\xi}  \bigg( \frac{{\rm e}^{-\frac{16\xi(0.5-t)}{3(q^2-1)}}}{\sqrt{(q^2-1)(0.5-t)}} + \frac{\kappa q^2 {\rm e}^{-\frac{16\xi(0.5+t)}{3(n^2-1)}}}{\sqrt{(n^2-1){(0.5+t)}}}\bigg).
\end{align}
%-----------
To ensure $a \ge 0.566$, we require
%------------
\begin{align}
    \xi \geq \frac{0.566}{12} \max\Big\{\frac{q^2-1}{0.5-t}, \frac{n^2-1}{0.5+t} \Big\},  
\end{align}
%------------
which implies
\begin{align}
    \xi \geq \frac{1}{10} \max\Big\{\frac{q^2}{0.5-t}, \frac{n^2}{0.5+t} \Big\}. 
\end{align}
This completes the proof.

\subsection{Proof of Proposition~\ref{Pr:Lambert}}\label{sec:PrLambert}

Using the approximation in \eqref{eq:approx_one},  the system of equations in \eqref{eq:MathcalG} reduces to
%-------------
\begin{align}
       \mathcal{F}_Q(t) = 0,\quad  \Rightarrow \quad 
      \frac{\gamma_{1,1}{\rm e}^{-\theta_{1} \Upsilon_1^2 (0.5-t)}}{\sqrt{0.5-t}}  = \frac{\gamma_{2,1}\kappa q^2{\rm e}^{-\theta_{1} \Upsilon_2^2 (0.5+t)}}{\sqrt{0.5+t}},
      \label{eq:approxf1eq}
\end{align}
%-------------
where $\theta_1 = 1/4$. To solve~\eqref{eq:approxf1eq}, we square both sides to eliminate the square roots, yielding 
%-------------
\begin{align}
      \frac{ {\rm e}^{ \frac{\Upsilon_1^2(t-0.5)}{2}} }{0.5-t} =\kappa^2 q^4\gamma_{2,1}^2\gamma_{1,1}^{-2} \frac{ {\rm e}^{ \frac{\Upsilon_2^2(-0.5-t)}{2}}}{0.5+t}, \quad \Rightarrow \quad
     \tilde{\kappa}^2  = \frac{ 0.5+t }{0.5-t}  {\rm e}^{ \frac{(\Upsilon_1^2+\Upsilon_2^2)t}{2}}, \label{eq:Lambert}
\end{align}
%-------------
where $\tilde{\kappa} ={\rm e}^{\frac{\Upsilon_1^2-\Upsilon_2^2}{8}} q^2\kappa \gamma_{2,1}/\gamma_{1,1}$.  The solution to~\eqref{eq:Lambert} is given in terms of the $r$-Lambert function~\cite[Theorem~3]{mezHo2017generalization}, as
%-------------
\begin{align}
    t = 0.5- \frac{2\mathcal{W}_{c}\Big(c \frac{\Upsilon_1^2+\Upsilon_2^2}{2} \Big)}{\Upsilon_1^2+\Upsilon_2^2},
\end{align}
%-------------
where $c =\tilde{\kappa}^2 {\rm e}^{-\frac{\Upsilon_1^2+\Upsilon_2^2}{8}}$. 
Hence, the optimal values $\Delta_1^*$ and $\Delta_2^*$ are obtained as
%-------------
\begin{subequations}
\label{eq:solution_lambert}
    \begin{align} 
    \Delta_1^* & = \Upsilon_1\sqrt{0.5-t^{*}}=\Upsilon_1\sqrt{  \frac{2\mathcal{W}_{c}\Big(-c \frac{\Upsilon_1^2+\Upsilon_2^2}{2} \Big)}{\Upsilon_1^2+\Upsilon_2^2}    }, \\
     \Delta_2^* & = \Upsilon_2\sqrt{0.5+t^{*}}=\Upsilon_2\sqrt{  1- \frac{ 2\mathcal{W}_{c}\Big(-c \frac{\Upsilon_1^2+\Upsilon_2^2}{2} \Big)}{\Upsilon_1^2+\Upsilon_2^2} }, 
\end{align}
\end{subequations}
%-------------
provided that $ \xi \geq 1/10\max\{q^2/(0.5-t),n^2/(0.5+t),\} $.  
Note that the expressions under the radicals in~\eqref{eq:solution_lambert} remain strictly positive for any $\xi$, and thus impose no further constraint on $\xi$.

\subsection{Proof of Proposition~\ref{Pr:MSE}}\label{Ap:proof-MSE}

The proof follows the same approach as in~\cite[Proposition~1]{Razavikia2024Ring}. 
To compute the MSE for the sum function $f = \sum_{k=1}^{K} s_k$, where $s_k \in \mathbb{Z}$, we first express the estimated value $\hat{f}$ in terms of the channel noise components. Specifically, 
% -----------
\begin{align}
  \hat{f}  = \mathscr{D}(r) = \mathscr{D}\Big(\sum_{k}\mathscr{E}(s_k) + z\Big) =  f + \mathscr{D}(z).  
 \label{eq:noisedecom}
\end{align}
% -----------
where $z$ denotes the additive channel noise.   Accordingly, the MSE can be written as
% -----------
\begin{align}
\nonumber
\mathbb{E}_z\big\{|f - \hat{f}|_2^2\big\} & = \mathbb{E}_z\big\{|\mathscr{D}(z)|^2\} =\mathbb{E}_z\big\{{|\underbrace{\mathscr{D}(\mathfrak{Re}(z))}_{:=e_{1}}|}^2\}  + q^2\mathbb{E}_z\big\{{|\underbrace{\mathscr{D}(\mathfrak{Im}(z))}_{:=e_{2}}|}^2\},
\end{align}
% -----------
where the last equality holds since the real and imaginary parts of $z$ are independent. Let $\Pr[s = j]$ denote the prior probability that $\sum_{k=1}^{K} s_k = j$.  Following the same steps as in~\cite[Appendix~B]{razavikia2024blind} and applying the law of total expectation~\cite{goldsmith2005wireless}, we obtain
% -----------
\begin{align}\label{eq:Expe1e2}
\mathbb{E}[e_1^2]  = \sum\nolimits_{j=0}^{N_{1,K}-1}\mathbb{E}[e_1^2|s=j]\Pr[s=j], 
\end{align}
% -----------
where $s=\sum_{k=1}^Ks_k$.  Since the prior distribution of $s$ follows the Gaussian approximation in~\eqref{eq:prior_dist}, for sufficiently large $K$ (i.e., $K \gg 1$), the tail probabilities of the normal distribution can be neglected. Hence, 
%-*-------------
\begin{align*}
    \Pr[s=j] = \frac{{\rm e}^{-\frac{(x_j-\mu_x)^2}{2\sigma_x^2}}}{\sqrt{2\pi \sigma_x^2}}  \approx 0, \quad \forall j \in \mathcal{J}, 
\end{align*}
%-*-------------
where $J = \{1,\ldots, q\} \cup \{N_{1,K}-q-1, \ldots, N_{1,K}-1\}$. Hence, the expression in \eqref{eq:Expe1e2} can be rewritten as
% -----------
\begin{align*}
    \mathbb{E}[e_1^2]  = \sum\nolimits_{j=q+1}^{N_{1,K}-q-2}\mathbb{E}[e_1^2|s=j]\Pr[s=j] + \hat{\delta}_1,
\end{align*}
% -----------
where $\hat{\delta}_1$ represents the approximation error. Furthermore, the conditional expectation $\mathbb{E}[e_1^2 \mid s = j]$ can be expanded as
% -----------
\begin{align}\label{eq:Expzet}
\mathbb{E}[e_1^2|s=j] & = 2\sum\nolimits_{m=1}^{N_{1,K}-j-1} m^2  \Xi(m), 
\end{align}
% -----------
where the function $\Xi(m)$ is defined by
% -----------
\begin{align}\label{eq:Zeta}
\Xi(m): & =  Q\Big(\frac{(2m-1)d_1 \eta}{\sqrt{2}\sigma}\Big) - Q\Big(\frac{(2m+1)d_1\eta}{\sqrt{2}\sigma}\Big),
\end{align}
% -----------
and $d_1$ denotes the in-phase symbol spacing.  
For $K \ge 2$, the $Q$-function tails satisfy 
$Q((2|q| + 1)d_1 / \sqrt{2}\sigma) \approx Q((2|q| - 1)d_1 / \sqrt{2}\sigma)$; thus, 
$\mathbb{E}[e_1^2 \mid s = j] \approx \sum_{m = 1}^{q} m^2 \Xi(m)$ 
for sufficiently large $K$ such that $\Xi(m) \approx 0$ for all $|m| > q$.  
Consequently, we obtain the following approximation for $\mathbb{E}[e_1^2]$:
% -----------
\begin{align}
       \nonumber
       \mathbb{E}[e_1^2] & =  \sum\nolimits_{j=q+1}^{N_{1,K}-q-2}2\sum\nolimits_{m=1}^{q} m^2  \Xi(m)\Pr[s=j] + \hat{\delta}_1^{'}= \underbrace{\sum\nolimits_{j=q+1}^{N_{1,K}-q-2}\Pr[s=j]}_{\approx 1} (2\sum\nolimits_{m=1}^{q} m^2  \Xi(m)), \\
       &= 2\sum\nolimits_{m=1}^{q} m^2  \Xi(m) + \delta_1 = 2\sum_{m=1}^{q}(2m -1)Q\Big(\frac{(2m-1)d_1}{\sqrt{2}\sigma}\Big)+ \delta_1,
       \label{eq:Errore_3}
\end{align}
% -----------
where the last equality follows by eliminating higher-order terms in~\eqref{eq:Zeta}.  
Following identical reasoning for $e_2$, we obtain
% -----------
\begin{align}\label{eq:E_q1}
    \mathbb{E}[e_2^2] &  = 2 \sum_{m=1}^{n}(2m -1)Q\Big(\frac{(2m-1)d_2 \eta }{\sqrt{2}\sigma}\Big)+ \delta_2,
\end{align} 
% ----------
where $\delta_2$ denotes the corresponding approximation error.  
Combining~\eqref{eq:Errore_3} and~\eqref{eq:E_q1} yields 
% -----------
\begin{align}
\nonumber
\text{MSE}= 2\sum_{m=1}^{q}\beta_m Q\Big(\frac{\beta_m d_1 \eta }{\sqrt{2}\sigma}\Big) + 2q^2 \sum_{m=1}^{n}\beta_m  Q\Big(\frac{\beta_md_2 \eta }{\sqrt{2}\sigma}\Big) + \delta,
\end{align}
% % -----------
where $\beta_m = 2m - 1$ and $\delta = \delta_1 + q^2 \delta_2$.  
It follows that the approximation error $\delta_1$ satisfies
 % -----------
\begin{align}
\nonumber
    |\delta_1| \leq 2\sum_{m=q+1}^{N_{1,K}-1}m^2\Xi[m] + 4 \Pr[s=q] \sum_{m=1}^{q}m^2\Xi[m] 
     \leq 2\sum_{m=q+1}^{N_{1,K}-1}m^2\Xi[m] + 4q^2 \Xi[1] \frac{{\rm e}^{-\frac{(q-\mu_x)^2}{2\sigma_x^2}}}{\sqrt{2\pi \sigma_x^2}}. 
\end{align}
 % -----------
Accordingly, the subsequent lemmas establish upper bounds on these error terms. 
%-------------
\begin{lem}\label{lem:upperBoundXi}
For any integer \(m \geq 1\), the following upper bound holds:
\begin{align}
    \Xi(m) \leq  \frac{ d_1 \eta}{\sqrt{\pi \sigma^2}} 
      e^{-\tfrac{d_1^2 \eta^2 (2m - 1)^2}{4\sigma^2}},
\end{align}
where  $\Xi(m)$ as defined in~\eqref{eq:Zeta}. 
\end{lem}
%-------------
%-------------
\begin{proof}
    See Appendix~\ref{Ap:lem:upperBoundXi}. 
\end{proof}
%-------------

%-------------
\begin{lem}
\label{lem:upperBoundSum}
Let $S$ denote the summation
\begin{align}
    S :=
 \sum_{m=q+1}^{M-1}
m^{2} \,\frac{4\,d_{1}\,\eta}{\sqrt{2\pi\,\sigma^{2}}}\,
{\rm e}^{-\frac{d_{1}^{2}\,\eta^{2}\,(2m-1)^{2}}{2\,\sigma^{2}}},
\end{align}
where \(q \ge 1\) is an integer.   Then,  $S$ is upper bounded above by
\begin{align}
\nonumber
    S & \leq \frac{4\sigma}{{d_{1}\eta}\sqrt{2\pi}}{\rm e}^{-\frac{d_{1}^{2}\eta^{2}}{2\sigma^{2}}(2q-1)^{2}}
\bigg(2q-1 + \frac{2\sigma^{2}}{{d_{1}^{2}\eta^{2}} (2q-1)} \bigg).
\end{align}
% where 
\end{lem}
%-------------
\begin{proof}
    See Appendix~\ref{Ap:upperBoundSum}. 
\end{proof}
%-------------
By applying the result of Lemma~\ref{lem:upperBoundXi}, we obtain
%---------------
\begin{align}
|\delta_1|  & \leq \sum_{m=q+1}^{N_{1,K}-1}m^2\frac{2d_1\eta}{\sigma\sqrt{\pi}} {\rm e}^{-\frac{d_1^2\eta^2(2m-1)^2}{4\sigma^2}} +  \frac{2d_1\eta q^2 }{\pi \sigma_x\sigma}  {\rm e}^{-\frac{(q-\mu_x)^2}{2\sigma_x^2} -\frac{d_1^2\eta^2}{4\sigma^2}}.
\end{align}
%---------------
Next, by applying Lemma~\ref{lem:upperBoundSum} to the last inequality, we obtain the following upper bound: 
%-----------------
\begin{align}
    \nonumber
    |\delta_1|  \leq \frac{4\sigma}{d_{1}\eta \sqrt{\pi}}
e^{-\tfrac{d_{1}^{2}\eta^{2}}{2\sigma^{2}}(2q - 1)^{2}}
\left( 2q - 1 + \frac{4\sigma^{2}}{d_{1}^{2}\eta^{2}(2q - 1)} \right) +  \frac{2d_1\eta q^2 }{\pi \sigma_x\sigma}  {\rm e}^{-\frac{(q-\mu_x)^2}{2\sigma_x^2} -\frac{d_1^2\eta^2}{2\sigma^2}} \leq A_1 {\rm e}^{-c_1(\frac{d_1q\eta}{\sigma})^2},
\end{align}
%-----------------
where $A_1= \mathcal{O}\!\left(\tfrac{d_1 q \eta}{\sigma}\right)$ and $c_1$ is a positive constant.  Similarly, it follows that
%-----------------
\begin{align}
    |\delta_2| \leq  A_2 {\rm e}^{-c_2(\frac{d_2n\eta}{\sigma})^2}, 
\end{align}
%-----------------
where $A_2 = \mathcal{O}\!\left(\tfrac{d_2 n \eta}{\sigma}\right)$ and $c_2 > 0$ is a constant.  Consequently, the overall approximation error satisfies
%-----------------
\begin{align}
    |\delta | \leq A_1 {\rm e}^{-c_1(\frac{d_1q\eta}{\sigma})^2}+ q^2A_2 {\rm e}^{-c_2(\frac{d_2n\eta}{\sigma})^2}.
\end{align}
%-----------------
Therefore, the proof of Proposition~\ref{Pr:MSE} is complete.

\subsection{Proof of Lemma~\ref{lem:upperBoundXi}}\label{Ap:lem:upperBoundXi}

By the definition of the \(Q\)-function, $\Xi(m)$ can be expressed as
%-----------------
\begin{align}
    \Xi(m) 
    = \int_{\tfrac{(2m-1)d_1 \eta}{\sqrt{2}\sigma}}^{\tfrac{(2m+1)d_1 \eta}{\sqrt{2}\sigma}} 
    \phi(t)\,dt, 
\end{align}
%-----------------
where $ \phi(x) =  {\rm e}^{-\frac{x^2}{2}}/{\sqrt{2\pi}}$.  Let the integration limits be 
\( a = \tfrac{(2m-1)d_1 \eta}{\sqrt{2}\sigma} \) 
and 
\( b = \tfrac{(2m+1)d_1 \eta}{\sqrt{2}\sigma} \).  
Since $\phi(t)$ is monotonically decreasing for $t \ge 0$, and $m \ge 1$ ensures $a \ge 0$, we have
%-----------------
\begin{align}
    \int_{a}^{b} \phi(t)\,dt 
    \leq \int_{a}^{b} \phi(a)\,dt 
    = (b - a)\,\phi(a).
\end{align}
%-----------------
Observing that \((b - a) = \tfrac{\sqrt{2} d_1 \eta}{\sigma}\) and substituting $a$ back into the expression yields
%--------------
\begin{align}
    \Xi(m) 
    \leq 
    \frac{\sqrt{2} d_1 \eta}{\sigma} 
    \, \phi\!\left(\frac{(2m - 1)d_1 \eta}{\sqrt{2}\sigma}\right)
    = \frac{ d_1 \eta}{\sqrt{\pi \sigma^2}} 
      e^{-\tfrac{d_1^2 \eta^2 (2m - 1)^2}{4\sigma^2}}.
\end{align}
%--------------
This completes the proof.

\subsection{Proof of Lemma~\ref{lem:upperBoundSum}}\label{Ap:upperBoundSum}

To simplify notation, let $\alpha = \tfrac{d_{1}^{2}\eta^{2}}{4\sigma^{2}}$.  
Since each term in the summation is positive, extending \(\sum_{m = q + 1}^{M - 1}\) to \(\sum_{m = q + 1}^{\infty}\) yields an upper bound:
%---------------
\begin{align}
    S 
    \leq 
    \frac{2 d_{1}\eta}{\sigma\sqrt{\pi}}
    \sum_{m = q + 1}^{\infty} 
    m^{2} e^{-\alpha (2m - 1)^{2}}.
\end{align}
%---------------
Next, we approximate the discrete sum by an integral.  
For a function \(f(x)\) that decreases for large \(x\), it holds that  
\(\sum_{m = q + 1}^{\infty} f(m) \leq \int_{q}^{\infty} f(x)\,dx\).  
Let \(f(x) = x^{2} e^{-\alpha (2x - 1)^{2}}\), then
%------------
\begin{align*}
    \sum_{m = q + 1}^{\infty} m^{2} e^{-\alpha (2m - 1)^{2}}
    \leq \int_{q}^{\infty} x^{2} e^{-\alpha (2x - 1)^{2}} dx.
\end{align*}
%------------
Applying the substitution \(y = 2x - 1\) gives \(x = \tfrac{y + 1}{2}\) and \(dx = \tfrac{1}{2} dy\).  
When \(x = q\), we have \(y = 2q - 1\).  
Hence,
\begin{align}
\int_{q}^{\infty} x^{2} e^{-\alpha (2x - 1)^{2}} dx
= \frac{1}{4}
\int_{2q - 1}^{\infty} (y + 1)^{2} e^{-\alpha y^{2}} dy \leq 
\int_{b}^{\infty} y^{2} e^{-\alpha y^{2}} dy,
\label{eq:upperintfirst}
\end{align}
where \(b = 2q - 1\), and the inequality follows from the fact that \((y + 1)^{2} \le 4y^{2}\) for \(y \ge 1\) (since \(b \ge 1\) for \(q \ge 1\)).   The remaining integral in~\eqref{eq:upperintfirst} can be evaluated via integration by parts:
%------------
\begin{align}
\label{eq:integraltermB}
\int_{b}^{\infty} y^{2} e^{-\alpha y^{2}} dy
= \frac{b}{2\alpha} e^{-\alpha b^{2}}
+ \frac{1}{2\alpha} \int_{b}^{\infty} e^{-\alpha x^{2}} dx.
\end{align}
%------------
The final integral in~\eqref{eq:integraltermB} can be upper bounded by a purely exponential term since \(q \ge 1\):
%------------
\begin{align}
\label{eq:expontialinteg}
\int_{b}^{\infty} e^{-\alpha x^{2}} dx
\leq 
\int_{b}^{\infty} e^{-\alpha b x}\,dx
= \frac{1}{\alpha b} e^{-\alpha b^{2}}.
\end{align}
%------------
Substituting~\eqref{eq:expontialinteg} into~\eqref{eq:integraltermB}, and then into~\eqref{eq:upperintfirst}, we obtain
%------------
\begin{align}
\int_{q}^{\infty} x^{2} e^{-\alpha (2x - 1)^{2}} dx
\leq 
\frac{e^{-\alpha b^{2}}}{2\alpha}
\left(b + \frac{1}{\alpha b}\right).
\label{eq:sumSintergalU}
\end{align}
%------------
Finally, invoking~\eqref{eq:sumSintergalU}, we derive the following upper bound on $S$:
%------------
\begin{align*}
S 
&\leq 
\frac{4\sigma}{d_{1}\eta \sqrt{\pi}}
e^{-\tfrac{d_{1}^{2}\eta^{2}}{2\sigma^{2}}(2q - 1)^{2}}
\left( 2q - 1 + \frac{4\sigma^{2}}{d_{1}^{2}\eta^{2}(2q - 1)} \right).
\end{align*}
%------------
This completes the proof.

%====================
\bibliographystyle{ieeetr}
\bibliography{IEEEabrv,Ref}

\end{document}